\documentclass[debug, journal]{rmaa}

\newcommand\rmxaa{RMAA}

\usepackage{paralist}

\usepackage{psfrag,color}

\usepackage[latin1]{inputenc}
\usepackage{afterpage,longtable,lscape}
\usepackage{rotating}
\usepackage{graphics}

\title{What Can We Learn About the Kinematics of Bright Extragalactic Planetary Nebulae?
} 

\author{
  M. G. Richer,\altaffilmark{1} 
  S.-H. Báez,\altaffilmark{2}
  J. A. López\altaffilmark{1}
  H. Riesgo\altaffilmark{1}
  Ma. T. García-Díaz\altaffilmark{1}}

\altaffiltext{1}{OAN, Instituto de Astronom\'\i a, Universidad Nacional Aut\'onoma de M\'exico, México.}

\altaffiltext{2}{Facultad de F\'\i sica e Inteligencia Artificial, Universidad Veracruzana, México.}

\shortauthor{Richer et al.}
\shorttitle{The Kinematics of Bright Extragalactic PNe}

\fulladdresses{
\item Sol-Haret Báez: Facultad de F\'\i sica e Inteligencia Artificial, Universidad Veracruzana, Circuito G. Aguirre Beltr\'an s/n, Zona Universitaria, Xalapa, Veracruz, M\'exico (solharet@hotmail.com).
\item María Teresa García-Díaz, José Alberto López, Michael G. Richer, and Hortensia Riesgo: Observatorio Astronómico Nacional, Instituto de Astronom\'\i a, Universidad Nacional Aut\'onoma de M\'exico, campus Ensenada, Apartado Postal 877, 22800 Ensenada, B. C., México (tere, jal, richer, hriesgo@astrosen.unam.mx).}

\listofauthors{M. G. Richer, S.-H. Báez, J. A. López, H. Riesgo, Ma. T. García-Díaz}
\indexauthor{Richer, M. G.}
\indexauthor{Báez, S.-H.}
\indexauthor{López, J. A.}
\indexauthor{Riesgo, H.}
\indexauthor{García-Díaz, Ma. T.}

\abstract{We present high resolution spectroscopy in the [\ion{O}{3}]$\lambda$5007 and H$\alpha$ lines of bright planetary nebulae in the Milky Way bulge and the dwarf galaxies M32, Fornax, Sagittarius, and NGC 6822 obtained at the Observatorio Astronómico Nacional in the Sierra San Pedro Mártir using the Manchester Echelle Spectrograph.  
We use the high signal-to-noise (S/N) observations of Milky Way bulge planetary nebulae to explore what kinematic information can be determined reliably when observing extragalactic planetary nebulae in the [\ion{O}{3}]$\lambda$5007 line at modest S/N.  We find that the intrinsic line widths measured in [\ion{O}{3}]$\lambda$5007 and H$\alpha$ are very similar.  Over the range of S/N available in this sample, the line width we measure is independent of the S/N.  Finally, deviations from a Gaussian line shape are small.  
Thus, the line width of the [\ion{O}{3}]$\lambda$5007 line in bright extragalactic planetary nebulae should reflect the kinematics of most of the mass in the ionized nebular shell.  
}

\resumen{
Presentamos espectroscopia de alta resoluci\'on en las líneas de [\ion{O}{3}]$\lambda$5007 y H$\alpha$ de nebulosas planetarias (NPs) brillantes en el bulbo de nuestra Vía Láctea as\'\i\ como las galaxias enanas M32, Fornax, Sagittarius y NGC 6822 obtenidas en el Observatorio Astronómico Nacional en la Sierra San Pedro Mártir con el espectrógrafo Manchester echelle.  
Utilizamos las observaciones profundas de las NPs galácticas para determinar la información disponible confiable en las observaciones mucho menos profundas de las NPs extragalácticas observadas en la línea de [\ion{O}{3}]$\lambda$5007.  Encontramos que la anchura intr\'\i nseca de las líneas en [\ion{O}{3}]$\lambda$5007 y H$\alpha$ son similares, que la anchura no depende del señal-a-ruido, dentro del intervalo cubierto por la muestra y que desviaciones con respecto a una forma gausiana son pequeñas.  
Concluimos que la anchura de la l\'\i nea de [\ion{O}{3}]$\lambda$5007 en NPs extragalácticas refleja de manera fiel la cinem\'atica de la mayor\'\i a de la masa de la cáscara ionizada.  
}

\addkeyword{Galaxy: Bulge}
\addkeyword{ISM: kinematics and dynamics}
\addkeyword{Planetary Nebulae}
\addkeyword{Stars: Evolution}

\begin{document}
\maketitle

\section{Introduction}\label{sec_introduction}

Past and present efforts to study the kinematics of galactic and extragalactic planetary nebulae have produced large and rapidly growing databases of these observations \citep[e.g.,][]{dopitaetal1985, dopitaetal1988, gesickizijlstra2000, medinaetal2006, arnaboldietal2008, lopezetal2009, richeretal2009}.  However, observations of galactic and extragalactic planetary nebulae often differ in fundamental ways.  Typically, galactic planetary nebulae are resolved by the (usually ground-based) spectrograph slit, whereas extragalactic planetary nebulae usually are not.  Furthermore, the kinematics are often measured from different emission lines in galactic (e.g., H$\alpha$ or [\ion{N}{2}]$\lambda$6584) and extragalactic (e.g., [\ion{O}{3}]$\lambda$5007) planetary nebulae.  It is not necessarily obvious, therefore, how to compare the kinematics of galactic and extragalactic planetary nebulae, so experiments that help us understand exactly what information is available when studying the kinematics of extragalactic planetary nebulae are worthwhile.

The existing literature \citep{dopitaetal1985, dopitaetal1988, zijlstraetal2006, arnaboldietal2008} as well as our own observations \citep{richeretal2009} indicate that the line profiles of extragalactic planetary nebulae are usually approximately Gaussian.  In almost all cases, these are spatially unresolved observations.  Obviously, the kinematic information available from such line profiles will be limited, even at high signal-to-noise \citep[S/N; e.g.,][]{morissetstasinska2008}.   A variety of studies exist of the effect of limited spatial resolution on kinematic studies using models as test cases \citep[e.g.,][]{gesickizijlstra2000, rozasetal2007}.  The experiments of \citet{morissetstasinska2008} demonstrate how different structures are visible in different emission lines and with different slit sizes or positions.  From their results, it is clear that smaller, precisely positioned slits allow the study of finer detail involving components of lower mass, a fact long exploited in observational studies \citep[e.g., see][for recent examples]{sabbadinetal2008, garciadiazetal2008, santandergarciaetal2008}.  None of these kinematic studies include the hydrodynamic effects that will occur in real nebulae, as do the models of \citet{villaveretal2002} and \citet{perinottoetal2004}.  The most detailed hydrodynamical study is perhaps that of \citet{schonberneretal2005} who emphasize the important differences between the motions of matter and shock fronts (see \citet{corradietal2007} for an application to observations).   

While instructive, these studies require particular assumptions for the construction of the models and it is not always clear how closely they match real planetary nebulae.  Furthermore, there is often a focus on extracting fine details rather than studying the bulk motion of the majority of the mass.  Considering the aforementioned experiments, studies of the line profiles of extragalactic planetary nebulae \citep{dopitaetal1985, dopitaetal1988, zijlstraetal2006, arnaboldietal2008} will most profitably focus upon the bulk motion of matter in their ionized shells, since the lack of spatial resolution and limited signal-to-noise (S/N) will render small-scale features difficult to discern, even if they have distinctive kinematics.  

At the modest S/N that observations achieve, two issues arise: the recovery of the available information and the interpretation of this information in terms of the kinematics.  Here, we focus on the former, investigating the empirical description of the line profile, rather than on the interpretation of the kinematic information that may be derived from it.  We aim to determine whether the information available may be recovered reliably and whether this information pertains to the entire nebular shell or to some small fraction of it.  Three aspects are most important.  First, extragalactic planetary nebulae will usually be studied in the [\ion{O}{3}]$\lambda$5007 line, since it is the brightest line in the optical spectrum (and often in the entire spectrum).  How well does this line probe the kinematics of the entire ionized mass?  Second, the faintness of the [\ion{O}{3}]$\lambda$5007 emission in extragalactic planetary nebulae means that the S/N will be modest.  How does the limited S/N affect the derived kinematics (e.g., the line width)?  Third, the modest S/N will often limit the information available to model the line profile to a simple function, such as a Gaussian \citep[e.g.,][]{dopitaetal1985, dopitaetal1988, arnaboldietal2008}.  To what extent does such a simple empirical description adequately reflect the kinematics of the ionized mass?

Here, we perform an experiment using observations of bright planetary nebulae in the Milky Way bulge (henceforth, Bulge).  Our intent is to use these observations to infer the limitations inherent to kinematic observations of extragalactic planetary nebulae.  We have chosen our sample of Bulge planetary nebulae in a way that we hope simulates populations of bright extragalactic planetary nebulae in environments without star formation (\S \ref{sec_obs_red}).  We obtain high resolution spectra in both the H$\alpha$ and [\ion{O}{3}]$\lambda$5007 lines.  We analyse the resulting data in the same way we would analyse those for extragalactic planetary nebulae (\S \ref{sec_analysis}).  In particular, we use our high S/N spectra to simulate extragalactic observations, normalizing the [\ion{O}{3}]$\lambda$5007 spectra to the total fluxes typical of extragalactic observations and adding noise, analyse these synthetic spectra, and compare the results to the original observations.  We then consider the three questions posed above (\S \ref{sec_results}).  We find that the [\ion{O}{3}]$\lambda$5007 line widths are similar to the H$\alpha$ line widths, that the observed [\ion{O}{3}]$\lambda$5007 line widths are not a function of S/N, at least for the S/N levels typical of spectra of extragalactic planetary nebulae, and that the line width is an adequate description of most of the emission observed, and so representative of the great majority of the ionized mass (\S \ref{sec_discussion}).  Hence, we conclude that the kinematics of the ionized shells of extragalactic planetary nebulae may be studied reliably using the [\ion{O}{3}]$\lambda$5007 line (\S \ref{sec_conclusions}).

\section{Observations and Reductions}\label{sec_obs_red}

Our sample of planetary nebulae in the Milky Way bulge is given in Tables \ref{table_kin_o3}-\ref{table_kin_ha} and was selected to simulate the properties of bright extragalactic planetary nebulae in environments without star formation.  As discussed in more detail in \citet{richeretal2008}, our selection criteria quickly converged to require the planetary nebulae (a) lie within $10^{\circ}$ of the galactic centre, (b) have a large reddening-corrected H$\beta$ flux, nominally $\log I(\mathrm H\beta) > -12.0$\,dex, (c) have a large [\ion{O}{3}]$\lambda 5007/\mathrm H\beta$ ratio, normally exceeding a value of 6, and (d) have been observed spectroscopically at low resolution with a detection of the [\ion{O}{3}]$\lambda$4363 line.  Our Bulge sample contains 86 objects.

We acquired our observations of Bulge planetary nebulae during eight observing runs spanning the period from 2003 June to 2007 August at the Observatorio Astron\'omico Nacional in the Sierra San Pedro M\'artir, Baja California, Mexico (OAN-SPM).  All of these objects were selected from existing spectroscopic surveys \citep{allerkeyes1987, webster1988, cuisinieretal1996, ratagetal1997, cuisinieretal2000, escuderocosta2001, escuderoetal2004, exteretal2004, gornyetal2004}.  Some of our observations of extragalactic planetary nebulae were acquired during the same runs, but also during three additional observing runs in 2001 September, 2002 July, and 2004 November.  More details of the observations will be provided elsewhere \citep{lopezetal2009, richeretal2009}.  

\begin{figure*}[!t]
\begin{center}
  \includegraphics[width=2\columnwidth]{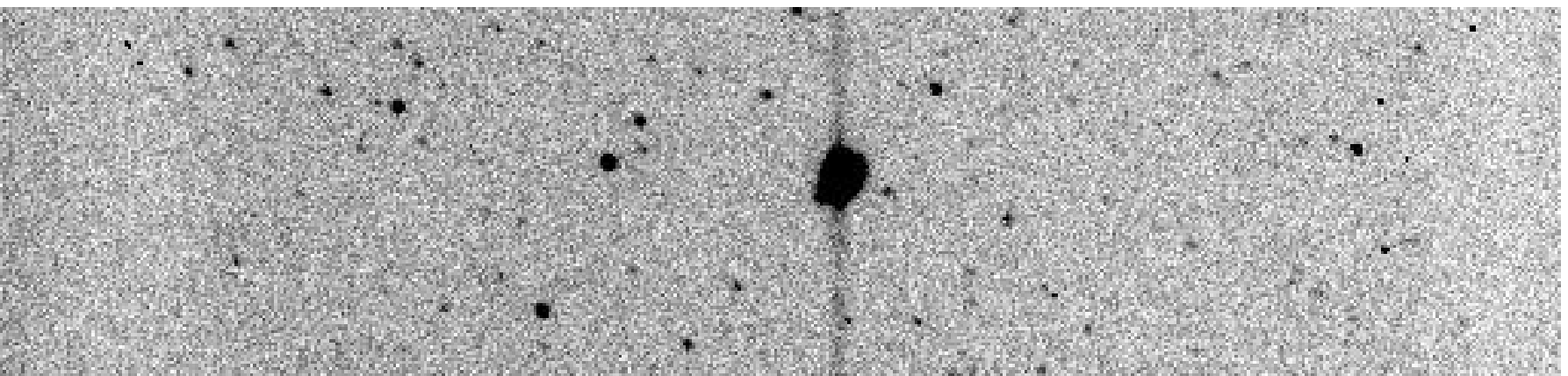}
  \includegraphics[width=2\columnwidth]{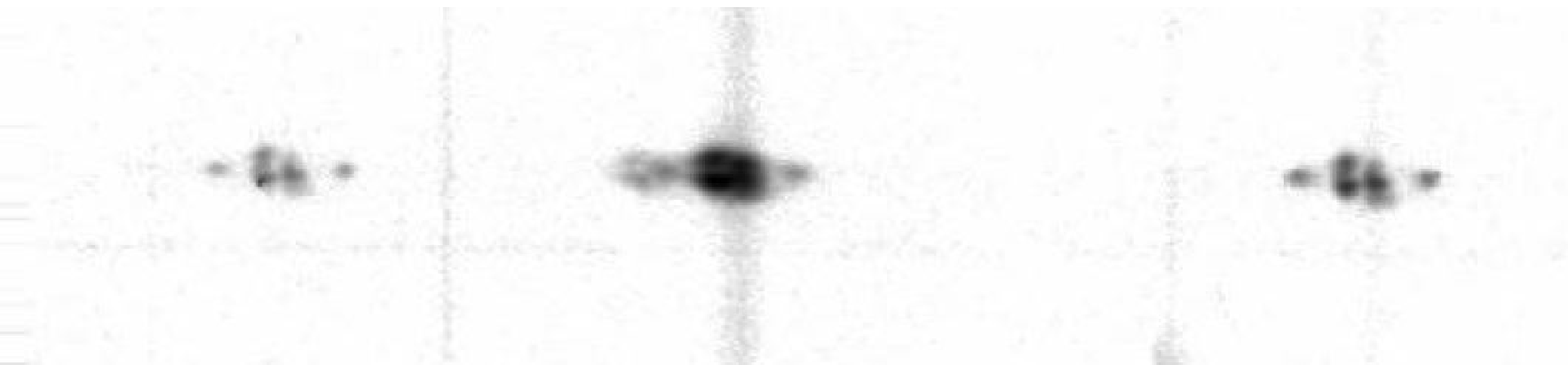}
  \includegraphics[width=0.75\columnwidth,angle=90]{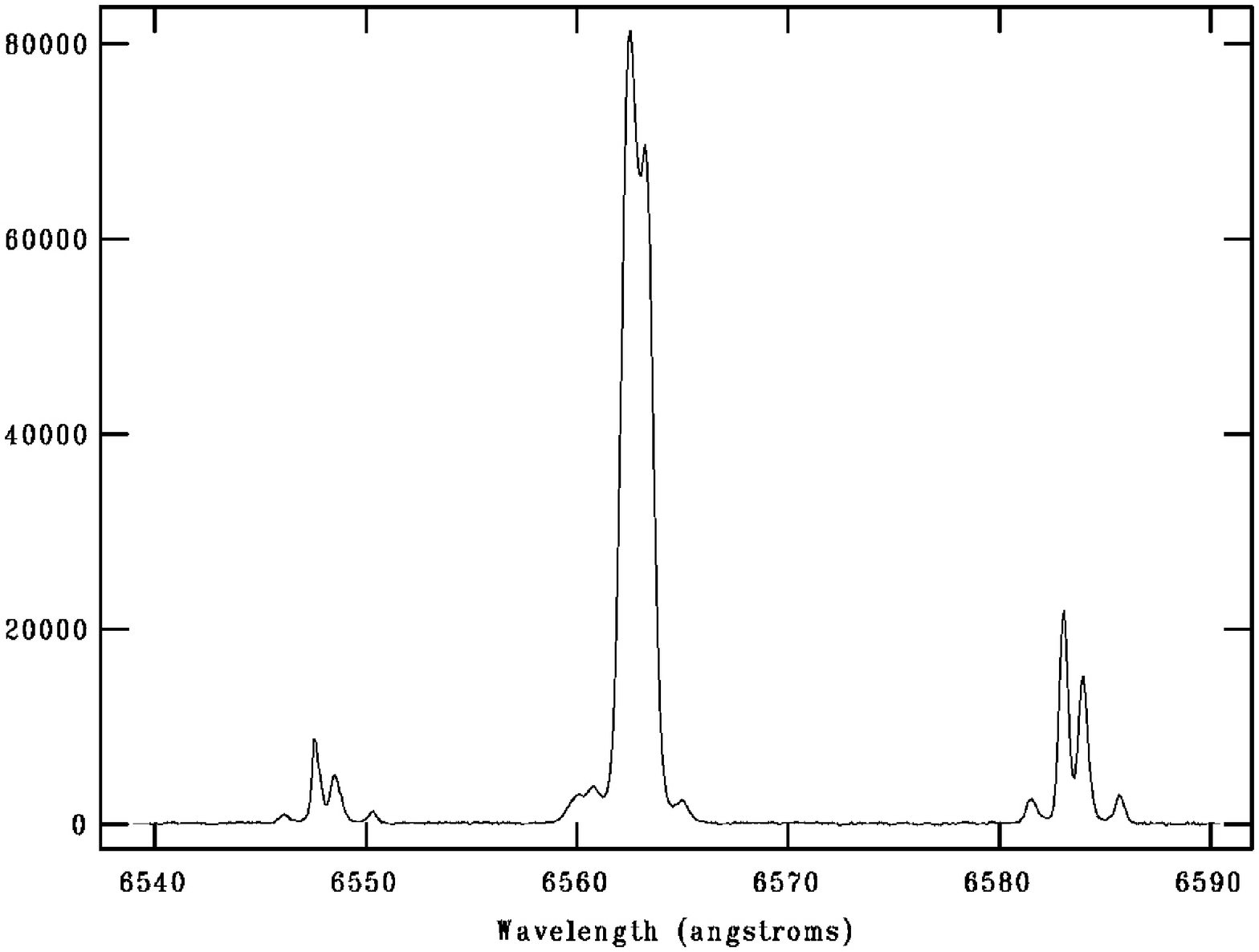}
  \includegraphics[width=0.75\columnwidth,angle=90]{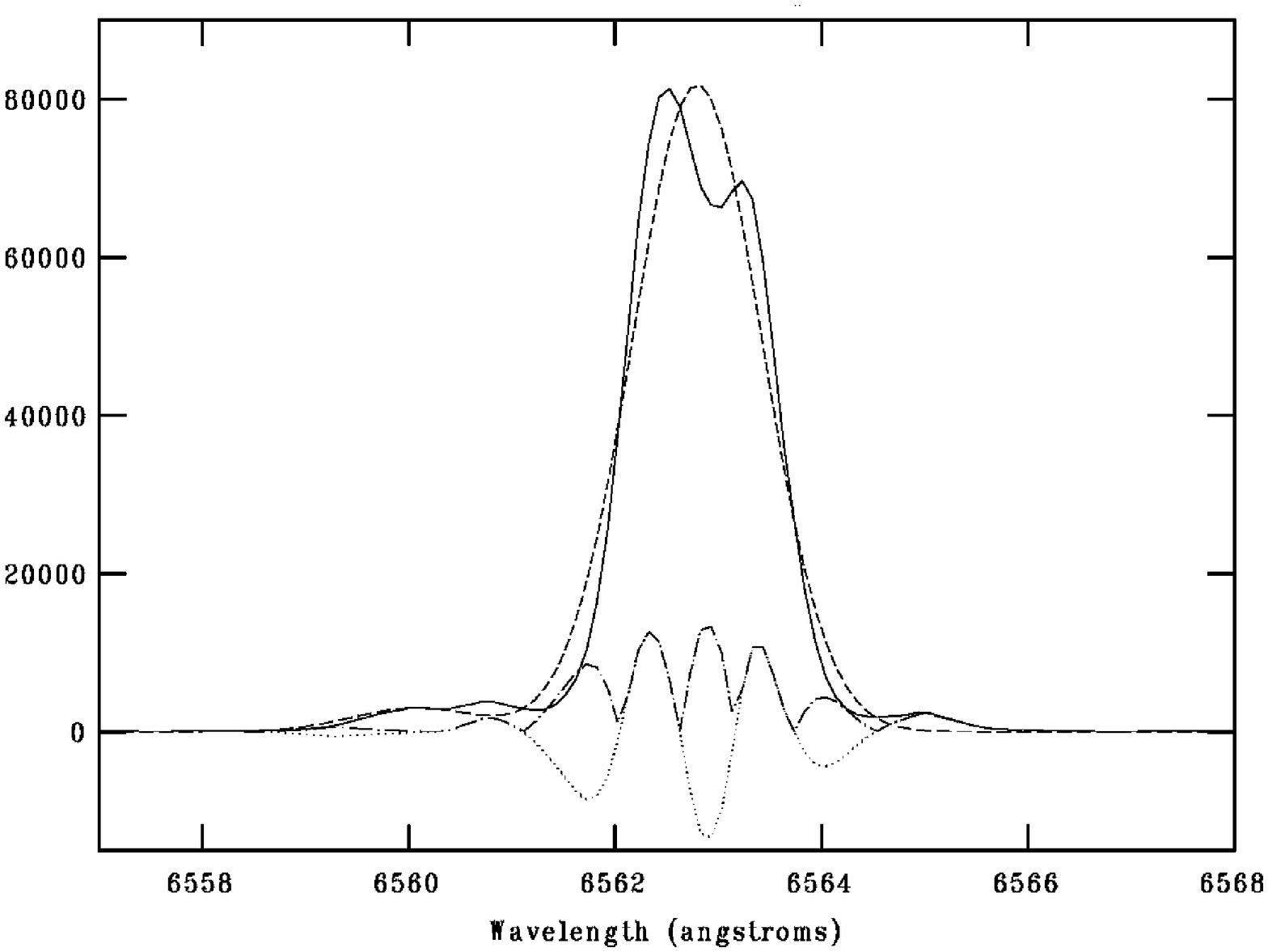}
  \end{center}
  \caption{We illustrate the analysis of our spectra using H 1-67 as an example.  The top image is the acquisition image, showing the slit superposed upon the object (north is up; east to the left).  The middle image is the rectified two-dimensional H$\alpha$ spectrum (north is up; blue to the left).  The structure of the H$\alpha$ line differs from that in the [\ion{N}{2}]$\lambda\lambda$6548,6583 lines because it is blended with \ion{He}{2}$\lambda$6560.  At lower left is the extracted, wavelength-calibrated, spatially unresolved, one-dimensional spectrum.  Finally, at lower right is the extracted spectrum (solid line), the fit of a single Gaussian component to the H$\alpha$ and \ion{He}{2}$\lambda$6560 lines (dashed line), the difference between the extracted spectrum and the fit (lower dotted line), and the absolute value of this difference (lower dot-dashed line).  The residual flux about the Gaussian fit is the sum of the flux under the dot-dashed line, and, for H 1-67, represents 17\% of the flux in the Gaussian component fit to the H$\alpha$ line.  For both 1-D spectra, the ordinate is in units of counts.}
  \label{fig_example_analysis}
\end{figure*}

High resolution spectra were obtained with the Manchester echelle spectrometer \citep[MES-SPM; ][]{meaburnetal1984, meaburnetal2003}.  The MES-SPM is a long slit echelle spectrometer, but uses narrow-band filters, instead of a cross-disperser, to isolate the orders containing emission lines of interest.  In this case, filters isolated orders 87 and 114 containing the H$\alpha$ and [\ion{O}{3}]$\lambda 5007$ emission lines, respectively.  All observations used a 150\,$\mu$m wide slit, equivalent to $1\farcs9$ on the sky.  Coupled with a SITe $1024\times 1024$ CCD with 24\,$\mu$m pixels binned $2\times 2$, the resulting spectral resolutions were approximately 0.077\,\AA/pix and 0.100\,\AA/pix at [\ion{O}{3}]$\lambda 5007$ and H$\alpha$, respectively (equivalent to 11\,km/s for 2.6\,pix FWHM).  Immediately before or after every object spectrum, exposures of a ThAr lamp were taken to calibrate in wavelength.  The internal precision of the arc lamp calibrations is better than $\pm 1.0$\,km/s. 

Typically, three spectra were obtained of each Bulge planetary nebula.  The first spectrum was a short exposure in [\ion{O}{3}]$\lambda 5007$ with a duration of 60-180s.  This was followed by a deep [\ion{O}{3}]$\lambda 5007$ spectrum, up to a maximum of 30 minutes, but chosen so as to avoid saturation.  The last spectrum was a deep H$\alpha$ spectrum, whose exposure time was chosen to achieve a S/N similar to that of the deep [\ion{O}{3}]$\lambda 5007$ spectrum, though it was also limited to a maximum of 30 minutes duration.  The purpose of the short [\ion{O}{3}]$\lambda 5007$ spectrum was to attempt to simulate the S/N in typical spectra of extragalactic planetary nebulae.  The deep [\ion{O}{3}]$\lambda 5007$ and H$\alpha$ spectra were obtained so as to detect kinematic details that are unobservable in typical spectra of extragalactic planetary nebulae.  For the extragalactic planetary nebulae, all of the spectra were of 30 minutes duration and, depending upon the S/N, one or two spectra were obtained.  It was not always possible to obtain multiple spectra during a single pointing.  

All of the Bulge planetary nebulae are resolved \citep{richeretal2008}.  In all cases, we attempted to center the slit on the object as carefully as possible (see the top image in Fig. \ref{fig_example_analysis}).  Normally, all of the spectra for a given object were obtained sequentially, which should help minimize positional mismatches.  For the extragalactic planetary nebulae, the slit was always oriented in the north-south direction, as it was for the vast majority of Bulge planetary nebulae.  

All of the spectra were reduced using the twodspec and specred packages of the Image Reduction and Analysis Facility\footnote{IRAF is distributed by the National Optical Astronomical Observatories, which is operated by the Associated Universities for Research in Astronomy, Inc., under contract to the National Science Foundation.} (IRAF).  For the Bulge objects, the data reduction followed the scheme recommended by \citet[][Appendix B]{masseyetal1992} for long slit spectroscopy.  We edited each spectrum of cosmic rays.  Then, we subtracted a nightly mean bias image from each object spectrum.  Next, we mapped positions of constant wavelength using the arc lamp spectra.  We then rectified the object spectra so that lines of constant wavelength fell exactly along the columns, a process that simultaneously applied a wavelength calibration (see the two-dimensional spectrum in Fig. \ref{fig_example_analysis}).  Finally, we extracted wavelength-calibrated, one-dimensional spectra for each object (see the one-dimensional spectrum in Fig. \ref{fig_example_analysis}).  We did not calibrate in flux.  

For the extragalactic planetary nebulae, the data reduction followed that outlined above for cosmic rays and bias.  We then extracted the source spectra and used these apertures to extract ThAr spectra from the lamp spectra.  The latter were used to calibrate in wavelength.  If two spectra were obtained, they were co-added after being calibrated in wavelength.  If they did not coincide exactly in wavelength, they were shifted to a common wavelength solution and then co-added.  Again, we did not calibrate in flux.  

\begin{figure*}[!t]
\begin{center}
  \includegraphics[height=\columnwidth,angle=90,bb=100 84 510 662,clip]{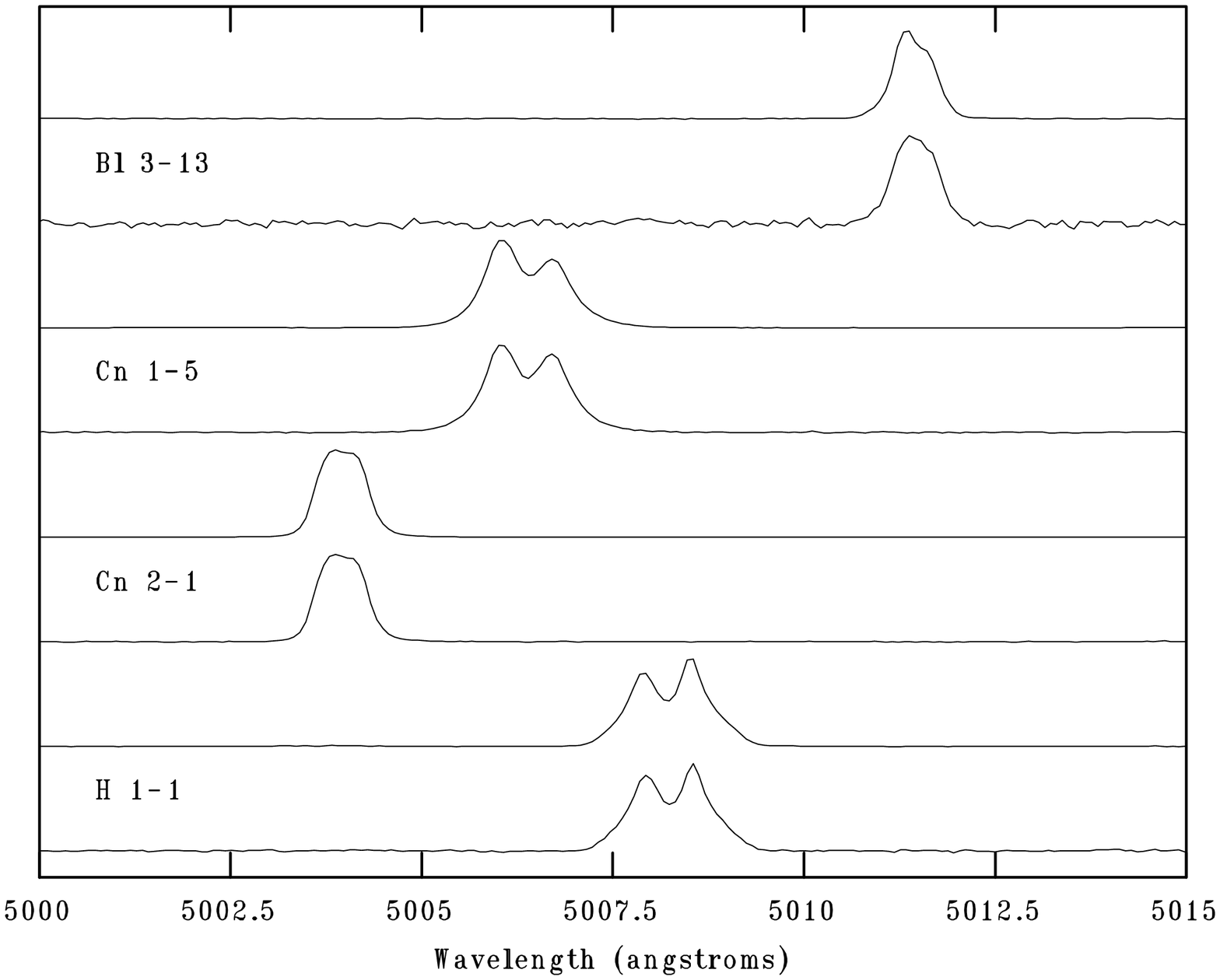}\quad\quad
  \includegraphics[height=\columnwidth,angle=90,bb=76 84 487 662,clip]{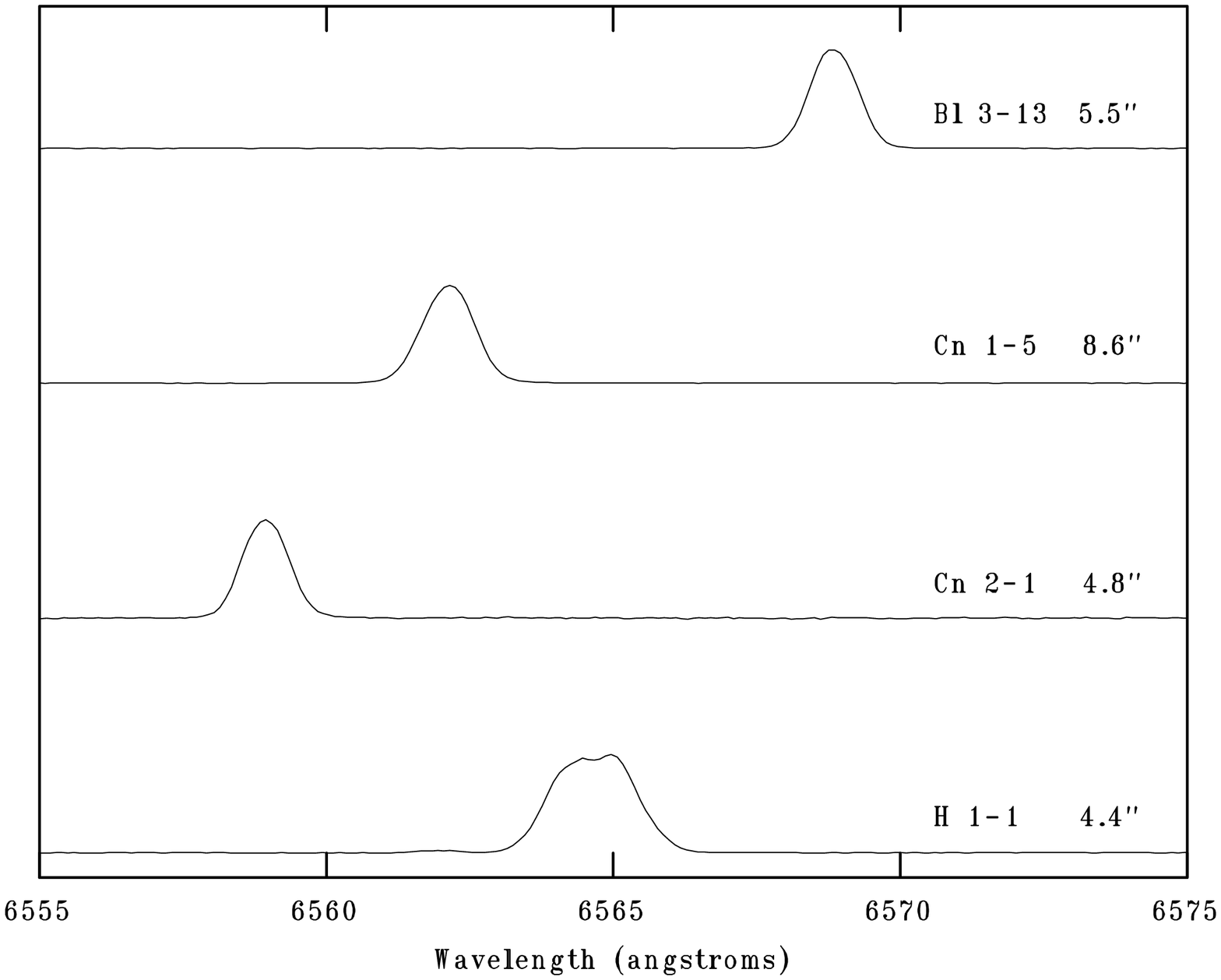}\\
  \includegraphics[height=\columnwidth,angle=90,bb=100 84 510 662,clip]{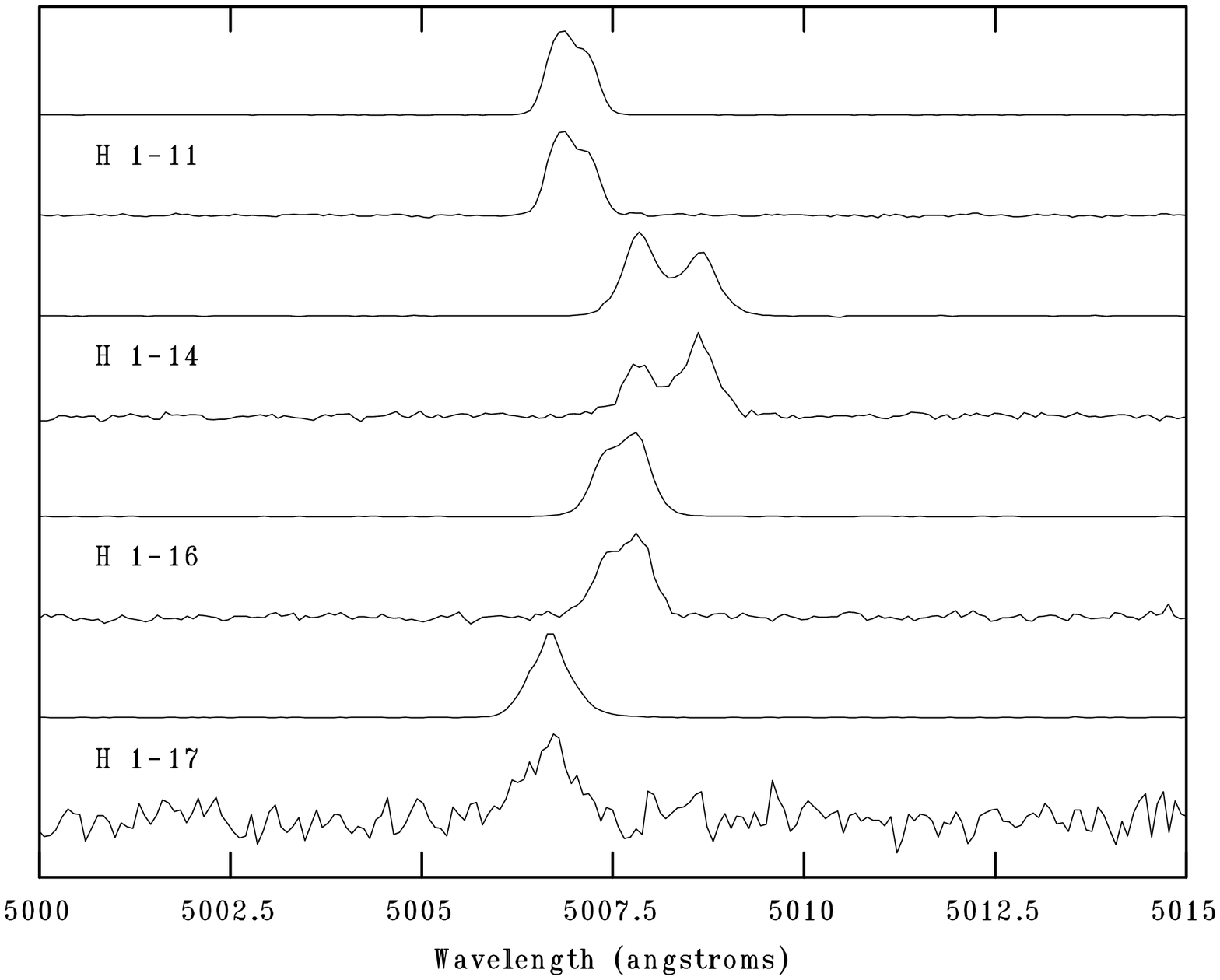}\quad\quad
  \includegraphics[height=\columnwidth,angle=90,bb=76 84 487 662,clip]{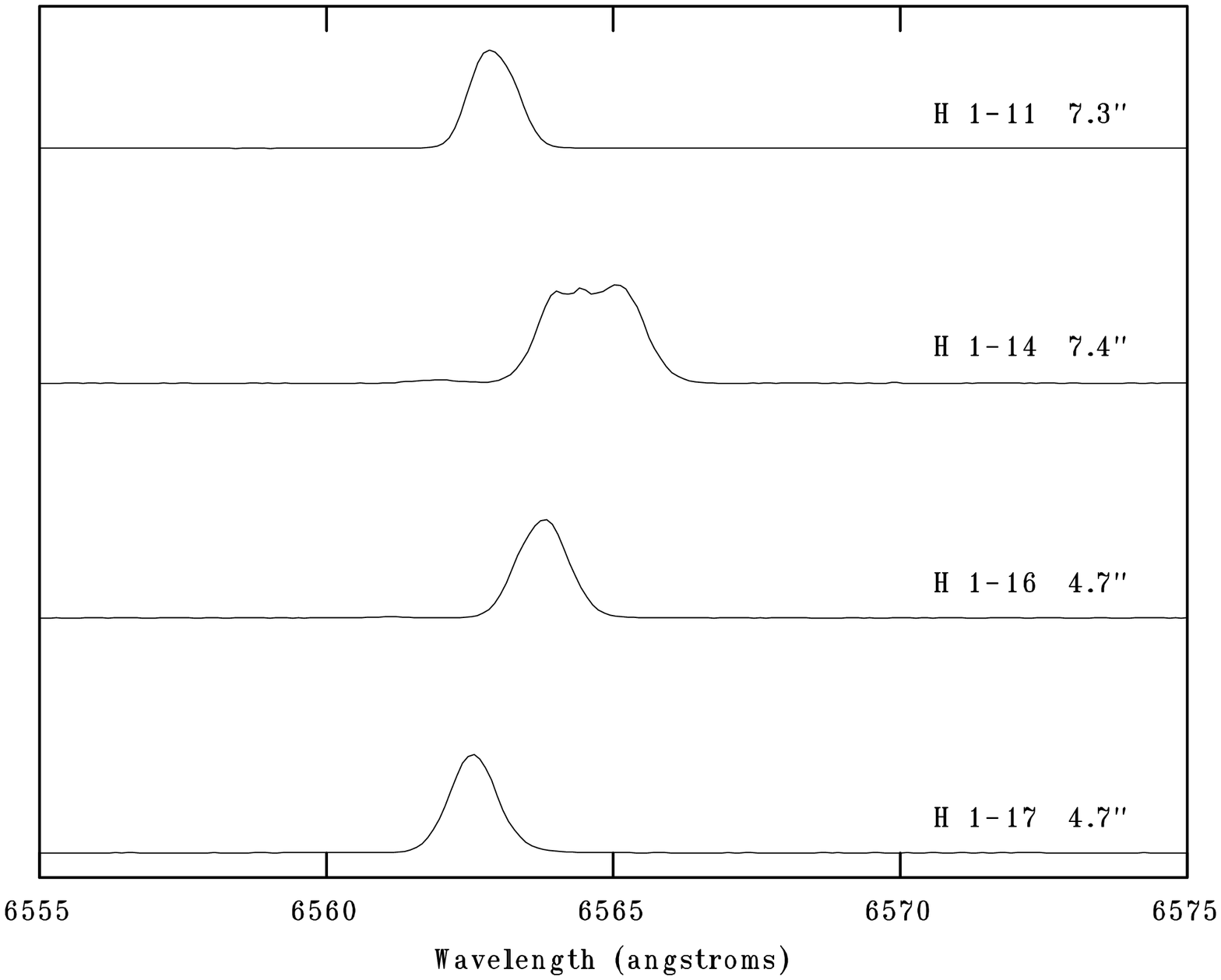}\\
  \includegraphics[height=\columnwidth,angle=90,bb=48 84 510 662,clip]{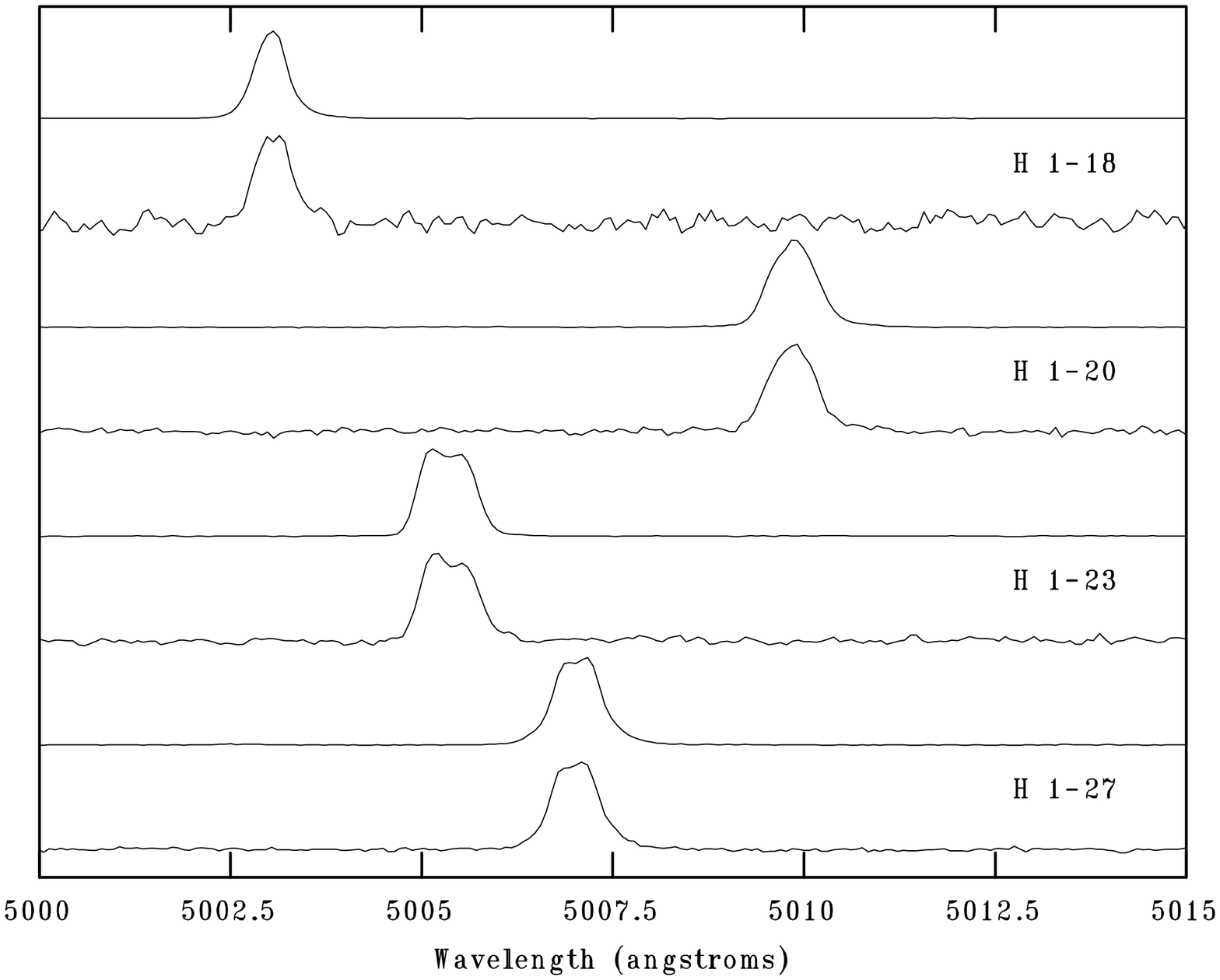}\quad\quad
  \includegraphics[height=\columnwidth,angle=90,bb=25 84 487 662,clip]{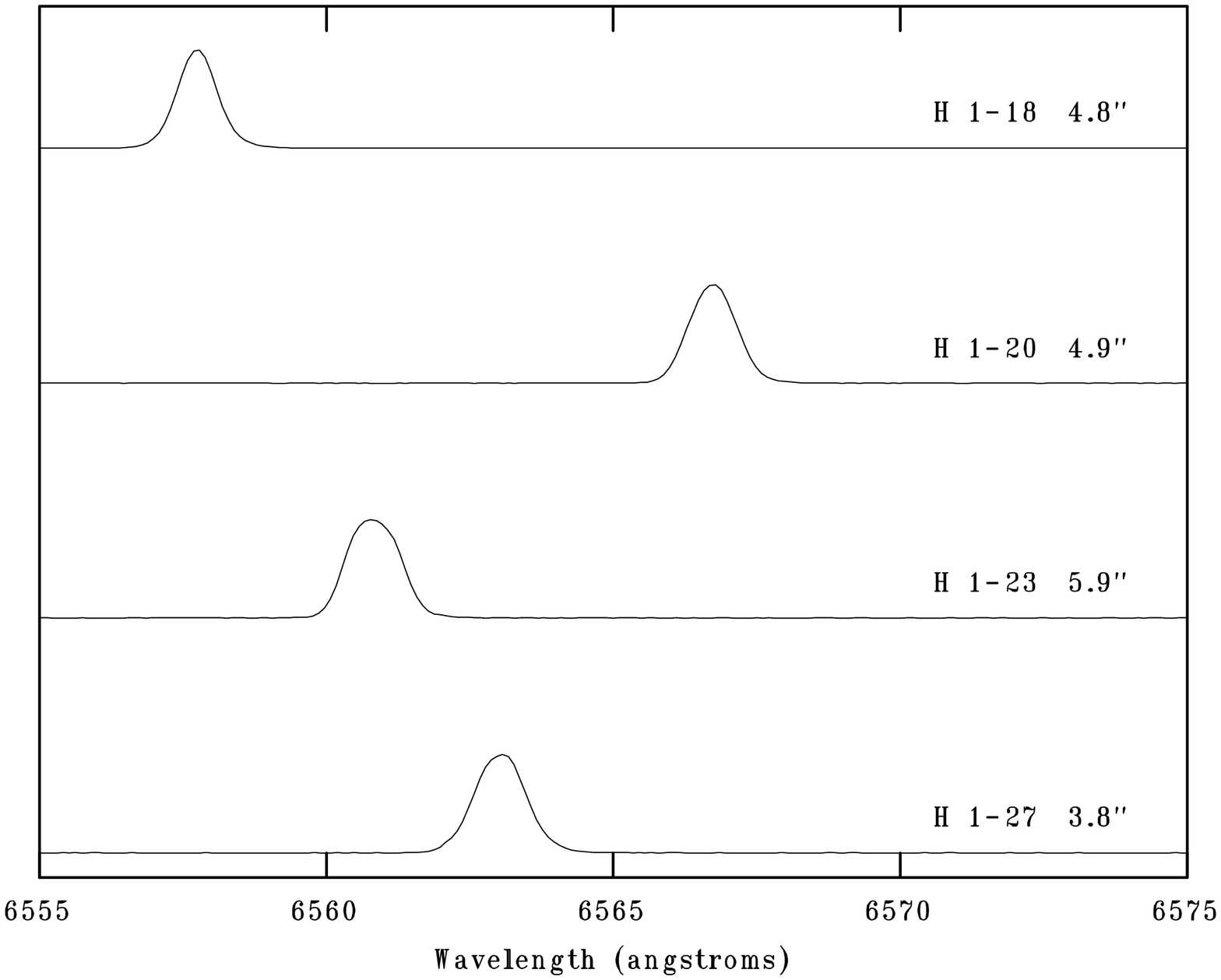}\\
  \end{center}
  \caption{We present the line profiles for our sample of planetary nebulae.  When both deep and shallow [\ion{O}{3}]$\lambda 5007$ spectra are available, both are shown (the shallow spectrum below the deep one).  Note how similar the two [\ion{O}{3}]$\lambda 5007$ spectra are.  At this scaling, the \ion{He}{2}\,$\lambda$6560 line is not easily visible in the H$\alpha$ spectra.  The wavelength scales are common in each column.  The H$\alpha$ panels include the diameter at 10\% of peak intensity from \citet{richeretal2008}.}
  \label{fig_lp_page1}
\end{figure*}


\begin{figure*}[!t]
\begin{center}
  \includegraphics[height=\columnwidth,angle=90,bb=100 84 510 662,clip]{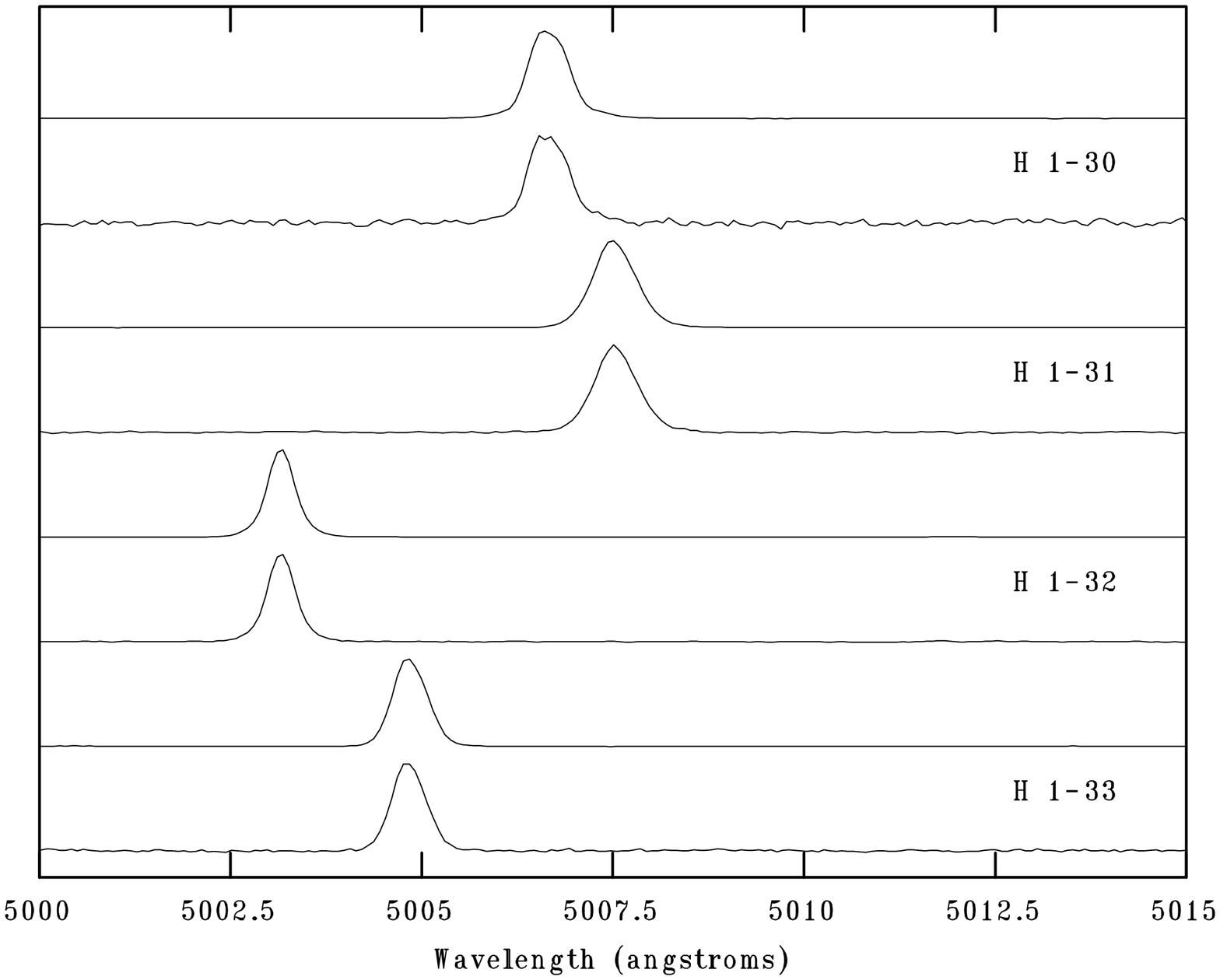}\quad\quad
  \includegraphics[height=\columnwidth,angle=90,bb=76 84 487 662,clip]{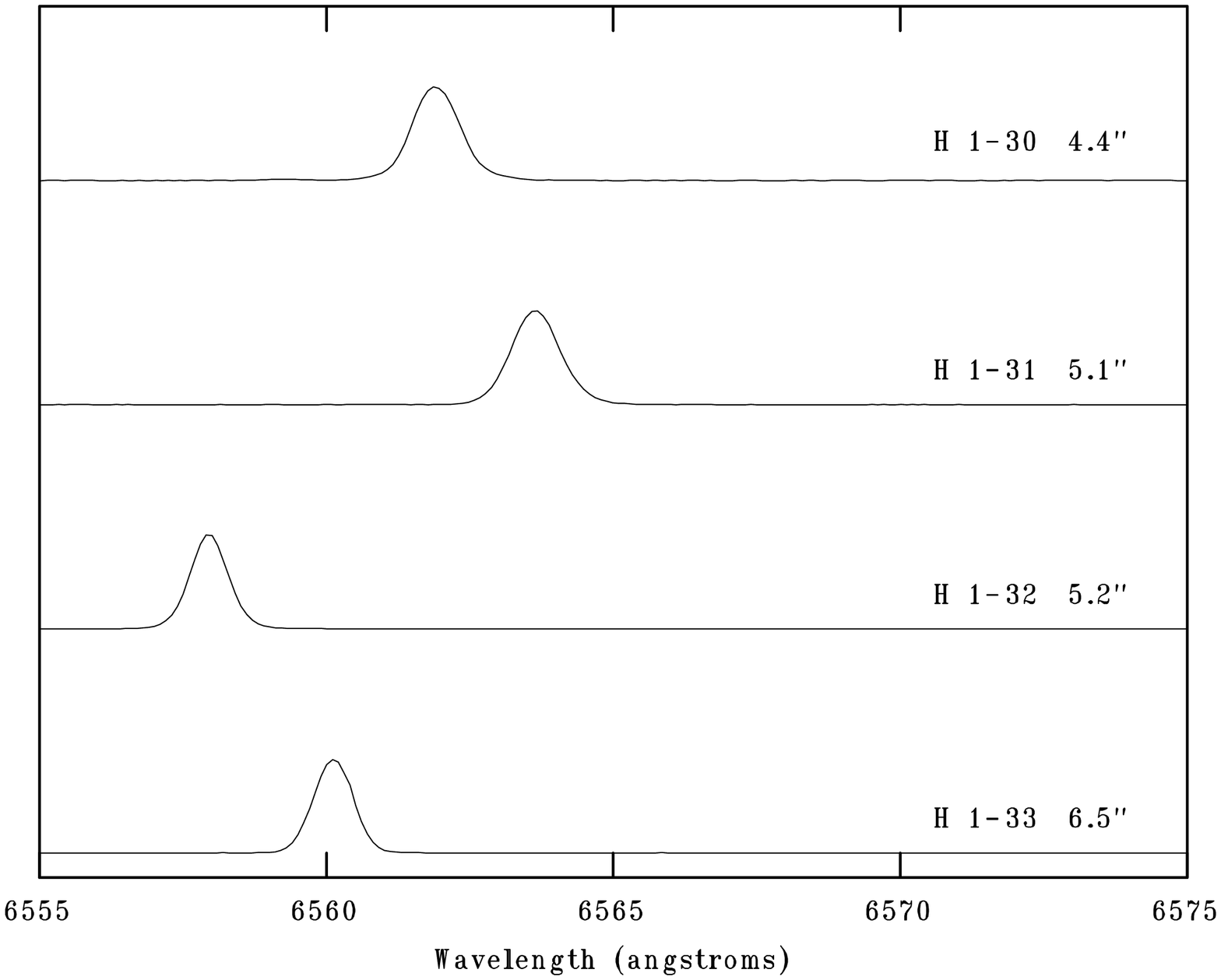}\\
  \includegraphics[height=\columnwidth,angle=90,bb=100 84 510 662,clip]{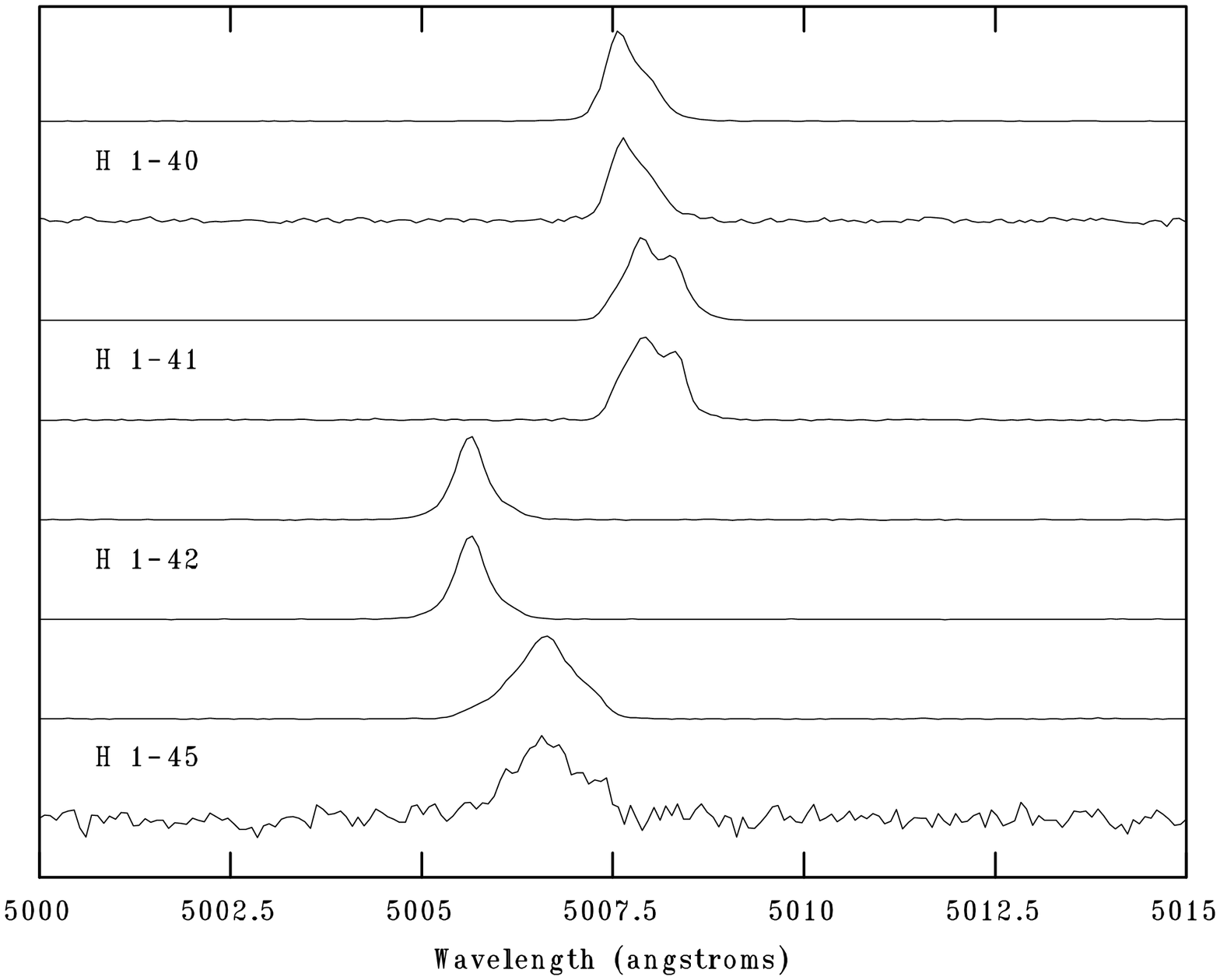}\quad\quad
  \includegraphics[height=\columnwidth,angle=90,bb=76 84 487 662,clip]{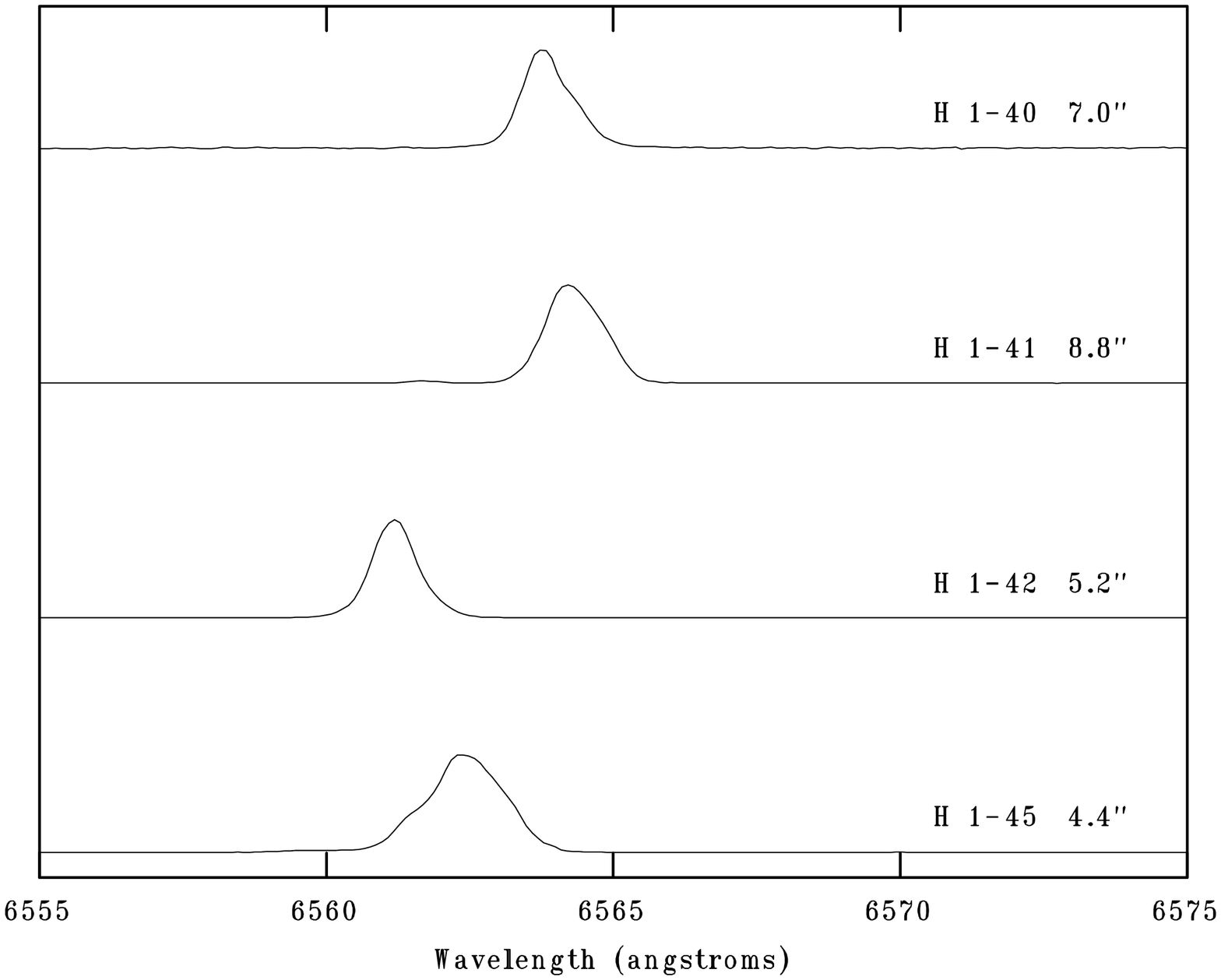}\\
  \includegraphics[height=\columnwidth,angle=90,bb=48 84 510 662,clip]{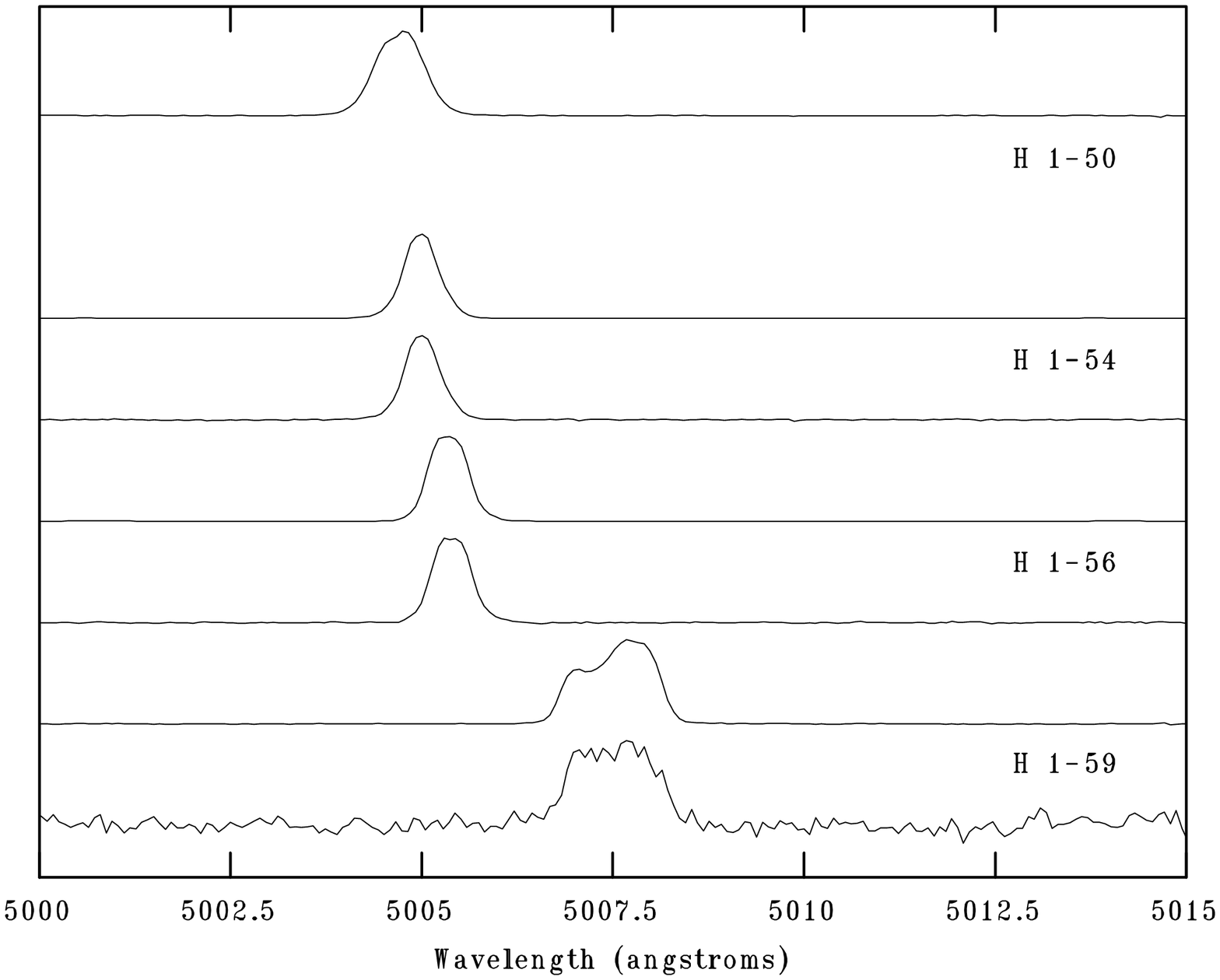}\quad\quad
  \includegraphics[height=\columnwidth,angle=90,bb=25 84 487 662,clip]{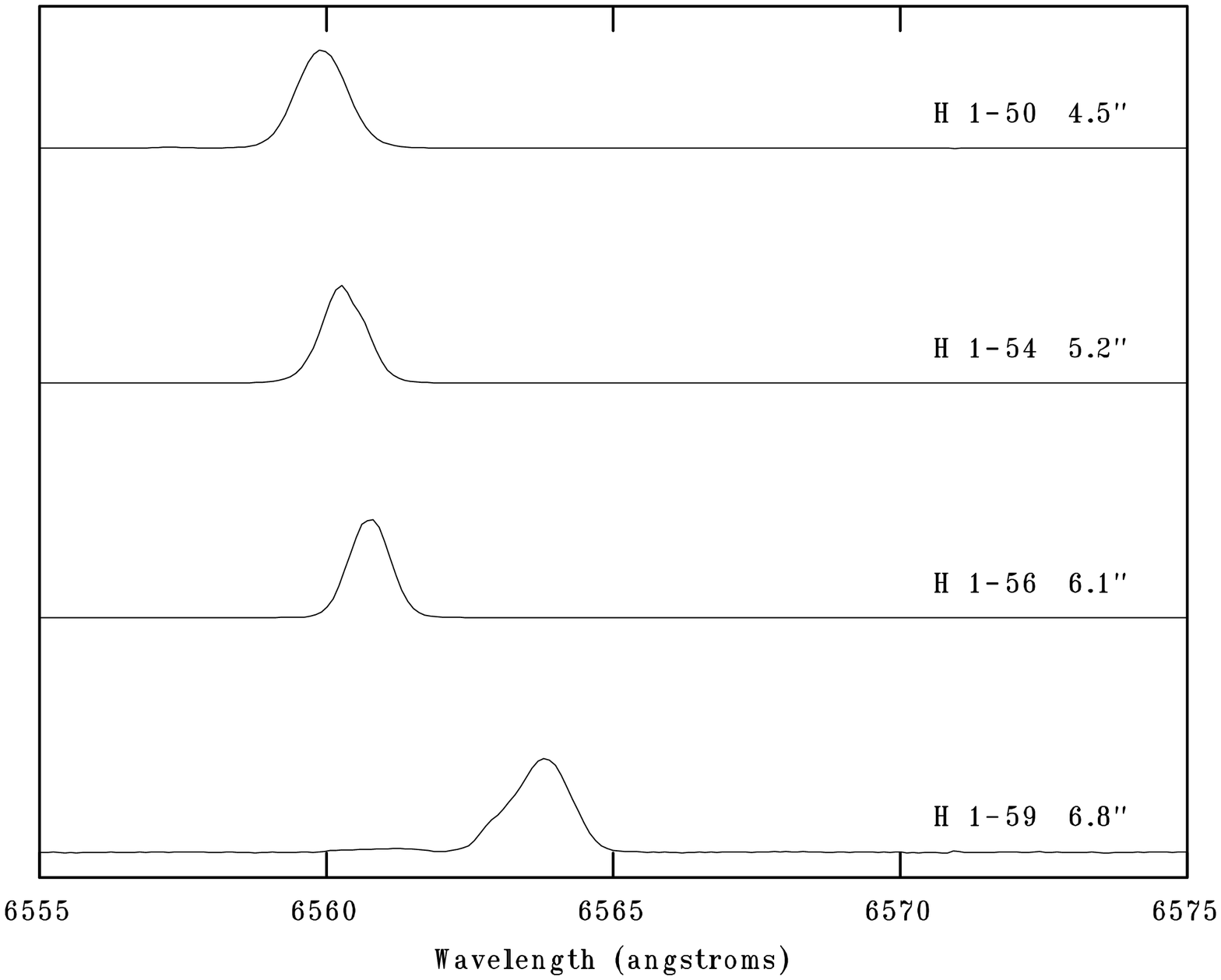}\\
  \end{center}
  \caption{For H 1-50, only a deep [\ion{O}{3}]$\lambda 5007$ spectrum was obtained.  See Fig. \ref{fig_lp_page1} for further details.}
  \label{fig_lp_page2}
\end{figure*}

\begin{figure*}[!t]
\begin{center}
  \includegraphics[height=\columnwidth,angle=90,bb=100 84 510 662,clip]{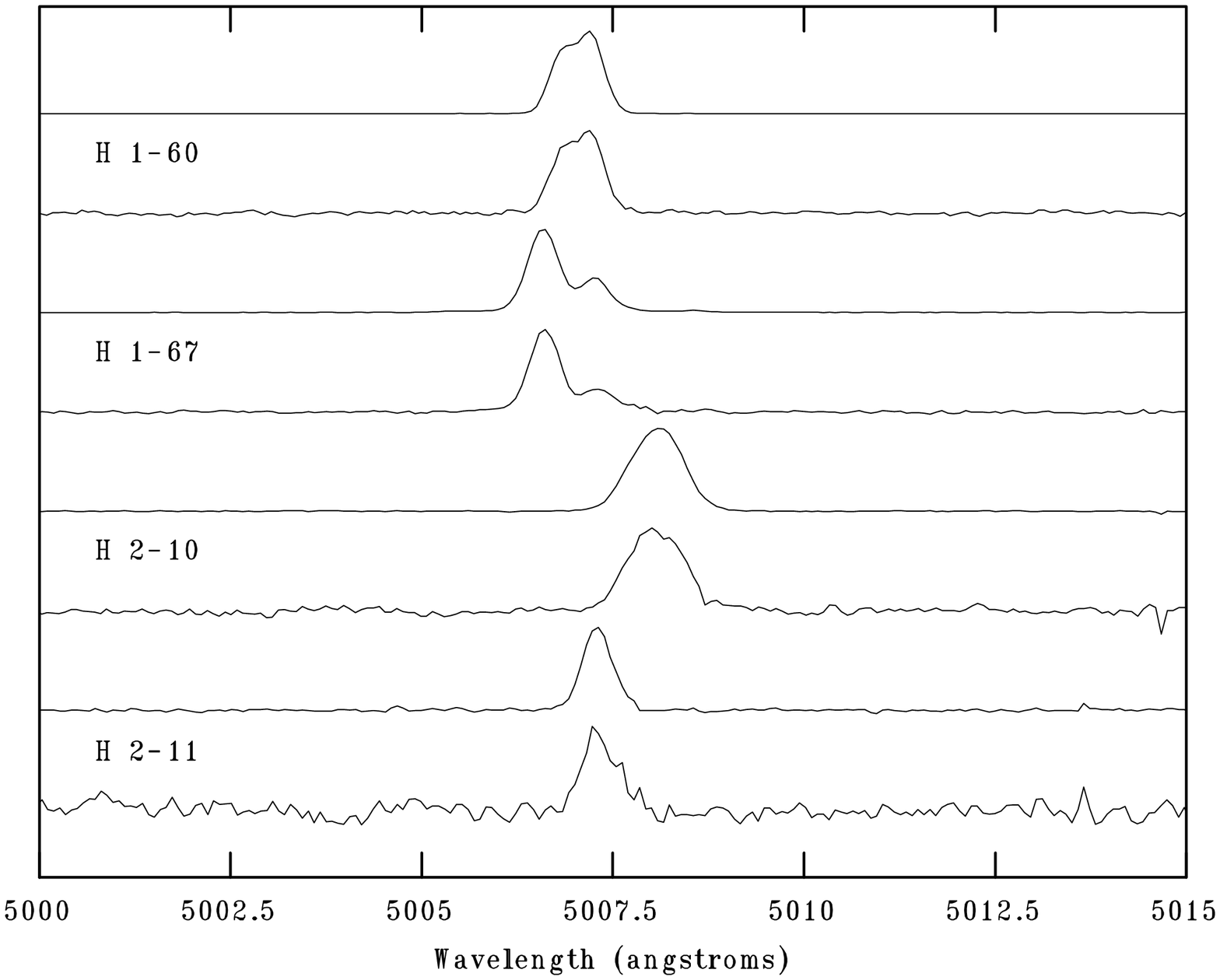}\quad\quad
  \includegraphics[height=\columnwidth,angle=90,bb=76 84 487 662,clip]{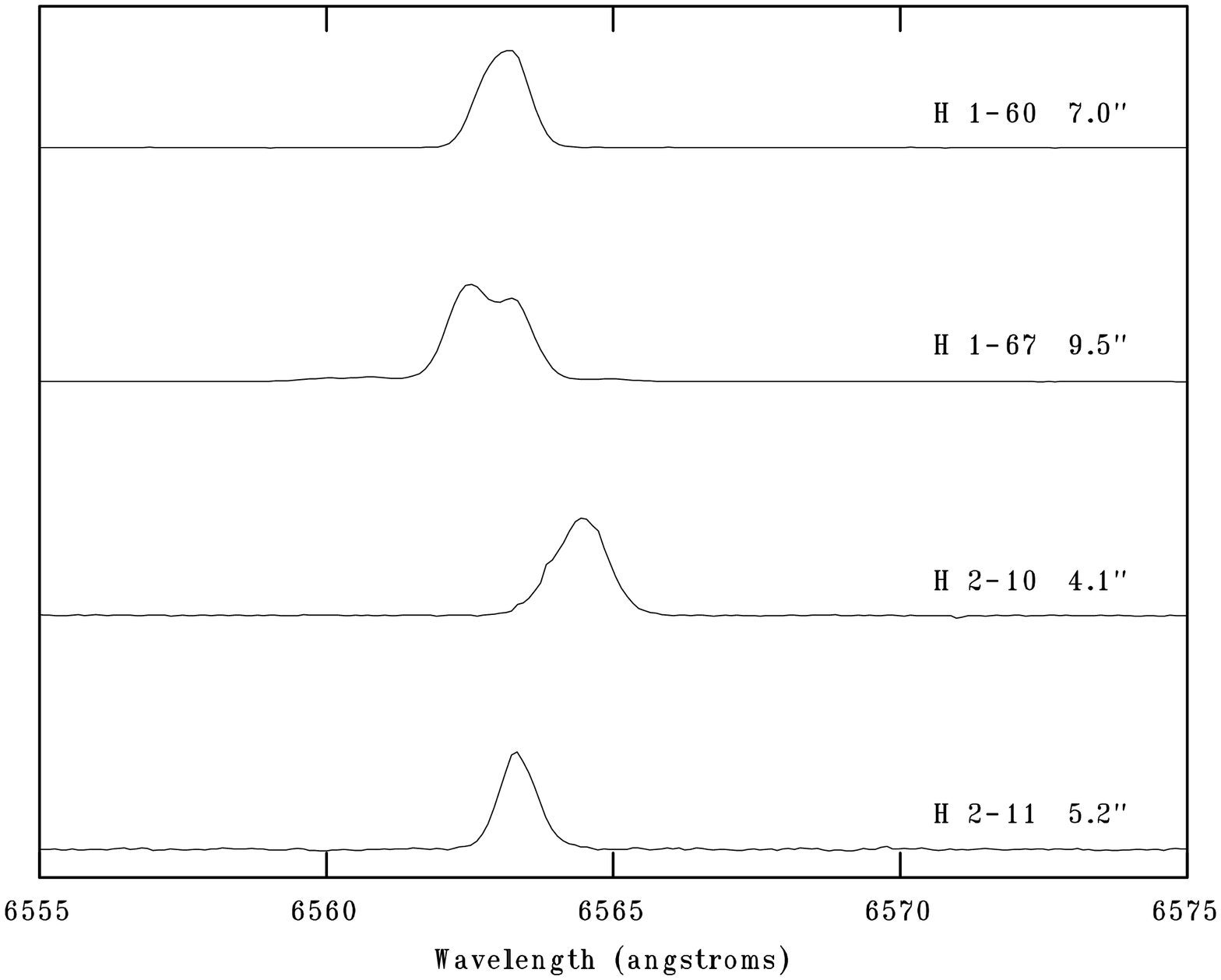}\\
  \includegraphics[height=\columnwidth,angle=90,bb=100 84 510 662,clip]{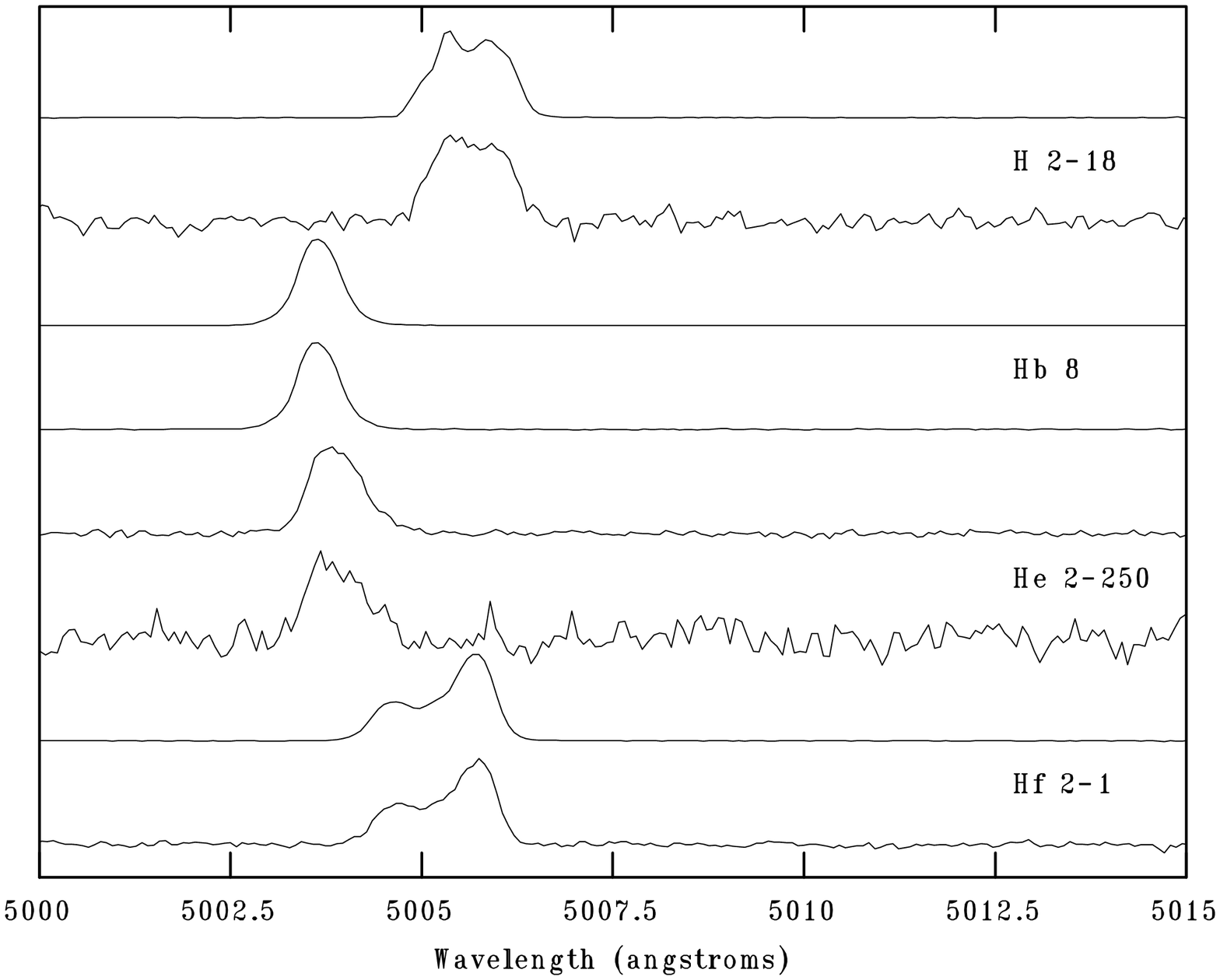}\quad\quad
  \includegraphics[height=\columnwidth,angle=90,bb=76 84 487 662,clip]{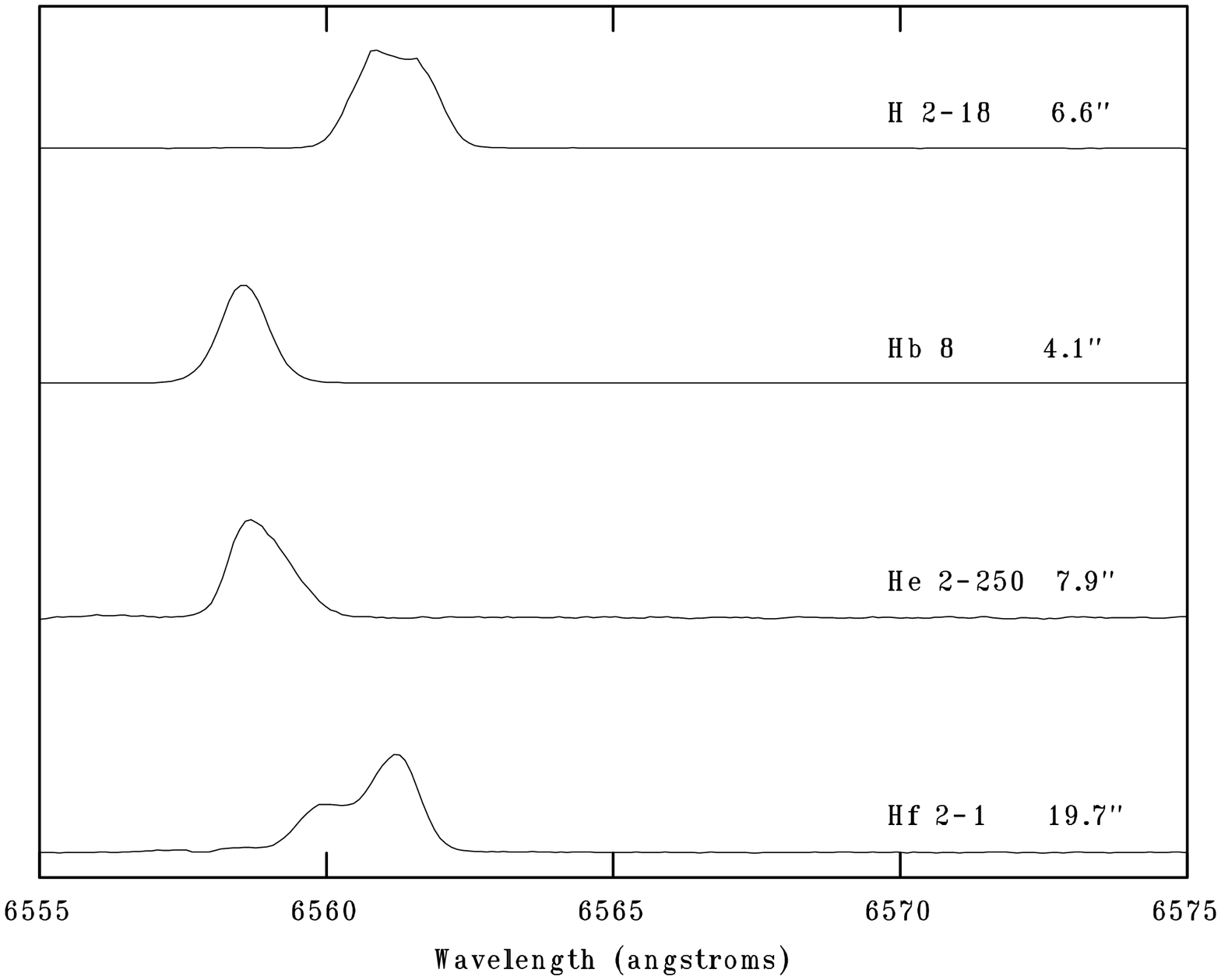}\\
  \includegraphics[height=\columnwidth,angle=90,bb=48 84 510 662,clip]{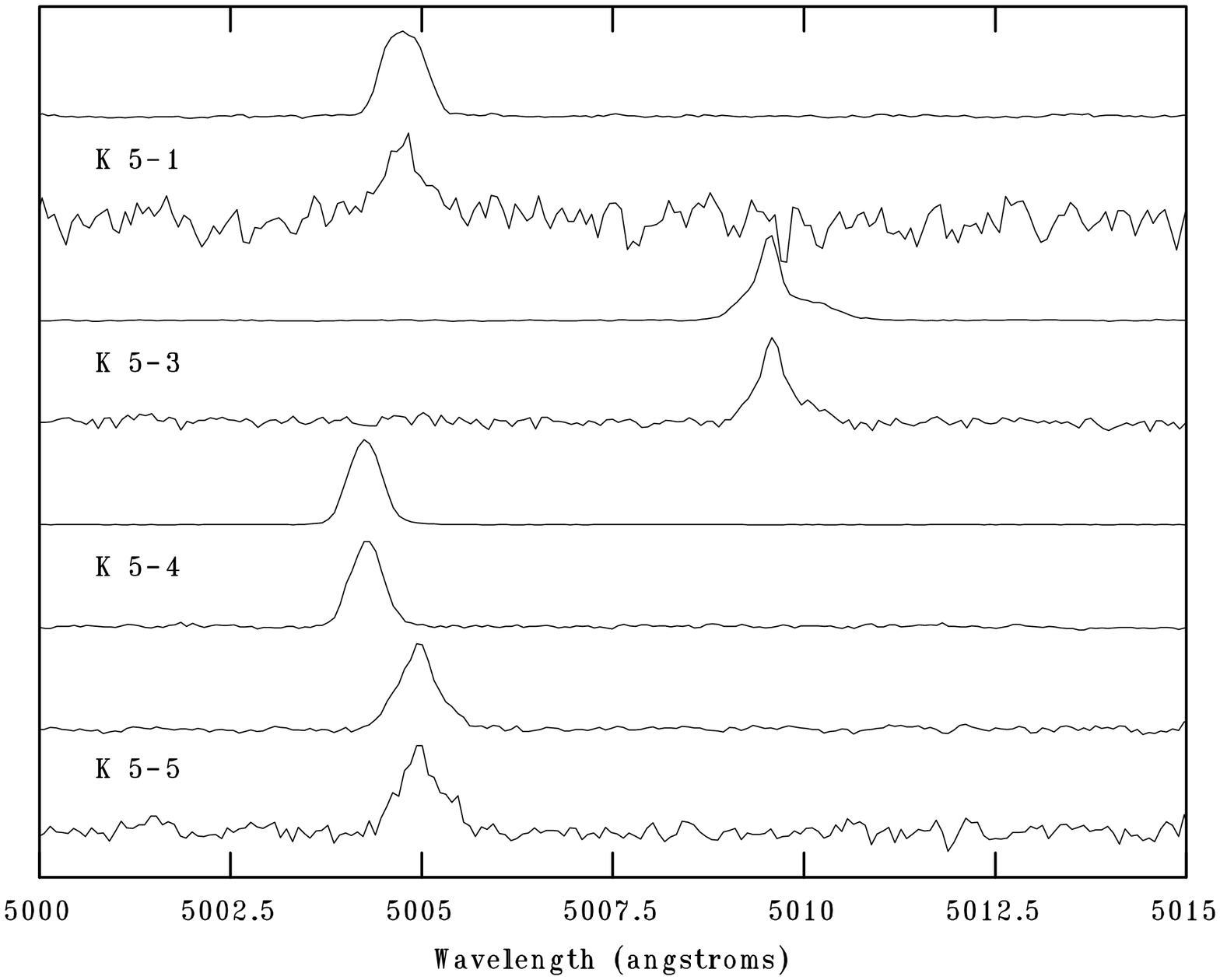}\quad\quad
  \includegraphics[height=\columnwidth,angle=90,bb=25 84 487 662,clip]{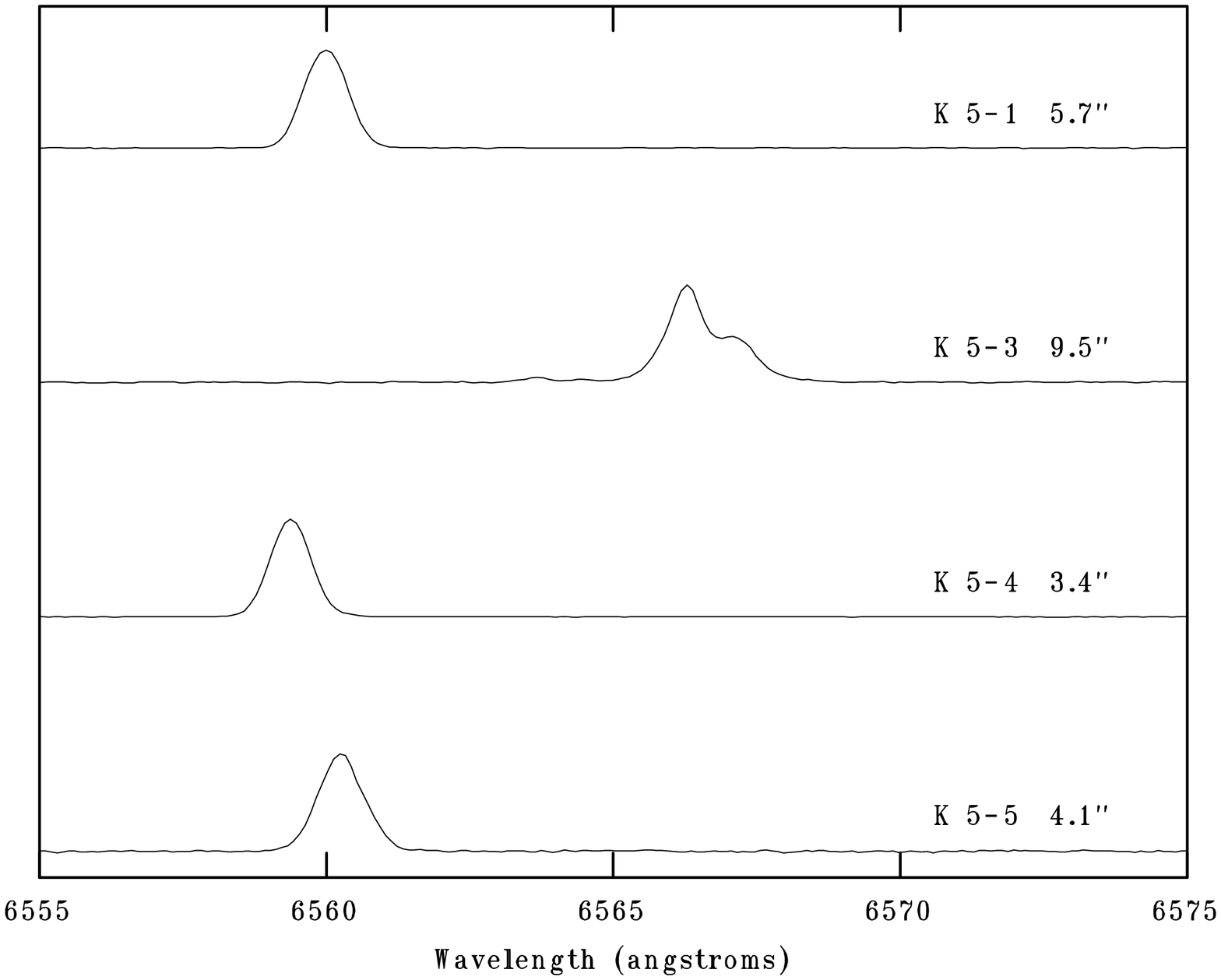}\\
  \end{center}
  \caption{See Fig. \ref{fig_lp_page1} for further details.}
  \label{fig_lp_page3}
\end{figure*}

\begin{figure*}[!t]
\begin{center}
  \includegraphics[height=\columnwidth,angle=90,bb=100 84 510 662,clip]{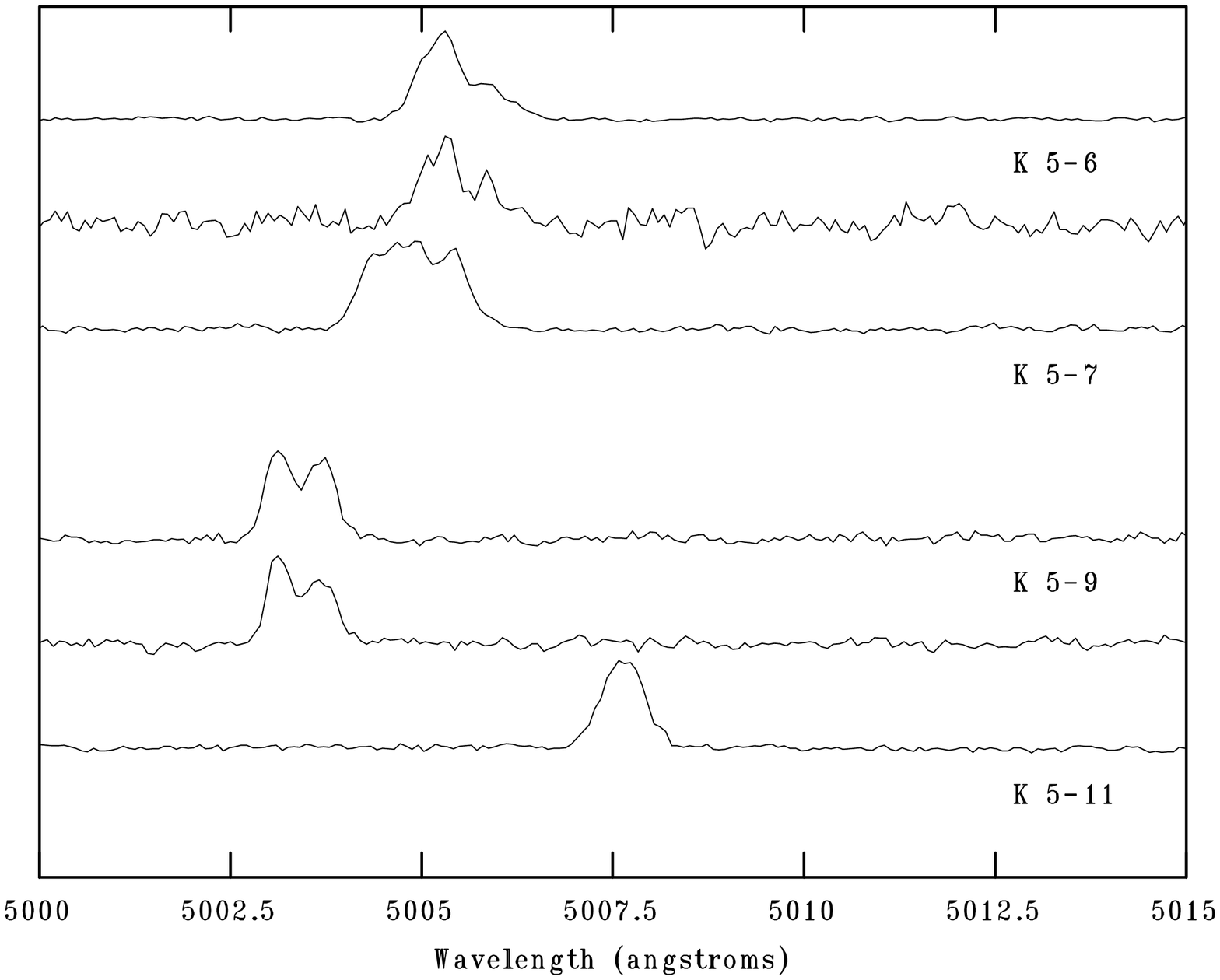}\quad\quad
  \includegraphics[height=\columnwidth,angle=90,bb=76 84 487 662,clip]{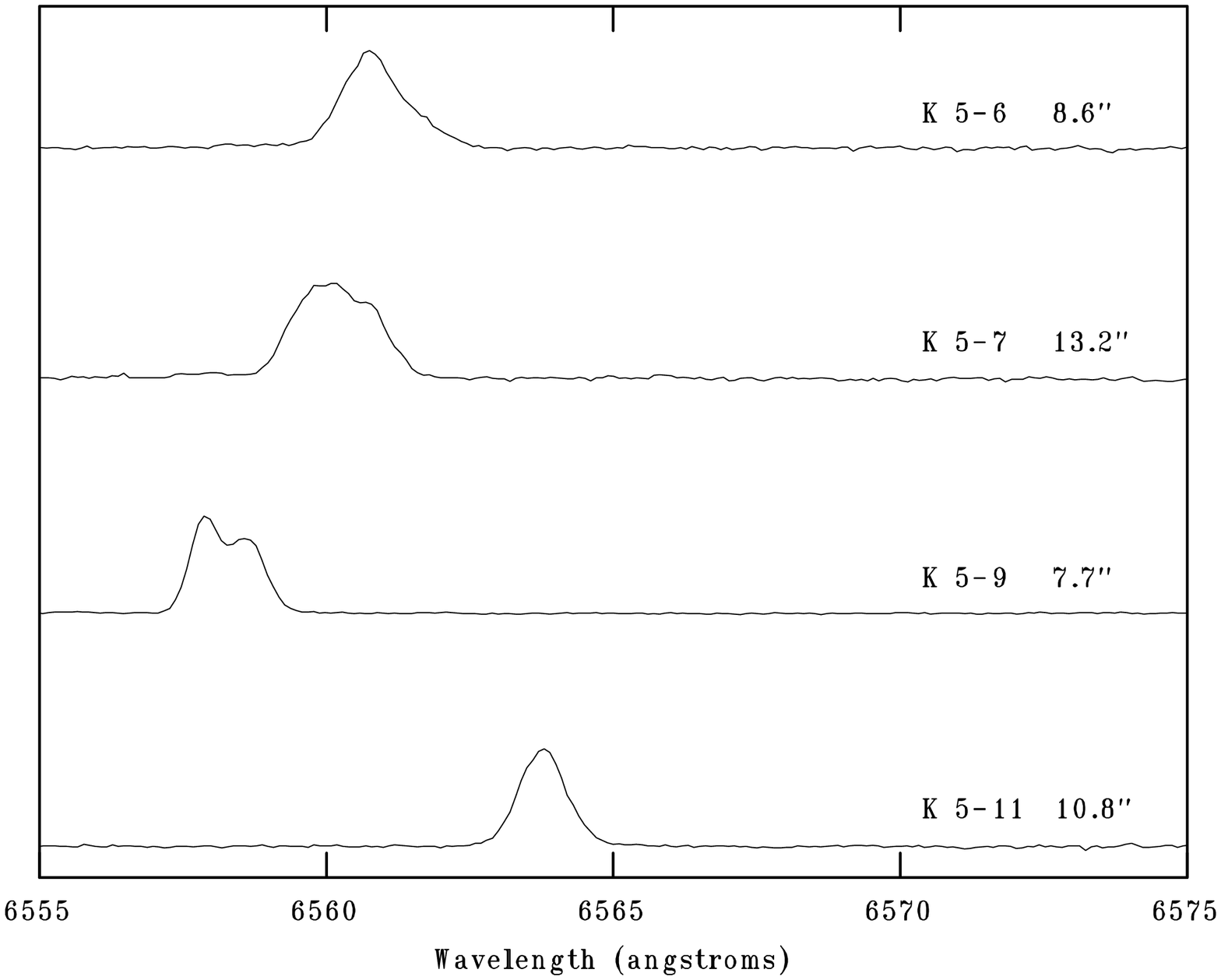}\\
  \includegraphics[height=\columnwidth,angle=90,bb=100 84 510 662,clip]{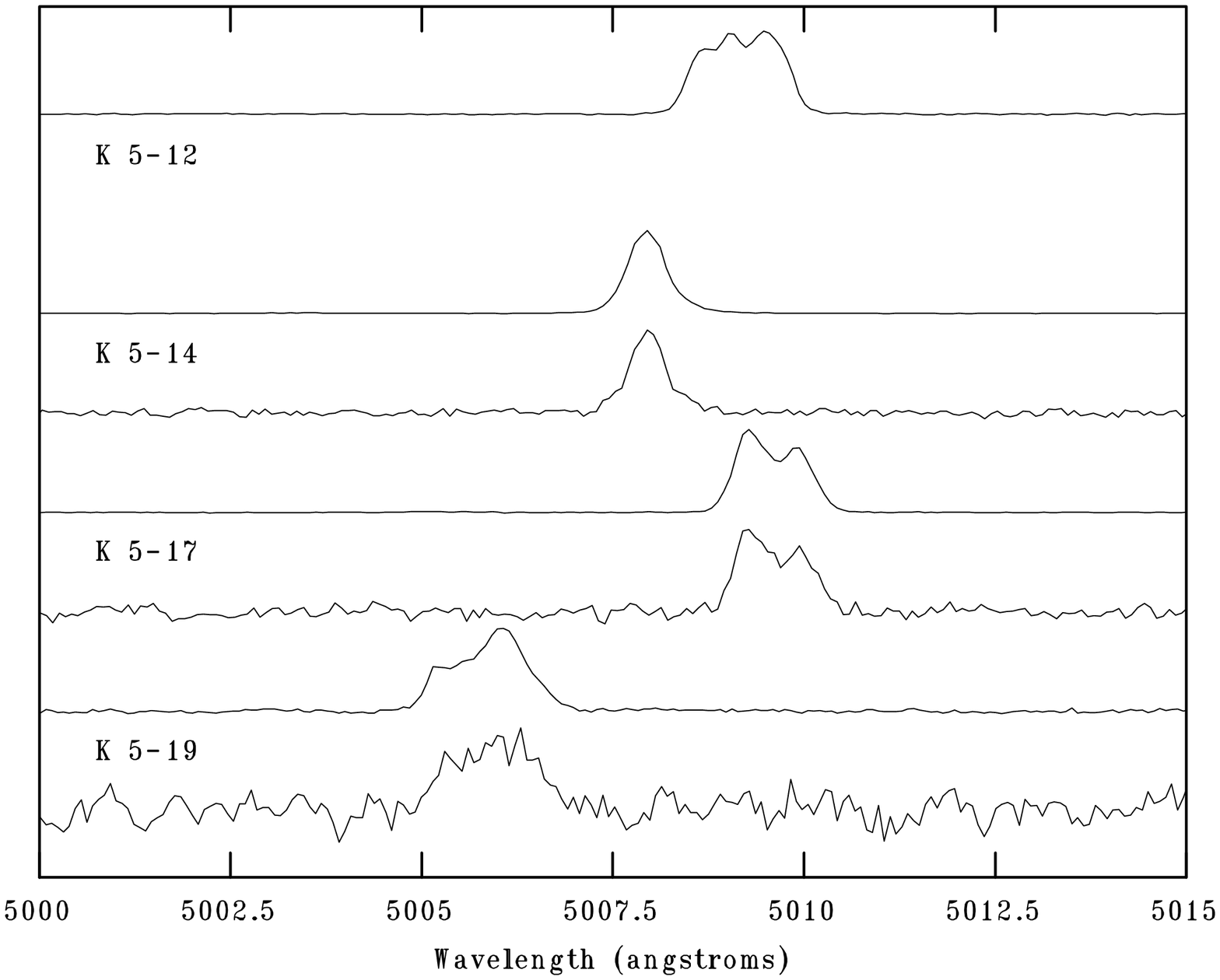}\quad\quad
  \includegraphics[height=\columnwidth,angle=90,bb=76 84 487 662,clip]{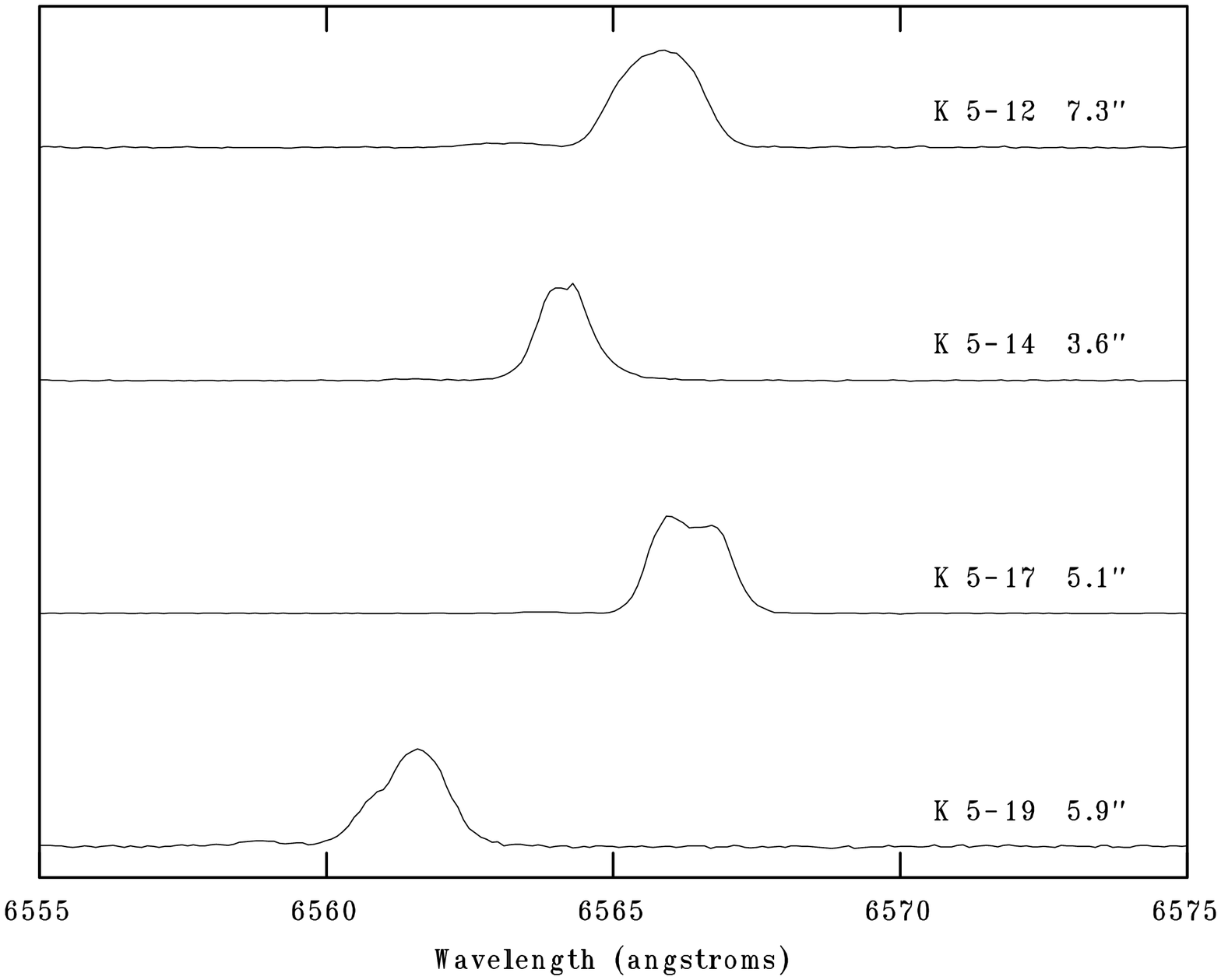}\\
  \includegraphics[height=\columnwidth,angle=90,bb=48 84 510 662,clip]{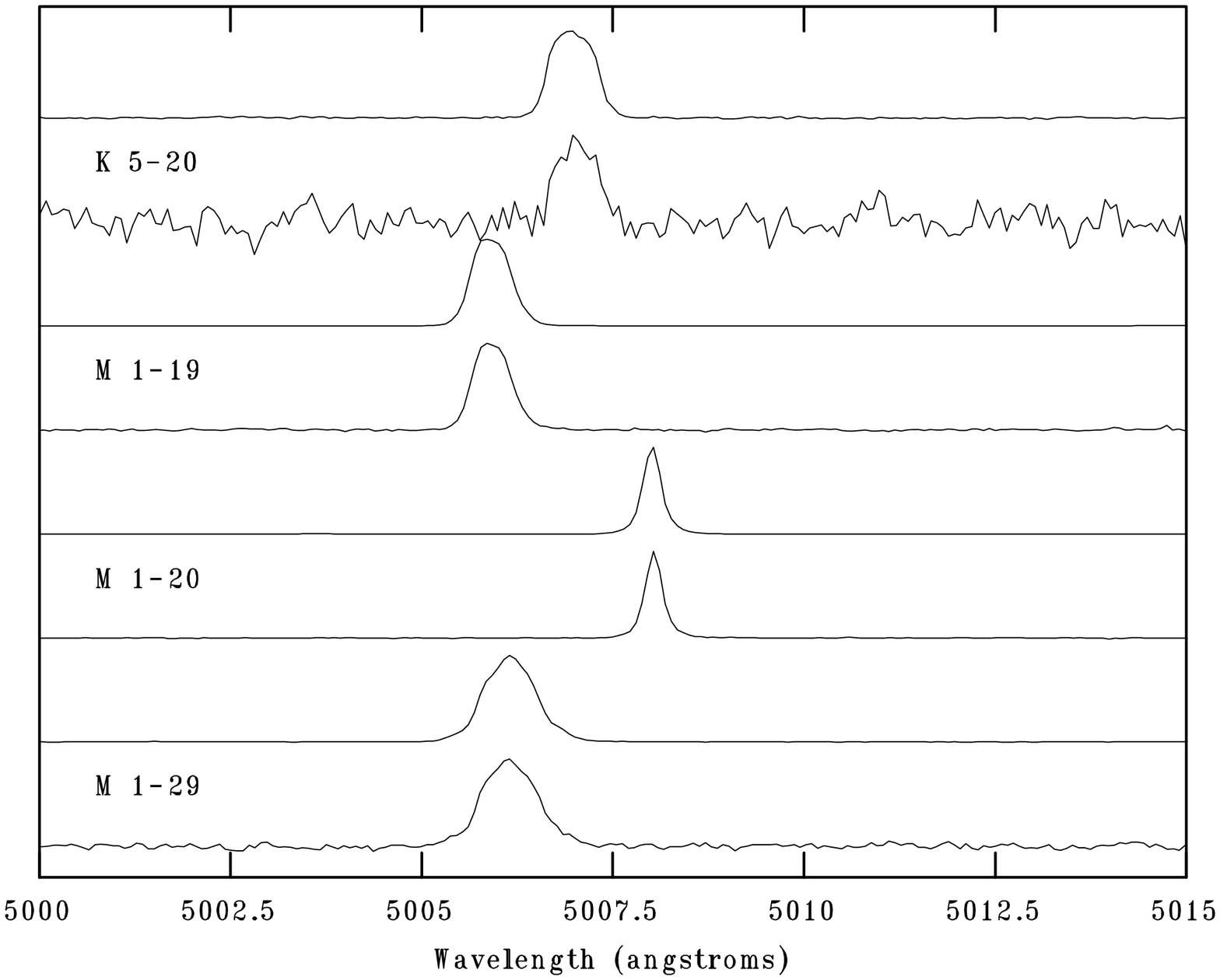}\quad\quad
  \includegraphics[height=\columnwidth,angle=90,bb=25 84 487 662,clip]{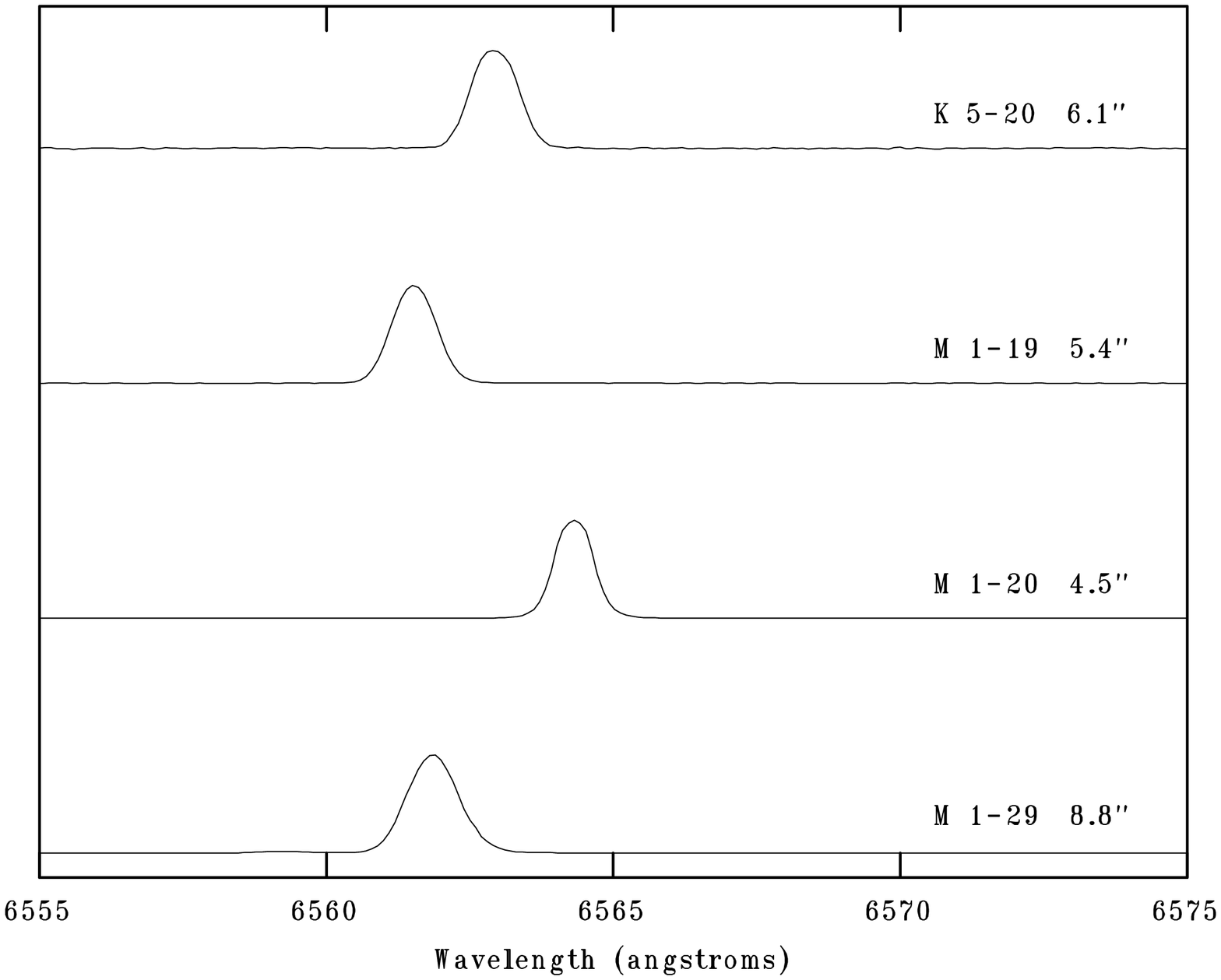}\\
  \end{center}
  \caption{For K 5-7 and K 5-12, only a deep [\ion{O}{3}]$\lambda 5007$ spectrum was obtained.  See Fig. \ref{fig_lp_page1} for further details.}
  \label{fig_lp_page4}
\end{figure*}

\begin{figure*}[!t]
\begin{center}
  \includegraphics[height=\columnwidth,angle=90,bb=100 84 510 662,clip]{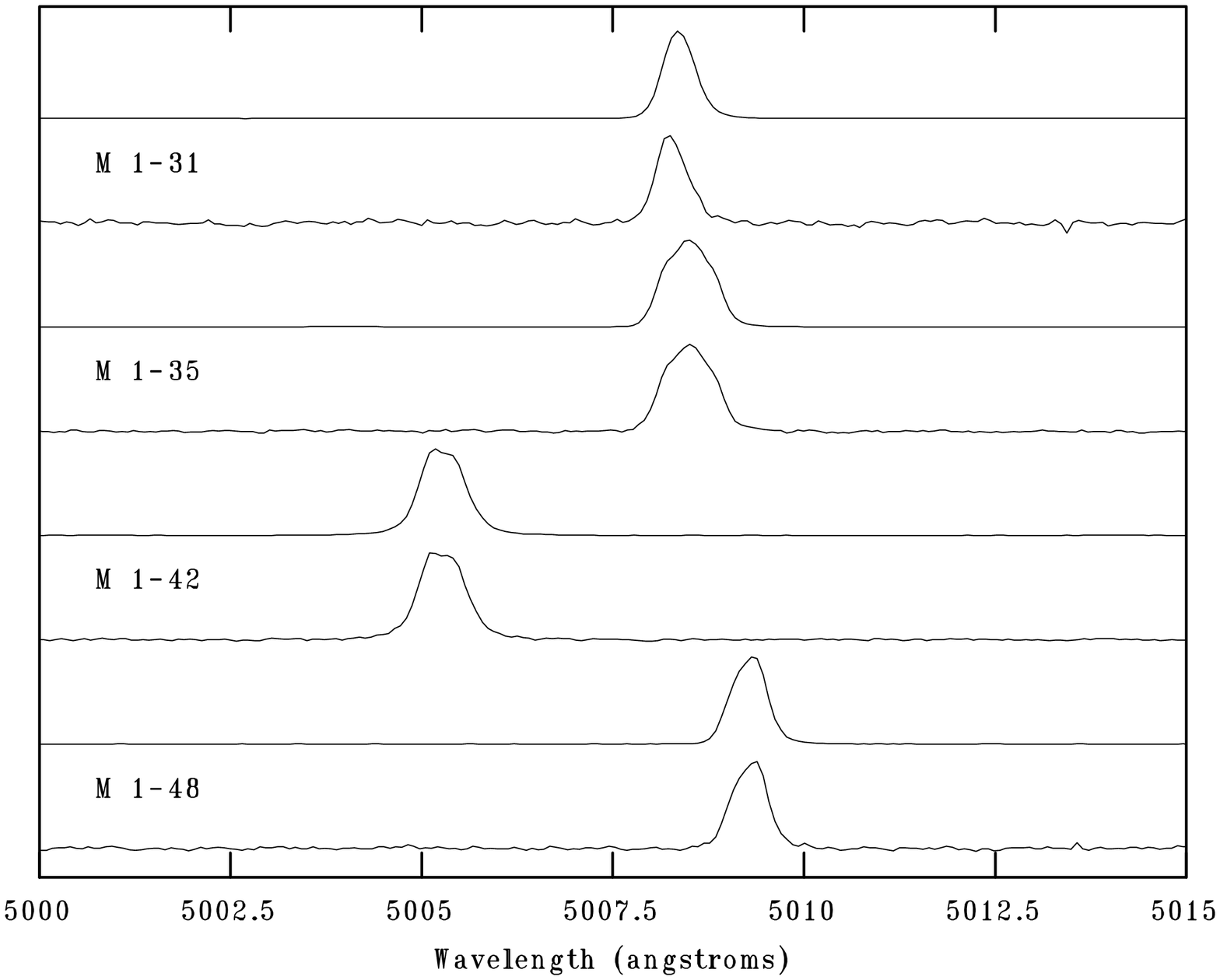}\quad\quad
  \includegraphics[height=\columnwidth,angle=90,bb=76 84 487 662,clip]{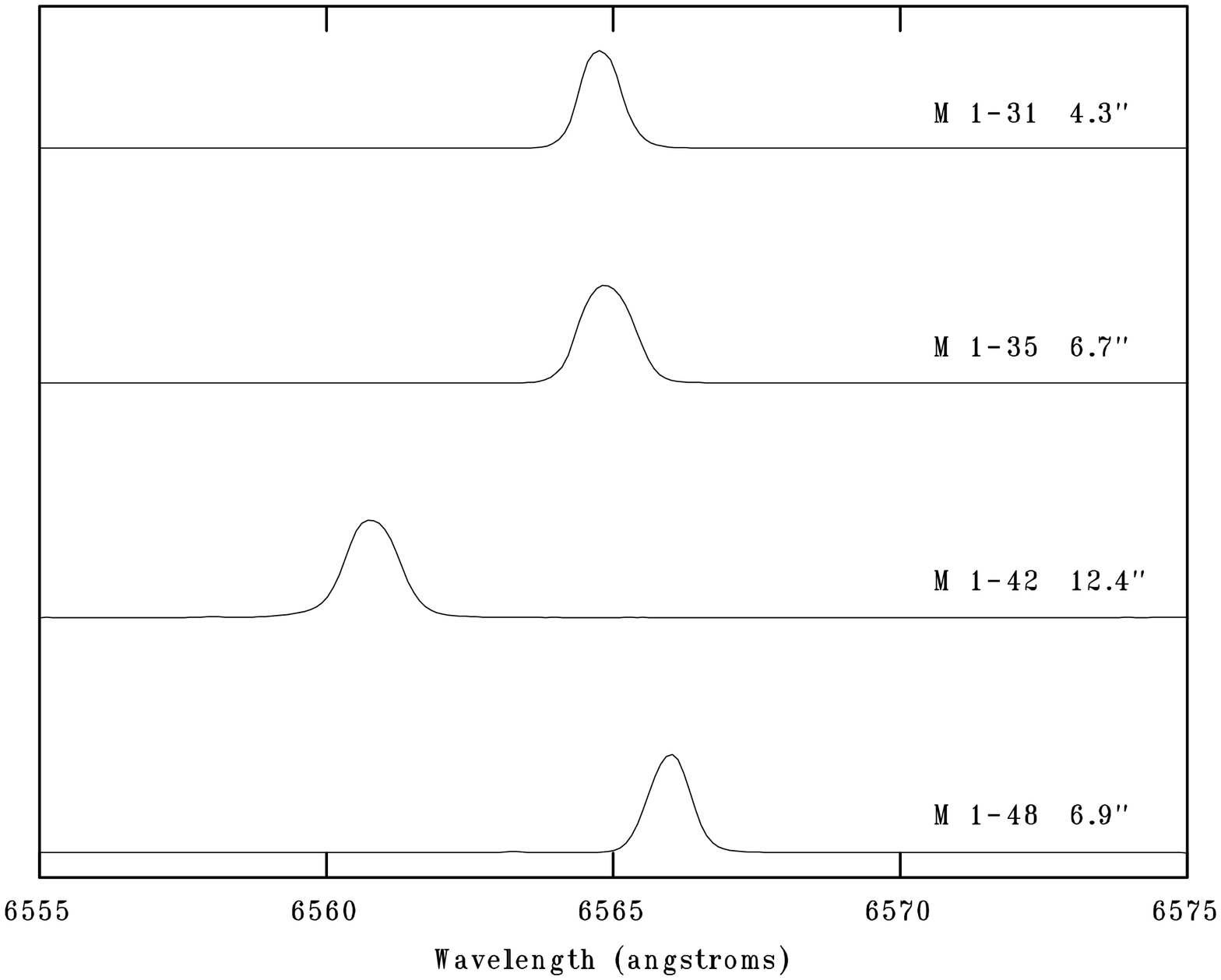}\\
  \includegraphics[height=\columnwidth,angle=90,bb=100 84 510 662,clip]{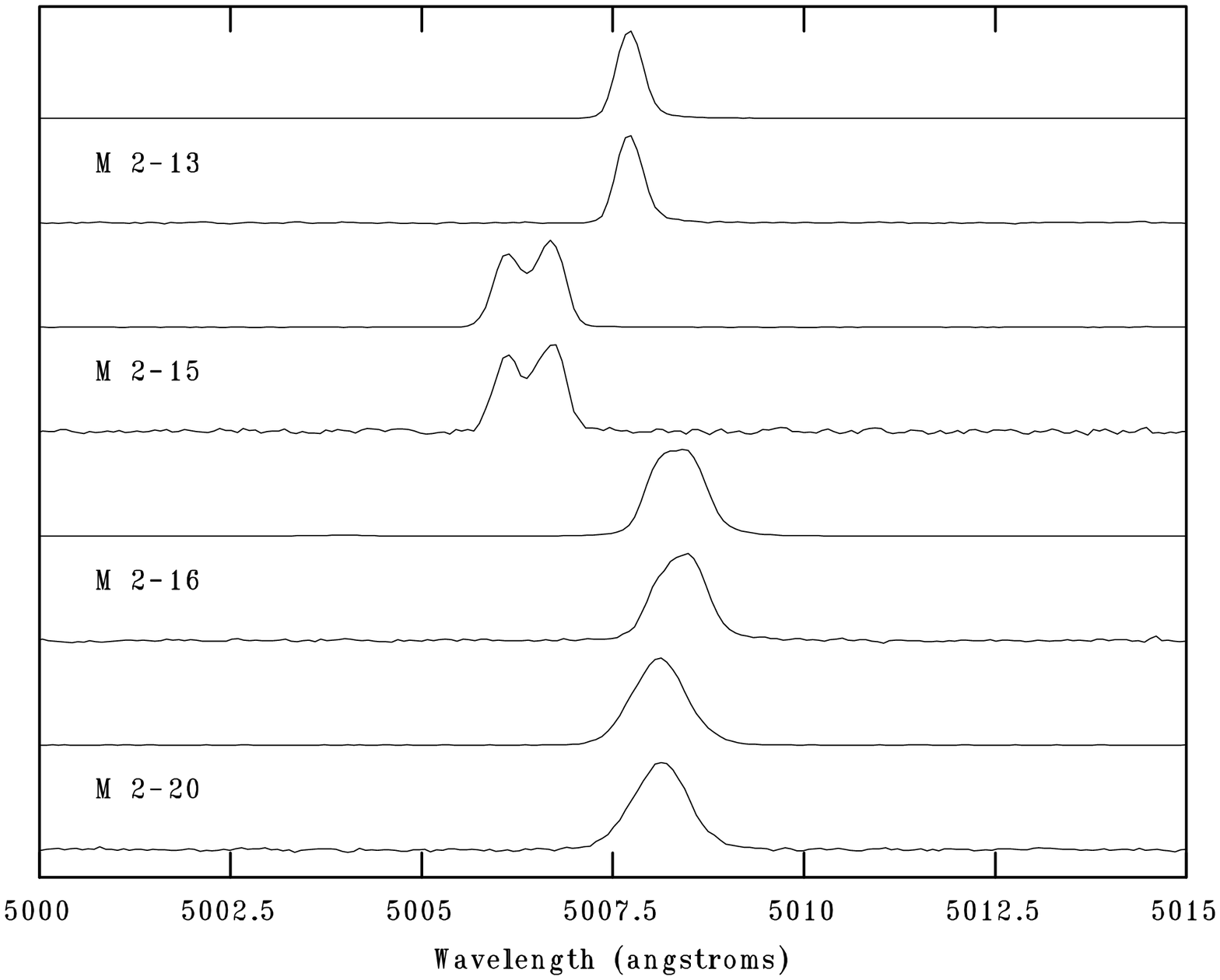}\quad\quad
  \includegraphics[height=\columnwidth,angle=90,bb=76 84 487 662,clip]{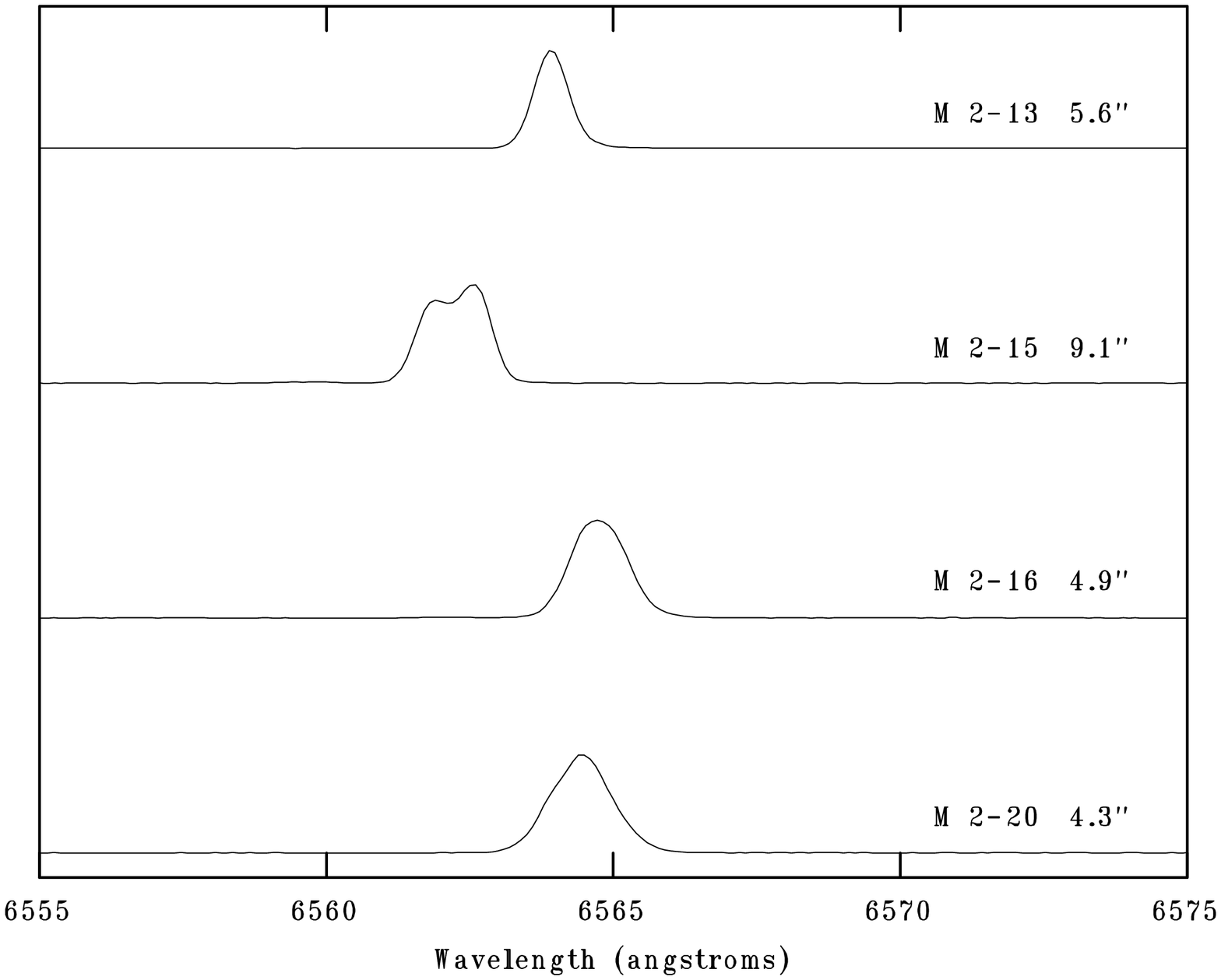}\\
  \includegraphics[height=\columnwidth,angle=90,bb=48 84 510 662,clip]{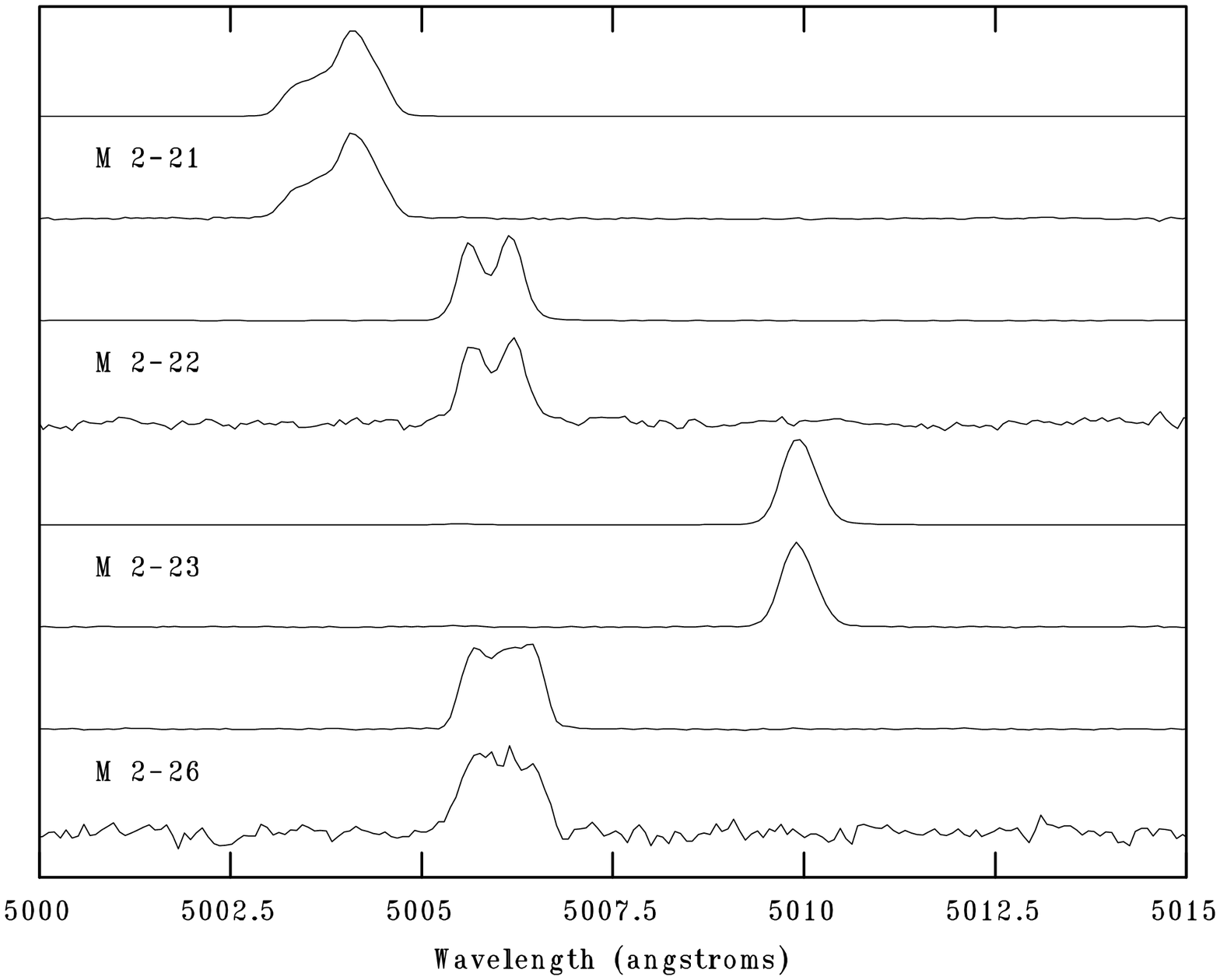}\quad\quad
  \includegraphics[height=\columnwidth,angle=90,bb=25 84 487 662,clip]{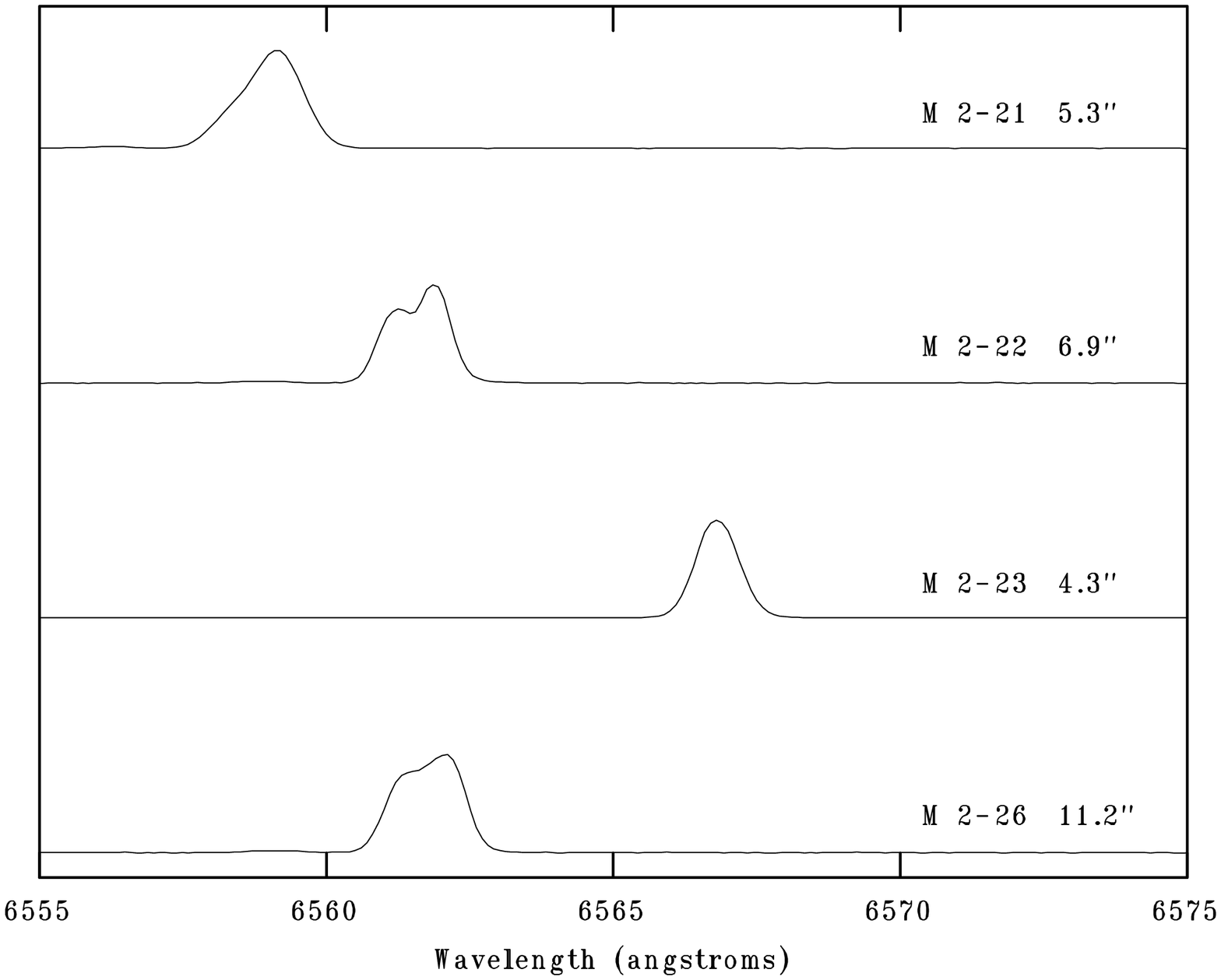}\\
  \end{center}
  \caption{See Fig. \ref{fig_lp_page1} for further details.}
  \label{fig_lp_page5}
\end{figure*}

\begin{figure*}[!t]
\begin{center}
  \includegraphics[height=\columnwidth,angle=90,bb=100 84 510 662,clip]{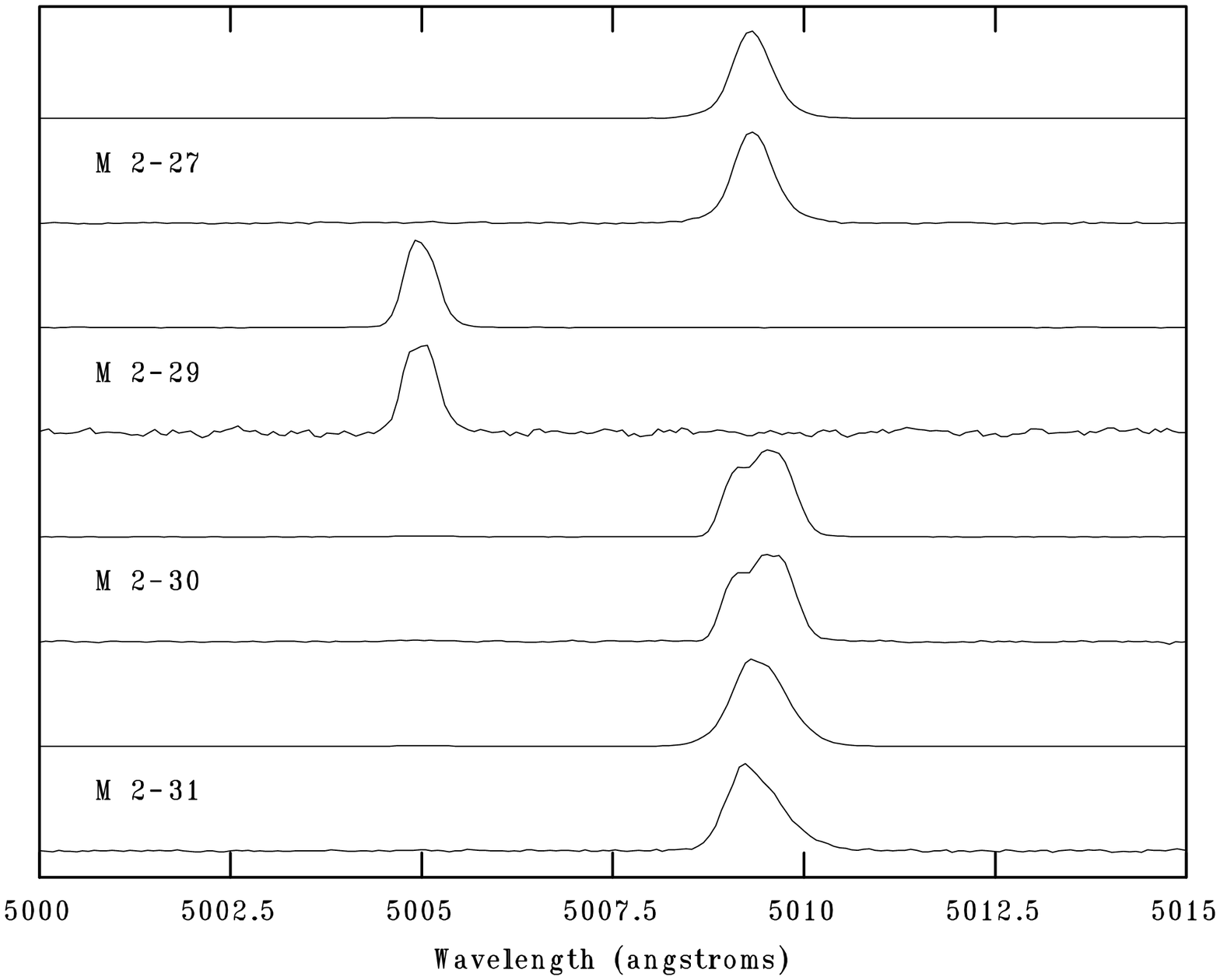}\quad\quad
  \includegraphics[height=\columnwidth,angle=90,bb=76 84 487 662,clip]{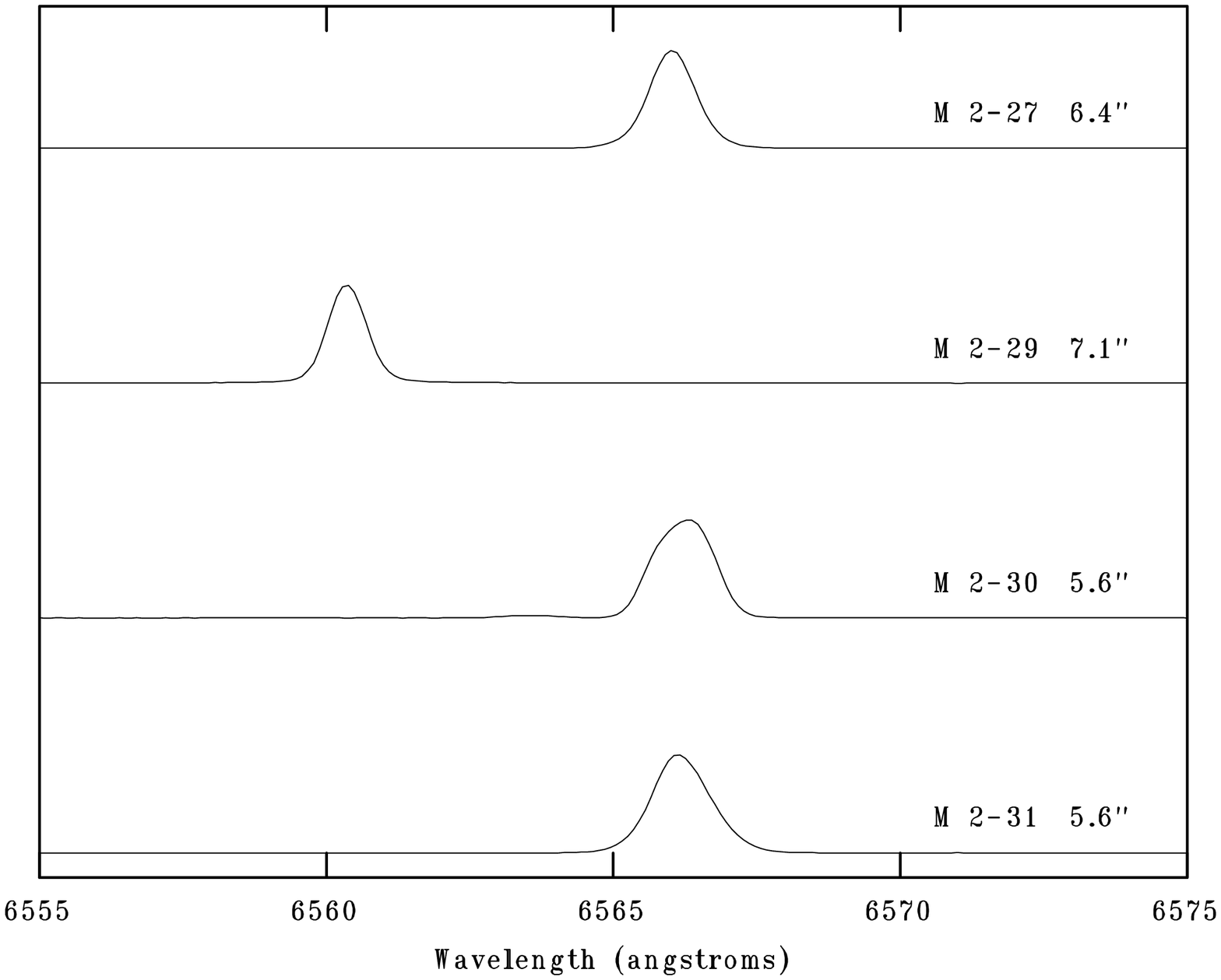}\\
  \includegraphics[height=\columnwidth,angle=90,bb=100 84 510 662,clip]{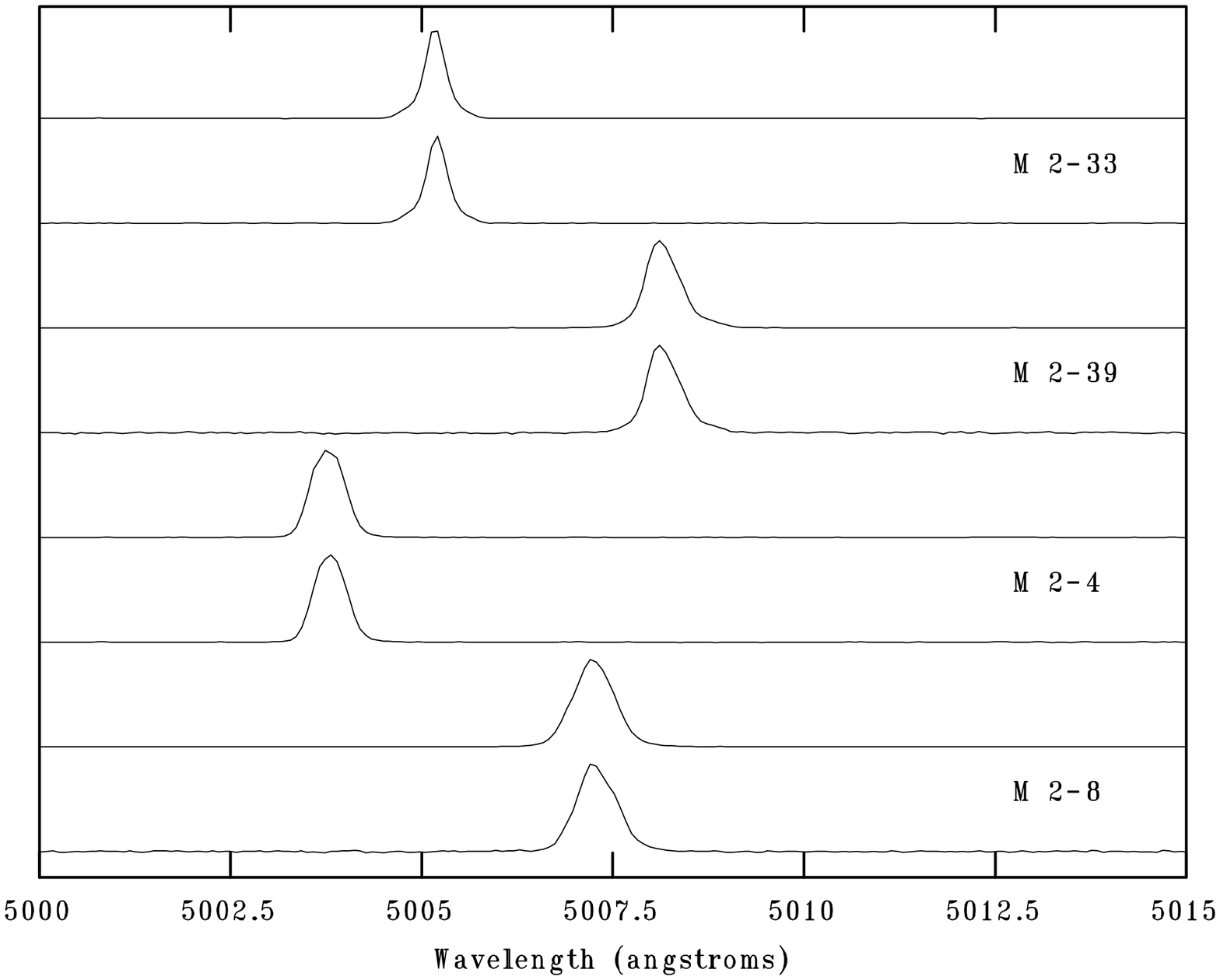}\quad\quad
  \includegraphics[height=\columnwidth,angle=90,bb=76 84 487 662,clip]{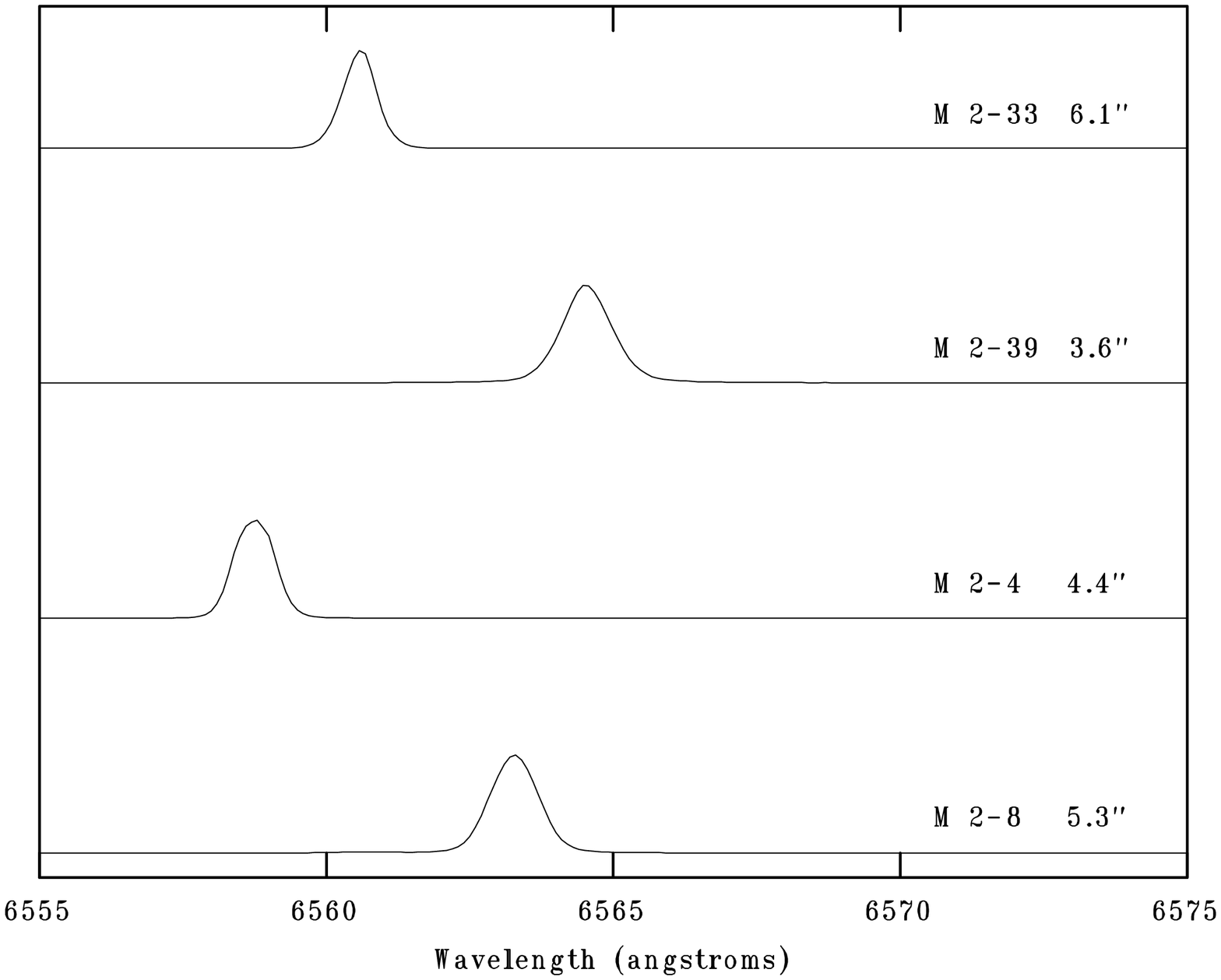}\\
  \includegraphics[height=\columnwidth,angle=90,bb=48 84 510 662,clip]{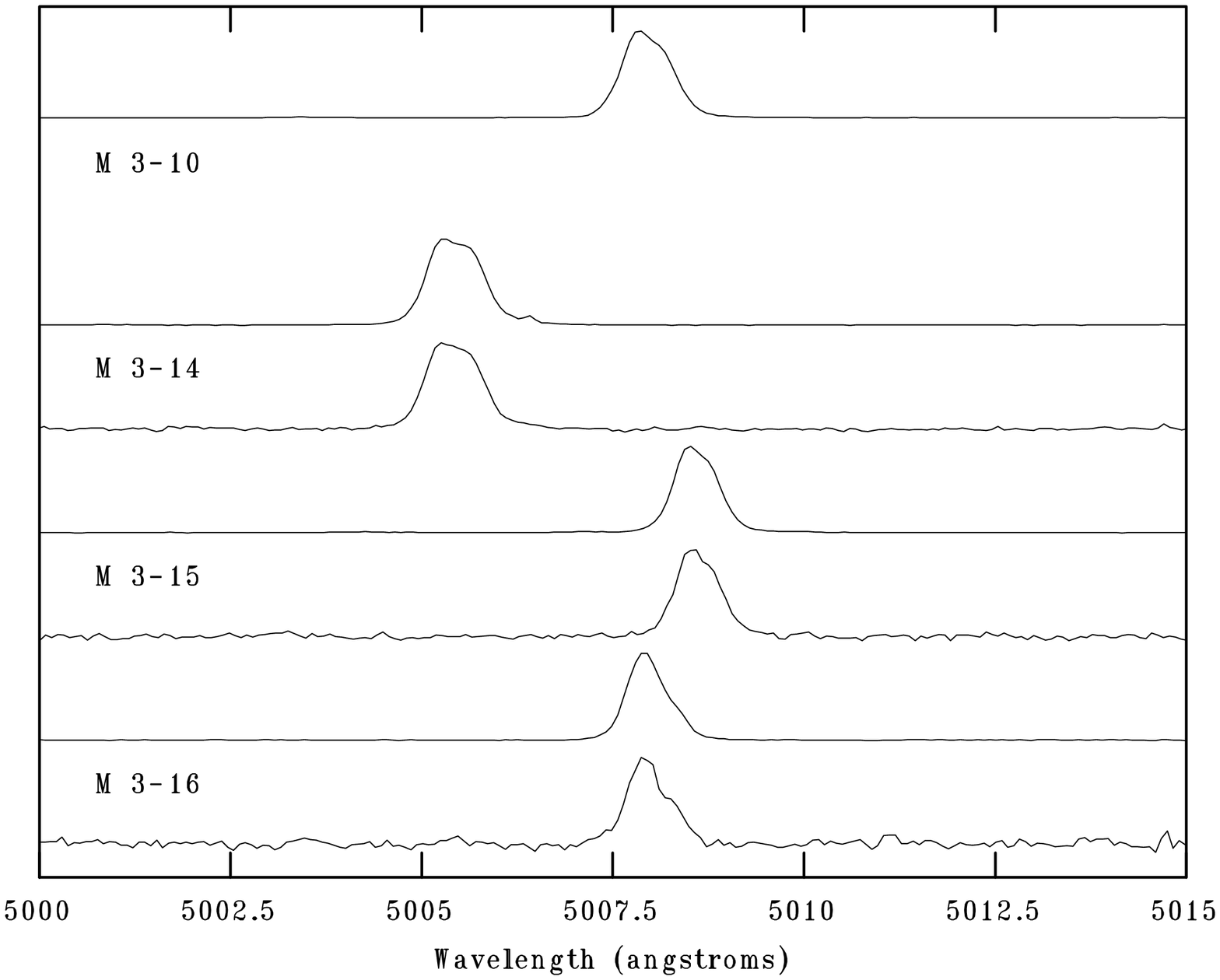}\quad\quad
  \includegraphics[height=\columnwidth,angle=90,bb=25 84 487 662,clip]{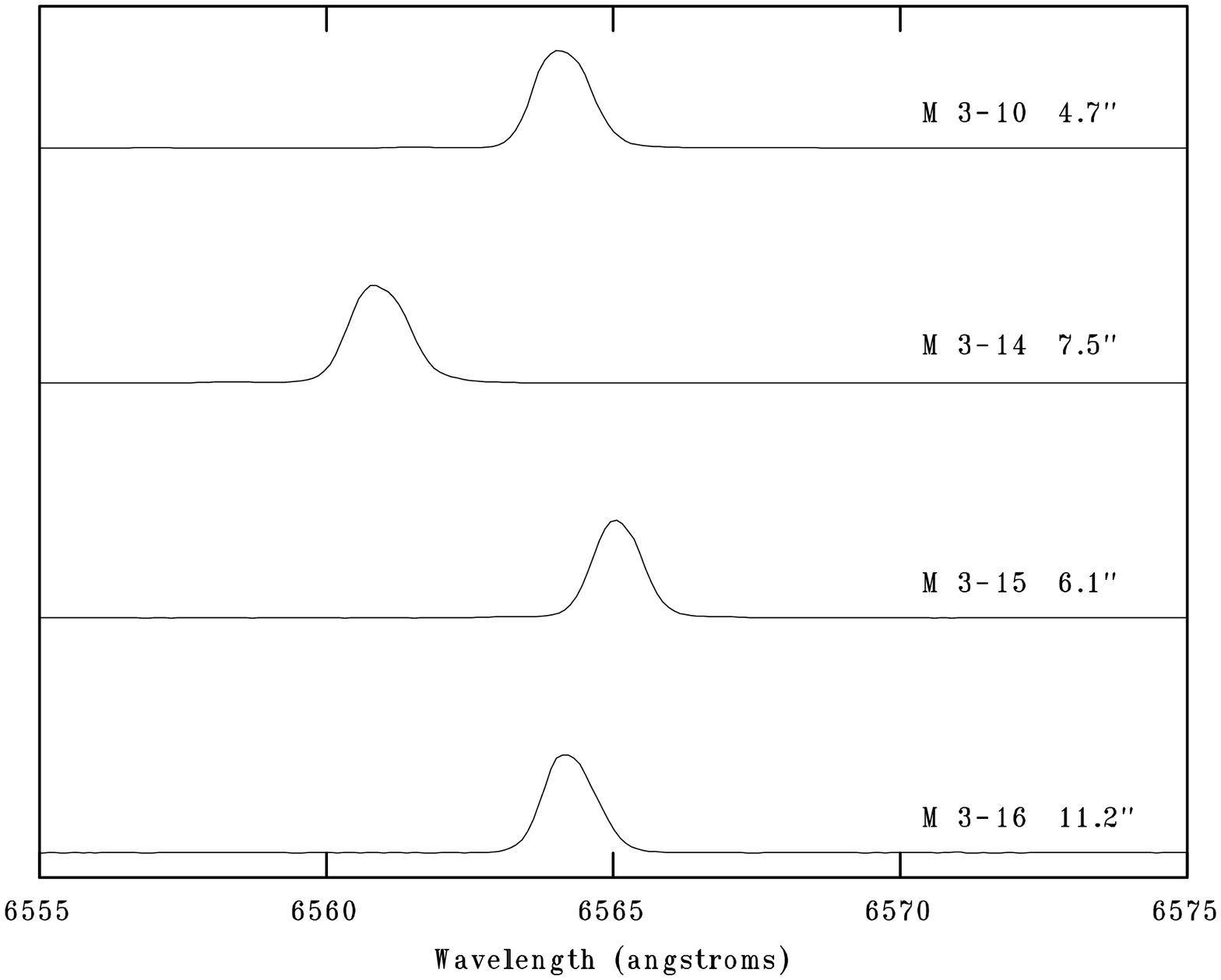}\\
  \end{center}
  \caption{For M 3-10, only a deep [\ion{O}{3}]$\lambda 5007$ spectrum was obtained.  See Fig. \ref{fig_lp_page1} for further details.}
  \label{fig_lp_page6}
\end{figure*}

\begin{figure*}[!t]
\begin{center}
  \includegraphics[height=\columnwidth,angle=90,bb=100 84 510 662,clip]{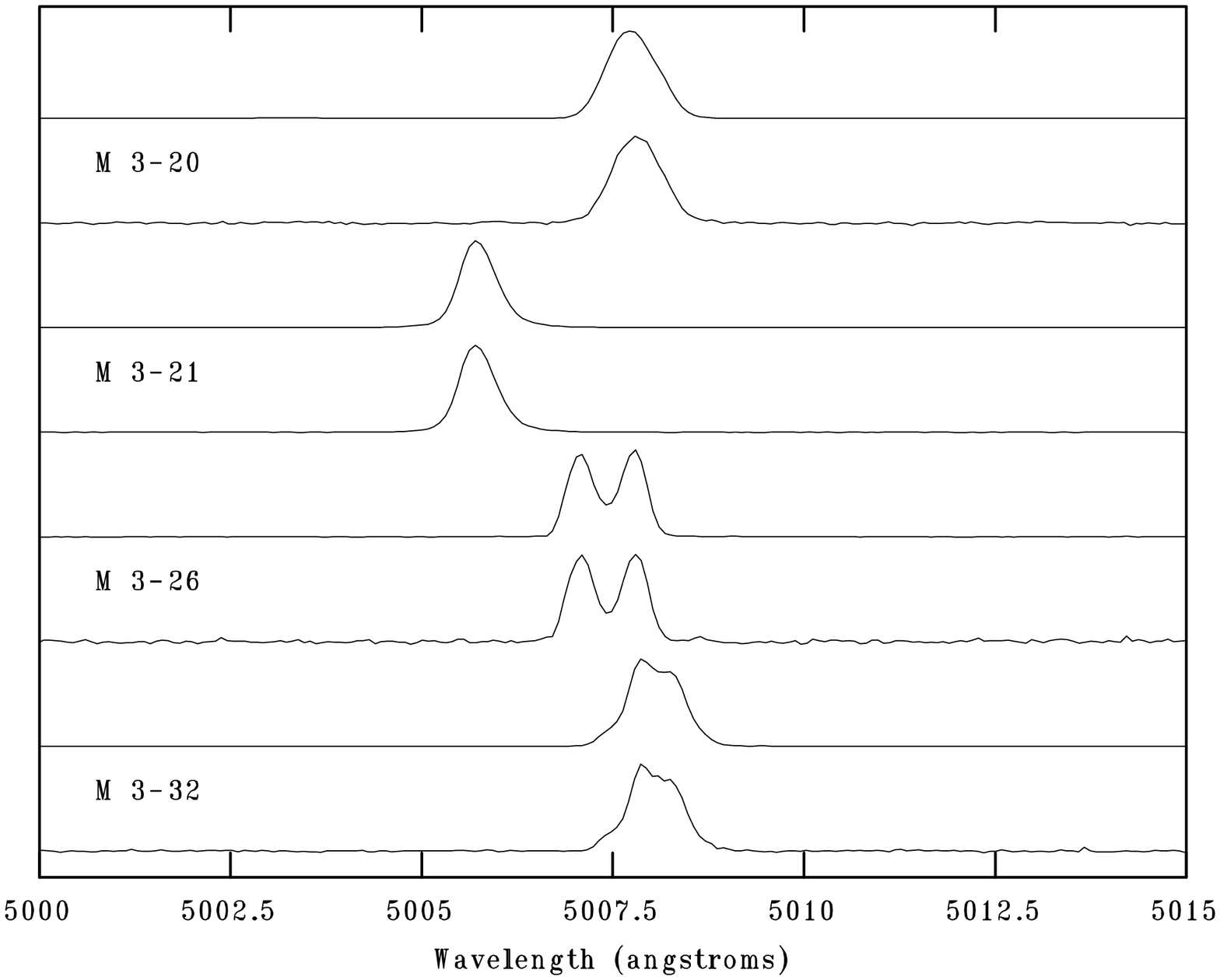}\quad\quad
  \includegraphics[height=\columnwidth,angle=90,bb=76 84 487 662,clip]{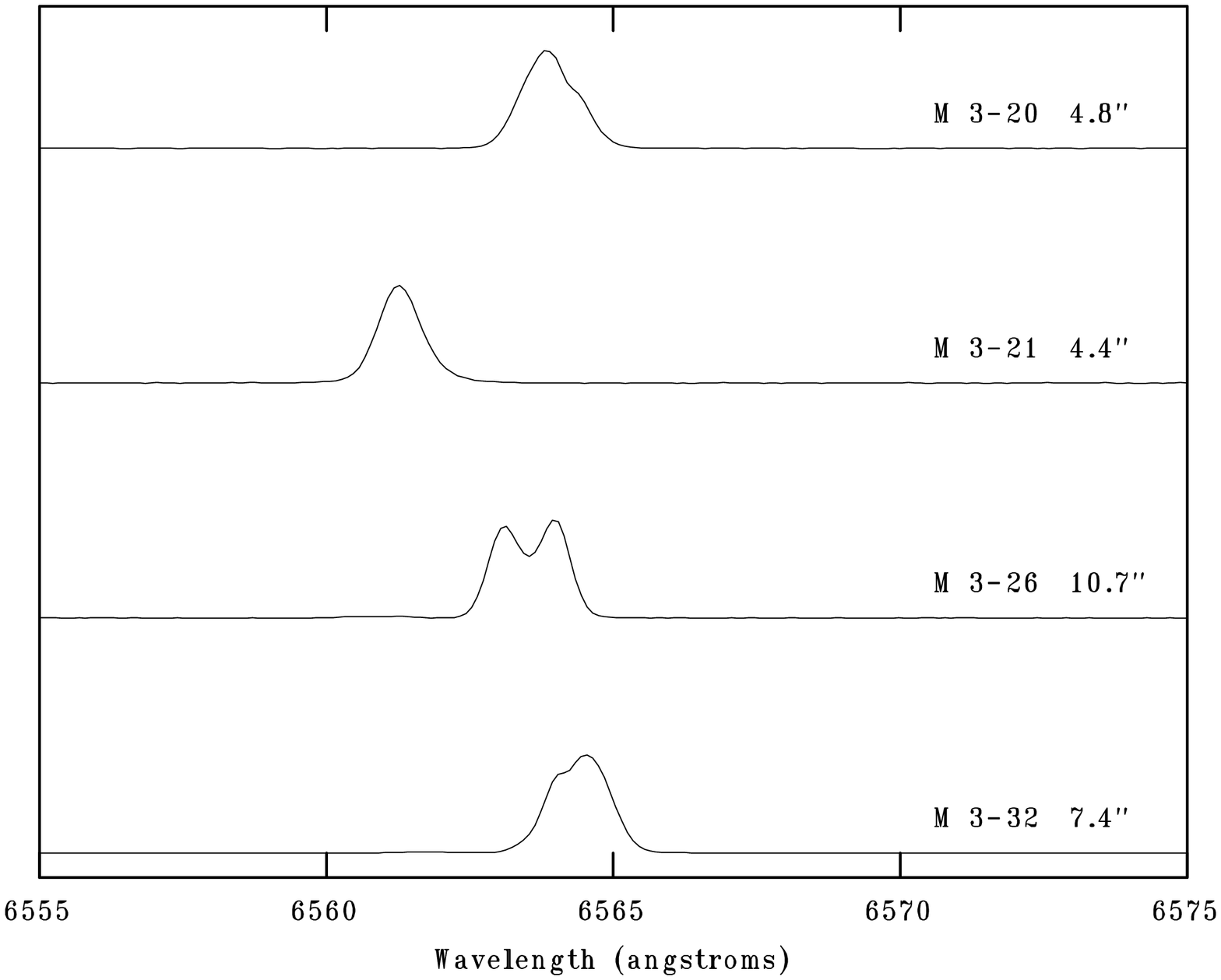}\\
  \includegraphics[height=\columnwidth,angle=90,bb=100 84 510 662,clip]{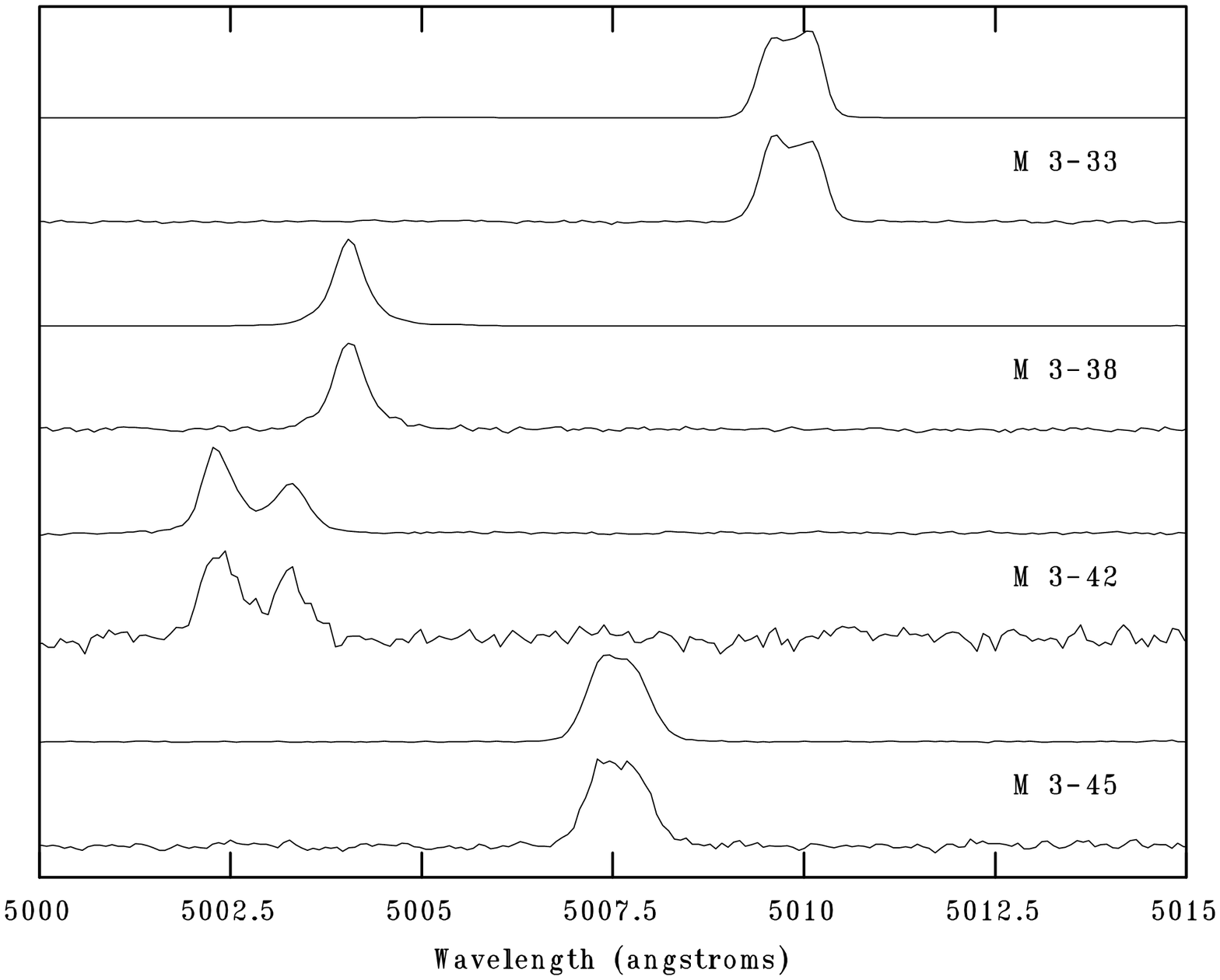}\quad\quad
  \includegraphics[height=\columnwidth,angle=90,bb=76 84 487 662,clip]{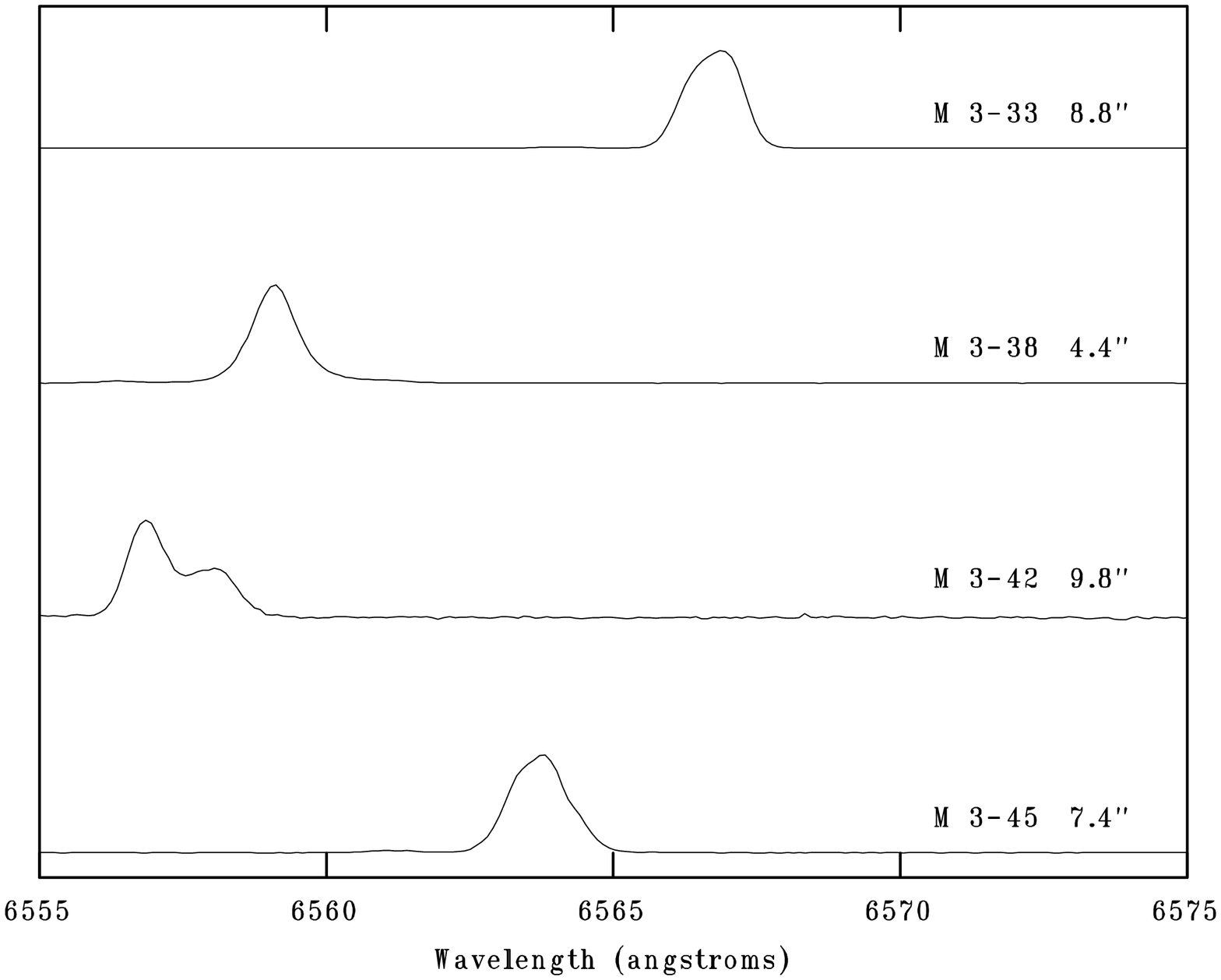}\\
  \includegraphics[height=\columnwidth,angle=90,bb=48 84 510 662,clip]{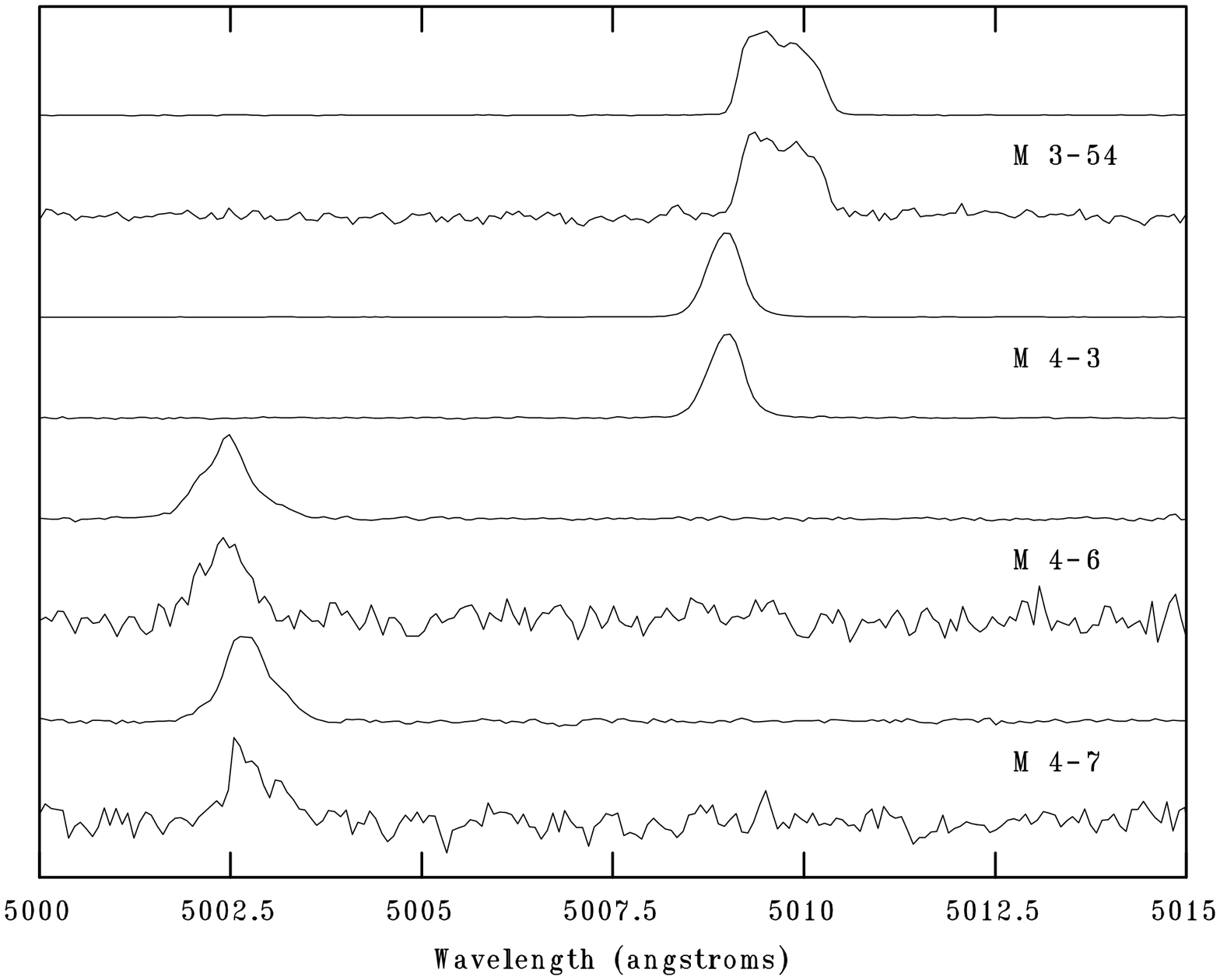}\quad\quad
  \includegraphics[height=\columnwidth,angle=90,bb=25 84 487 662,clip]{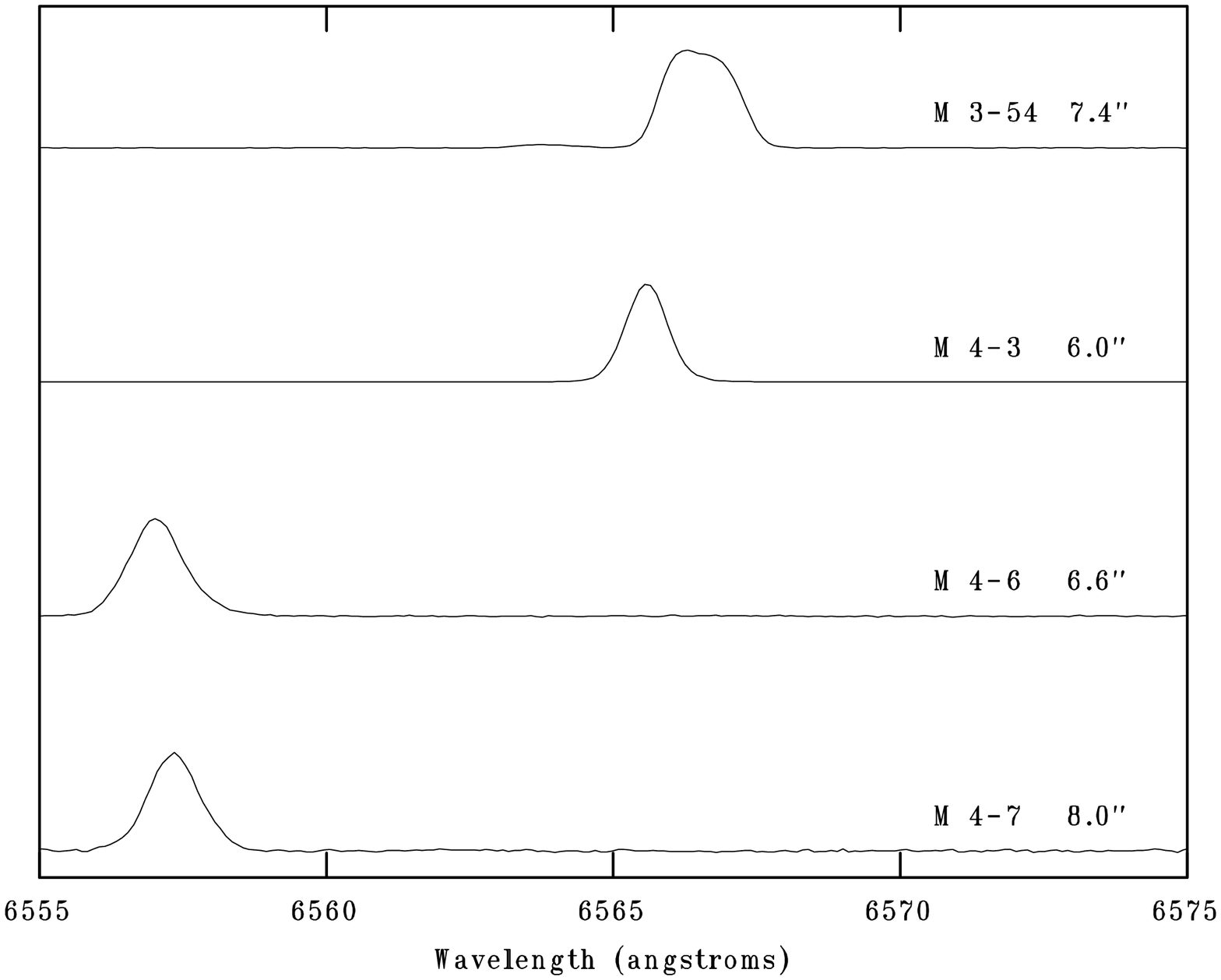}\\
  \end{center}
  \caption{See Fig. \ref{fig_lp_page1} for further details.}
  \label{fig_lp_page7}
\end{figure*}
  
\begin{figure*}[!t]
\begin{center}
  \includegraphics[height=\columnwidth,angle=90,bb=76 84 487 662,clip]{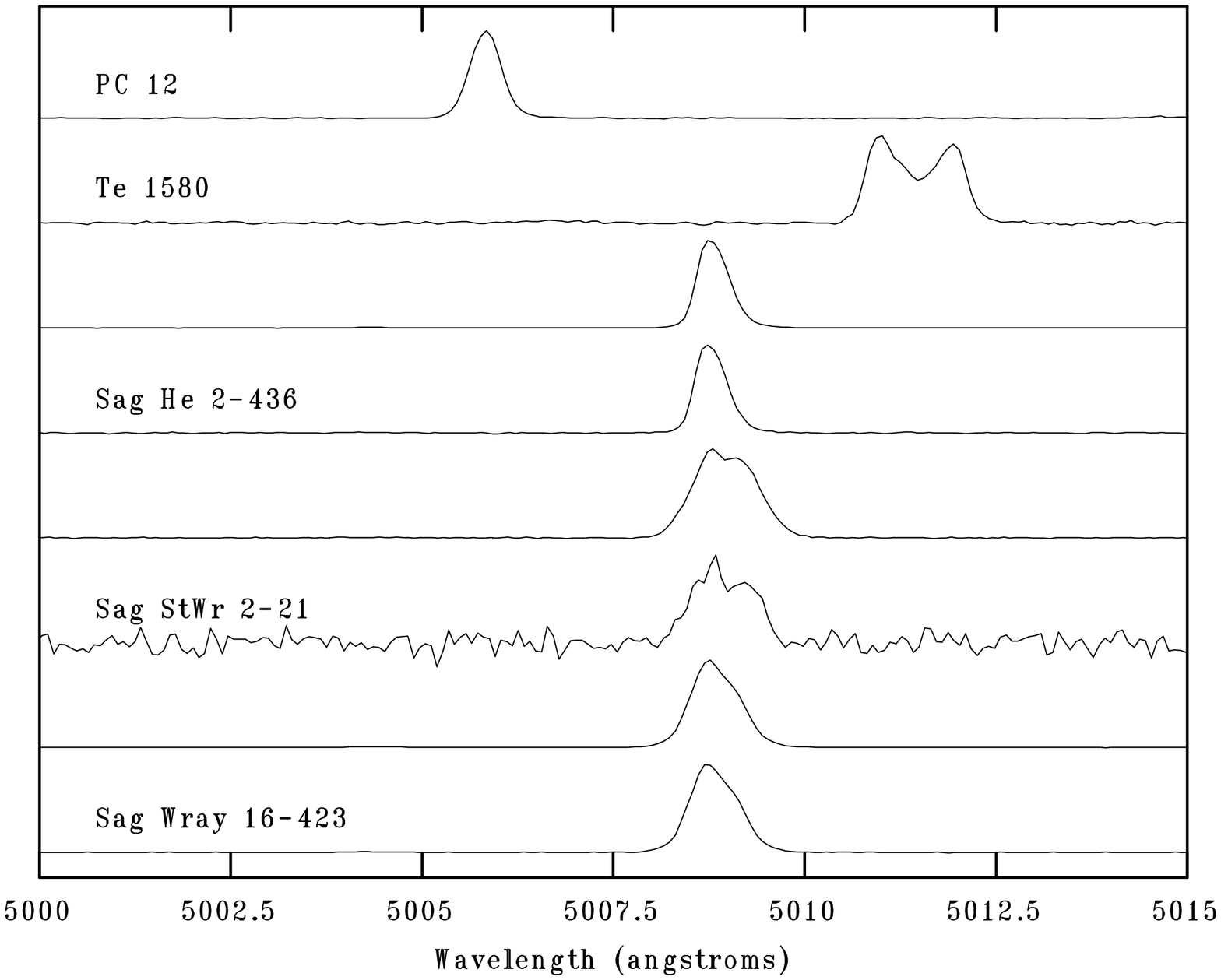}\quad\quad
  \includegraphics[height=\columnwidth,angle=90,bb=76 84 487 662,clip]{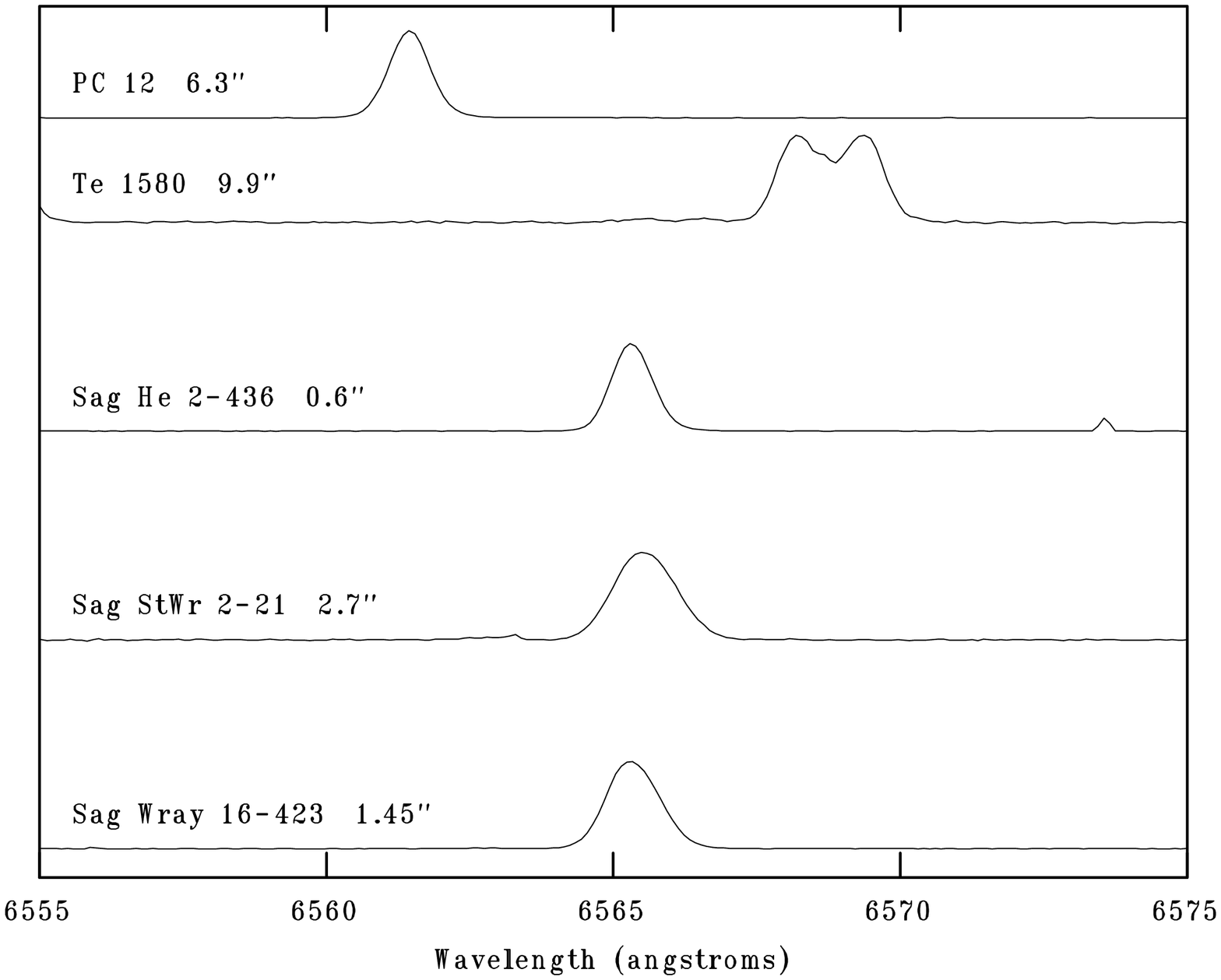}\\
  \includegraphics[height=\columnwidth,angle=90,bb=48 84 510 662,clip]{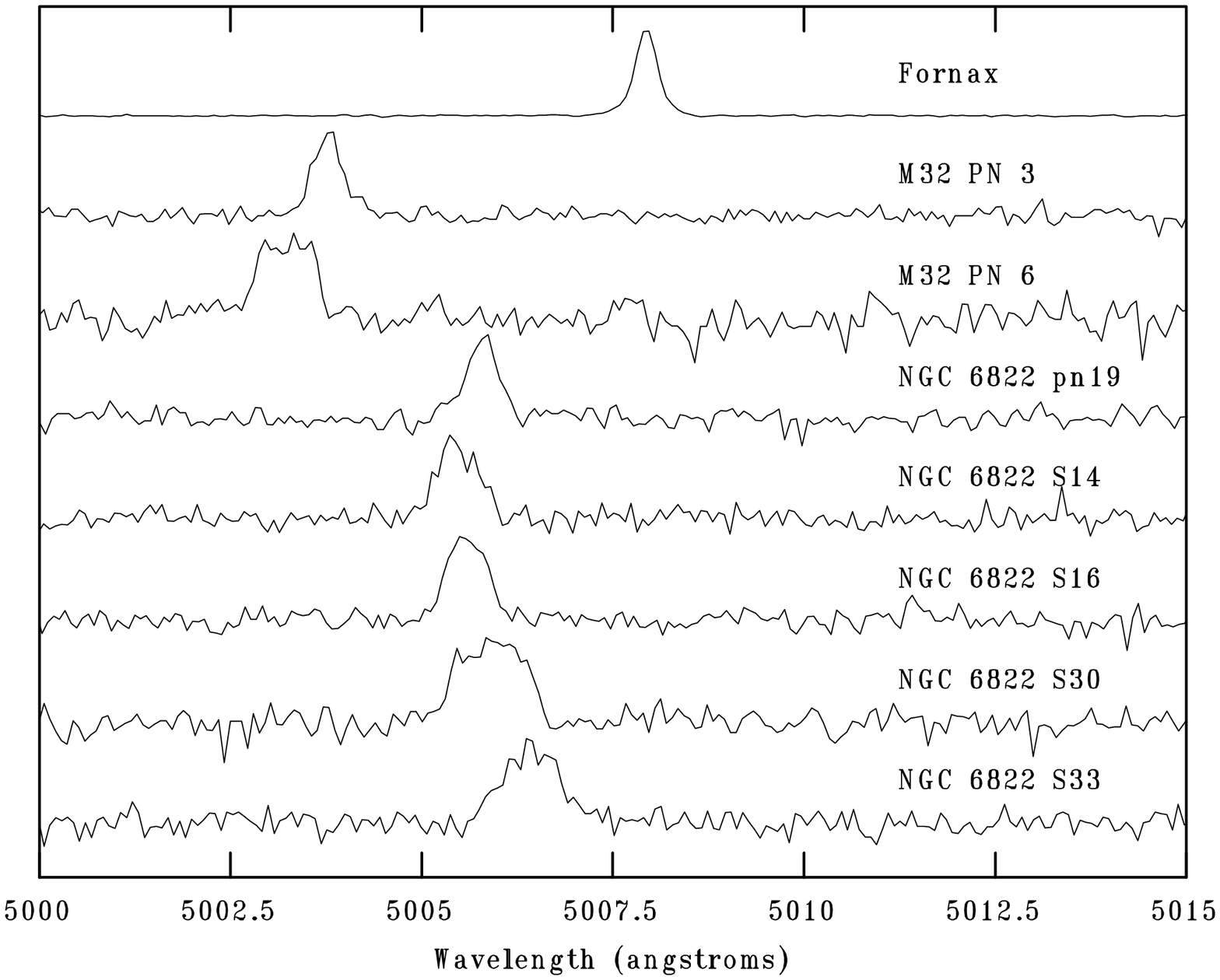}\quad\quad
  \includegraphics[height=\columnwidth,angle=90,bb=48 84 510 662,clip]{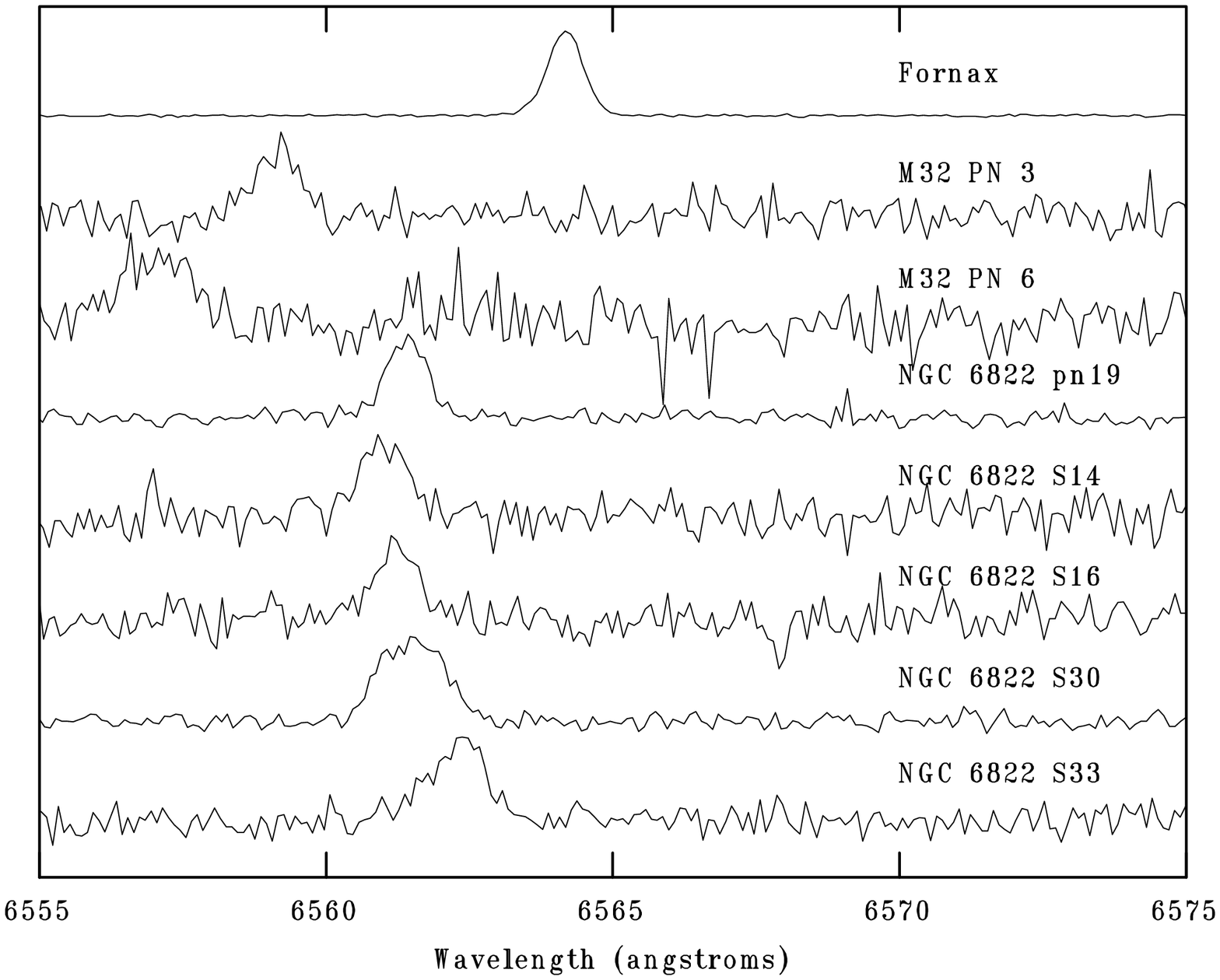}\\
  \end{center}
  \caption{For the Bulge planetary nebulae Te 1580 and PC 12, only a deep [\ion{O}{3}]$\lambda 5007$ spectrum was obtained.  Deep and shallow spectra were obtained of the three planetary nebulae in the Sagittarius dwarf spheroidal \citep[diameters from][]{zijlstraetal2006}.  Only deep spectra were obtained for the other extragalactic planetary nebulae.  See Fig. \ref{fig_lp_page1} for further details.}
  \label{fig_lp_page8}
\end{figure*}

\begin{figure}[!t]
  \includegraphics[width=0.7\columnwidth,angle=90]{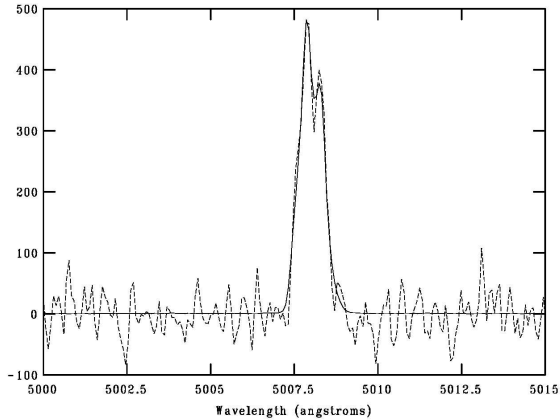}
  \caption{The synthetic spectra of extragalactic planetary nebulae (dashed line) are constructed using the deep spectra (solid line) of Bulge planetary nebulae.  The deep [\ion{O}{3}]$\lambda 5007$ spectrum is scaled to the typical flux for the spectra of extragalactic planetary nebulae, then a background continuum is added.  Here, the deep spectrum of H 1-41 (scaled; solid line) is compared with the synthetic spectrum generated from it.  The ordinate is in units of counts.}
  \label{fig_h141_syn}
\end{figure}

\begin{figure}[!t]
  \includegraphics[width=0.7\columnwidth,angle=90]{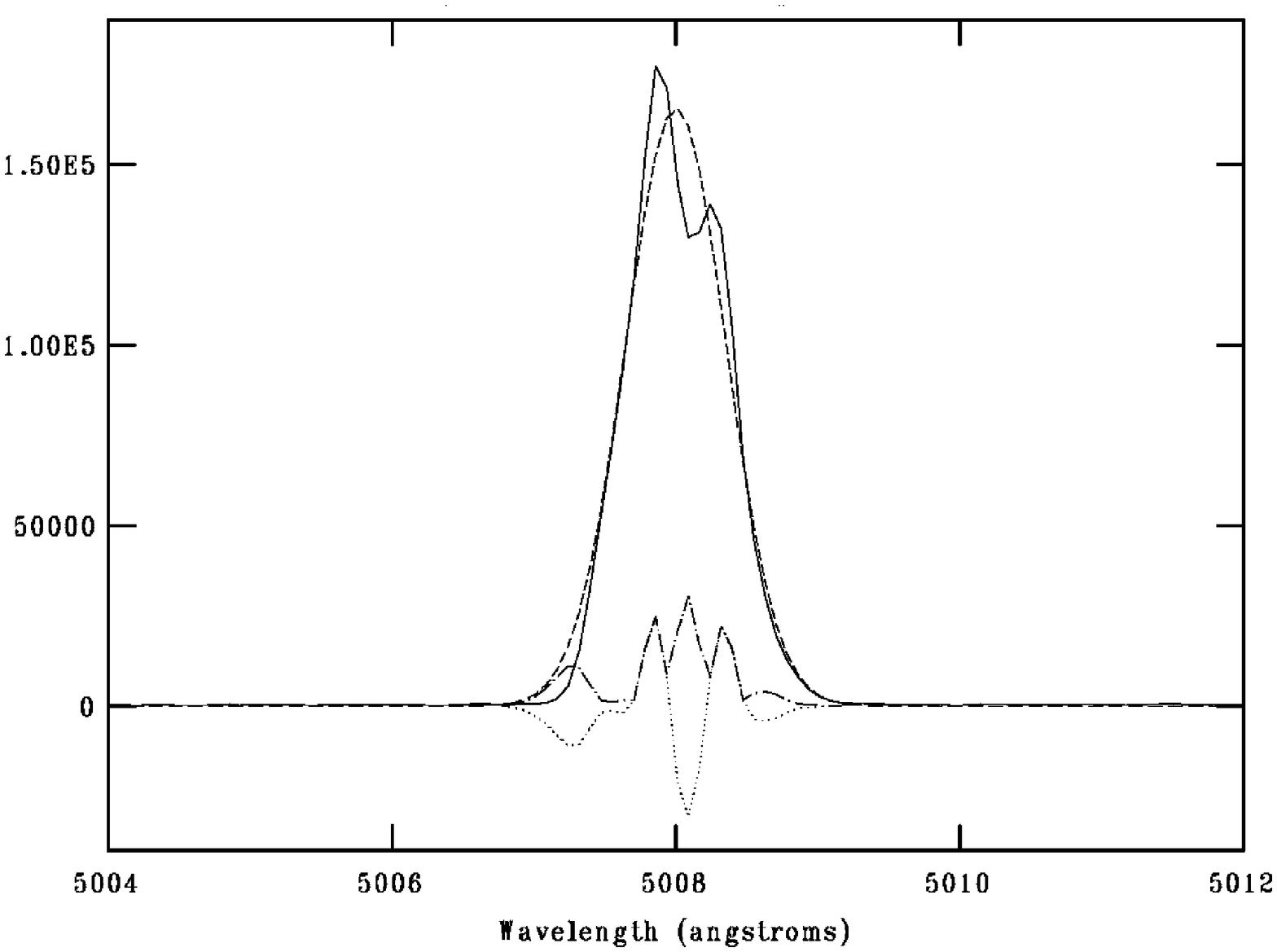}
  \caption{To compute the residual flux, the Gaussian fit from INTENS (dashed line) is subtracted from the original line profile (solid line).  The difference (lower dotted line) is computed and its absolute value (lower dash-dotted line) is determined.  The flux contained in the absolute value curve is summed to obtain the residual flux.  Here, the residual flux represents 12\% of the flux in the Gaussian component.  Again, we use H 1-41 as an example.  The ordinate is in units of counts.}
  \label{fig_resid_h141}
\end{figure}

Figs. \ref{fig_lp_page1}-\ref{fig_lp_page8} present the one-dimensional line profiles for all of the objects in our sample.  For the Bulge planetary nebulae, we present the shallow and deep [\ion{O}{3}]$\lambda 5007$ spectra as well as the H$\alpha$ spectra.  For the extragalactic planetary nebulae, we present the spectra in both [\ion{O}{3}]$\lambda 5007$ and H$\alpha$.   

\section{Analysis of the Line Profiles}\label{sec_analysis}

The line profiles of extragalactic planetary nebulae usually cannot be distinguished statistically from a Gaussian shape \citep[][see also Fig. \ref{fig_lp_page8}]{dopitaetal1985, dopitaetal1988, zijlstraetal2006, arnaboldietal2008}.  While it may be somewhat surprising, the majority of the line profiles for the Bulge planetary nebulae in Figs. \ref{fig_lp_page1}-\ref{fig_lp_page8} are not too different from a Gaussian in shape.  These observations motivate the analysis that follows.  

We analyzed the one-dimensional line profiles with a locally-implemented software package \citep[INTENS;][]{mccalletal1985} to determine the radial velocity, flux, and profile width (FWHM; full width at half maximum intensity) as well as the uncertainties ($1 \sigma$) in these parameters.  This software fits the emission line profile with a sampled Gaussian function and models the continuum as a straight line (see the last panel in Fig. \ref{fig_example_analysis}).  Thus, this analysis assumes that the lines have a Gaussian shape and that they are superposed on a flat continuum.  In the case of the H$\alpha$ line, the \ion{He}{2}\,$\lambda$6560 line may also be present.  In this case, a fit is made simultaneously to both lines and the continuum, but assuming  that the widths of both lines are identical.  

In addition to analysing all of the observed spectra of the Bulge planetary nebulae with INTENS, we also used them to construct synthetic [\ion{O}{3}]$\lambda 5007$ spectra of extragalactic planetary nebulae.  To construct these synthetic spectra, we re-normalized each deep [\ion{O}{3}]$\lambda 5007$ spectrum of our Bulge planetary nebulae to the total flux typically observed in bright extragalactic planetary nebulae, about 5300 counts \citep{richeretal2009}.  We then added a typical extragalactic background continuum spectrum to these re-normalized spectra.  (The background used was that for the Fornax PN, once the emission line was removed.)  We then analysed these synthetic spectra with INTENS in the same way as for the observed spectra.  Figs. \ref{fig_example_analysis} and \ref{fig_h141_syn} demonstrate this transformation and analysis for H~1-67 and H~1-41, respectively.  

We end our analysis of the deep H$\alpha$ and [\ion{O}{3}]$\lambda 5007$ spectra by determining the flux that is not represented by the Gaussian fit.  To obtain this residual flux, we subtract the Gaussian fit from the original line profile, take the absolute value of the residual, and sum the residual over the line.  This procedure measures the flux that deviates from a Gaussian line profile.  Figs. \ref{fig_example_analysis} and \ref{fig_resid_h141} illustrate this procedure for H 1-67 and H 1-41, respectively.  Since extragalactic planetary nebulae have line profiles that are approximately Gaussian in shape \citep[Fig. \ref{fig_lp_page8}, also][]{dopitaetal1985, dopitaetal1988, arnaboldietal2008}, our intention here is to estimate the fraction of the flux in our deep observations of Bulge planetary nebulae that would be missed in observations of extragalactic planetary nebulae.  

Clearly, a Gaussian function will be a poor approximation to the shape of the double-peaked line profiles in Figs. \ref{fig_lp_page1}-\ref{fig_lp_page8}.  However, even for those cases, the FWHM of the Gaussian is similar to the separation of the outer 50\% intensity points for the two peaks, except when the line profiles are very asymmetric.  This would not have been the case had we adopted a width based upon some lower intensity fraction, such as the 10\% that is often adopted \citep[e.g.,][]{dopitaetal1985, dopitaetal1988}.  Since our goal is to retain an analysis as close to the method we would employ for extragalactic planetary nebulae, whose line profiles are nearly Gaussian, we have refrained from applying a double-Gaussian fit to the double-peaked line profiles.  A consequence of our single-Gaussian method is that the residual fluxes will be over-estimated for the profiles that differ strongly from a Gaussian shape.  Even so, the residual fluxes are modest (Fig. \ref{fig_frac_res}).

The results of this analysis for all of the spectra of the Bulge planetary nebulae (observed and synthetic) are given in Tables \ref{table_kin_o3} and \ref{table_kin_ha}.  Both tables present the observed fluxes, observed line widths (FWHM), intrinsic line widths, and, for the deep H$\alpha$ and [\ion{O}{3}]$\lambda 5007$ spectra, the residuals with respect to the Gaussian fit.  For the shallow [\ion{O}{3}]$\lambda 5007$ spectrum and the synthetic extragalactic spectrum, Tables \ref{table_kin_o3}-\ref{table_kin_ha} present the observed fluxes and line widths (FWHM).  With the exception of the residuals with respect to the Gaussian fit, Tables \ref{table_kin_o3} and \ref{table_kin_ha} present the uncertainties in all quantities for each object.  All of these uncertainties are the formal uncertainties (one sigma) from INTENS.  The results for the extragalactic planetary nebulae will be presented elsewhere \citep{richeretal2009}.

In order to derive the intrinsic line widths, the observed line widths must be corrected for several effects that broaden the lines, and all of which are assumed to contribute to the observed line width in quadrature.  The effects that broaden the true, intrinsic profile are instrumental ($\sigma_{inst}$), thermal ($\sigma_{th}$), and fine structure ($\sigma_{fs}$) broadening,

\begin{equation}
\sigma^2_{obs} = \sigma^2_{true} + \sigma^2_{inst} + \sigma^2_{th} + \sigma^2_{fs}\ .\label{eq_broadening}
\end{equation}

\noindent  The first term, $\sigma^2_{true}$, is the true, intrinsic line width resulting from the kinematics of the planetary nebula.  The instrumental profile has a measured FWHM of 2.5-2.7 pixels, for which we adopted  FWHM of 2.6 pixels for all objects ($\sim 11$\,km/s FWHM).  We compute the thermal broadening from the usual formula \citep[][eq. 2-243]{lang1980}, adopting rest wavelengths of 6562.83\AA\ and 5006.85\AA\ for H$\alpha$ and [\ion{O}{3}]$\lambda 5007$, respectively, the electron temperatures available in the literature (preferentially from the [\ion{O}{3}]$\lambda\lambda$4363,5007 lines, but from [\ion{N}{2}]$\lambda\lambda$5755,6584 lines otherwise), and assuming no turbulent velocity.  The resulting thermal broadening (FWHM) at $10^4$\,K amounts to 0.47\AA\ (21.4\,km/s) and 0.089\AA\ (5.3\,km/s) for H$\alpha$ and [\ion{O}{3}]$\lambda 5007$, respectively.  The fine structure broadening, $\sigma_{fs}$, was taken to be 3.199\,km/s (FWHM 7.53\,km/s) for H$\alpha$ and zero for [\ion{O}{3}]$\lambda 5007$ \citep{garciadiazetal2008}.  

The analysis of line broadening in Eq. \ref{eq_broadening} is strictly correct only if all components are Gaussian in shape.  Otherwise, a full component deconvolution should be used.  Only the thermal and fine structure broadening are truly Gaussian.  However, the instrumental profile is only very slightly more square than a Gaussian with the CCD binning used\footnote{Details are available on the observatory website.}, so treating it as Gaussian should not introduce any significant error, particularly in the case of H$\alpha$ where the thermal and fine structure broadening are more important.  The intrinsic line profile for each object, however, may deviate from a Gaussian shape by amounts that vary depending upon the object's structure and kinematics.  
The use of Eq. \ref{eq_broadening}, rather than a full component deconvolution, would be more worrisome were we trying to recover fine details of the line profile or if the lines are intrinsically very narrow.  However, the modest S/N of the line profiles of extragalactic planetary nebulae precludes the reliable recovery of detailed line profiles, justifying the simplicity of Eq. \ref{eq_broadening}.

It is not simple to interpret the resulting FWHM of the intrinsic line width, $\Delta V$, for the Bulge objects (\S \ref{sec_introduction})

\begin{equation}
\Delta V = 2.3556\sigma_{true}\ . 
\end{equation}

\noindent 
The observed intrinsic line width is a luminosity-weighted velocity width for the mass projected within the spectrograph slit, i.e., it represents the spatially-integrated projected outflow velocity of the emitting ions along the line of sight. Note that this velocity is different from the expansion velocity \citep{schonberneretal2005}. 

The intrinsic line width that we measure should typically exceed the luminosity-weighted line width for the entire object.  The spectrograph slit was centered on each Bulge planetary nebula, each of which is resolved.  Therefore, matter near the edges of the objects is excluded from the observations and this matter is likely to have projected velocities similar to the systemic velocity.  Consequently, it is likely that our observations miss some matter at the systemic velocity for each object, so the line profile we measure for the matter included within the slit will be slightly larger than the true luminosity-weighted line width.  The results presented by \citet{gesickizijlstra2000} and \citet{rozasetal2007} support these arguments.  Their simulations of thin, expanding, spherical shells indicate that the line widths we measure may over-estimate the integrated line widths for the entire objects by up to approximately 15\%, but that the exact amount will depend upon the fraction of the object covered by the slit and by the real matter and velocity distributions.  For the extragalactic planetary nebulae that are not resolved (StWr 2-21 in Sagittarius is the only exception), the line width should be similar to the emission-weighted line width, but the real matter or velocity distributions may also affect this somewhat \citep{schonberneretal2005, rozasetal2007}.

\begin{figure*}[!t]
\begin{center}
  \includegraphics[width=2\columnwidth,angle=0]{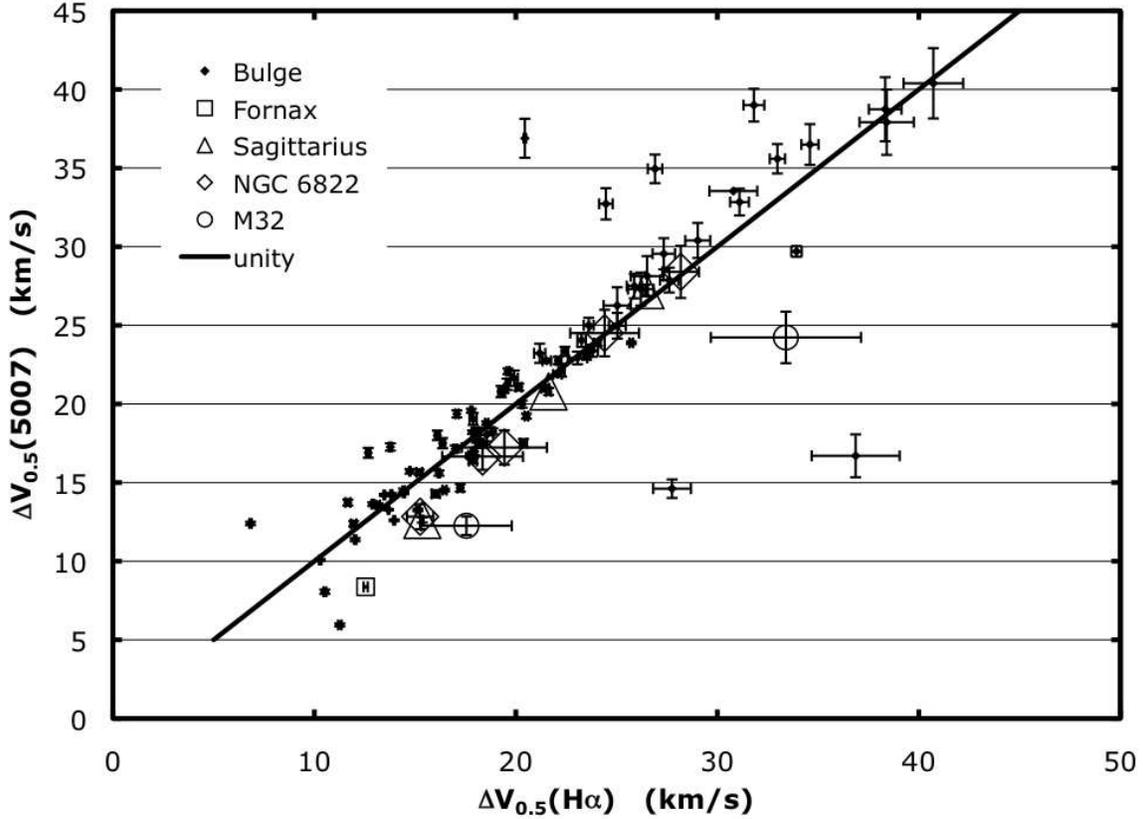}
  \caption{We find an excellent correlation between the intrinsic H$\alpha$ and [\ion{O}{3}]$\lambda 5007$ line widths for both Bulge and extragalactic planetary nebulae.  The solid line indicates the locus of equal line widths.  Therefore, the typical projected outflow velocity deduced from the [\ion{O}{3}]$\lambda 5007$ line accurately reflects that of the ionized mass in these nebulae.  Note that the error bars on the Bulge symbols make these symbols resemble stars when the uncertainties are small.}
  \label{fig_o3vsha}
\end{center}
\end{figure*}

Fortunately, we are not concerned with interpreting the line width in what follows.  However, since it is clear that this line width will be similar to twice the typical projected outflow velocity, Tables \ref{table_kin_o3} and \ref{table_kin_ha} present half of the line width in velocity units for each object, i.e.,   

\begin{equation}
\Delta V_{0.5} = 0.5\Delta V = 1.1778\sigma_{true}, 
\end{equation}  

\noindent which we adopt henceforth as our measure of the kinematics of the zone containing the emitting ion (the entire ionized shell in the case of H$\alpha$).  

\section{Results}\label{sec_results}

The relation between the line widths in H$\alpha$ and [\ion{O}{3}]$\lambda 5007$ is shown in Figure \ref{fig_o3vsha}.  Clearly, there is an excellent correlation in almost all cases.  The two Bulge planetary nebulae for which the H$\alpha$ line width substantially exceeds the [\ion{O}{3}]$\lambda 5007$ line width, M 3-42 and K 5-3, are objects with very asymmetric line profiles for which INTENS fit to only one component of the [\ion{O}{3}]$\lambda 5007$ profile.  The narrower thermal width of the [\ion{O}{3}]$\lambda 5007$ line favours this error.  Note that the shallow [\ion{O}{3}]$\lambda 5007$ spectra have similar intrinsic line widths to those measured in H$\alpha$.  There are also two Bulge planetary nebulae for which the opposite occurs, Cn 1-5 and H 1-59.  The first is one of the few objects for which the H$\alpha$ and [\ion{O}{3}]$\lambda 5007$ spectra were obtained on different nights.  The pointings are slightly different, with the [\ion{O}{3}]$\lambda 5007$ observation slightly better centered, so it is possible that the [\ion{O}{3}]$\lambda 5007$ observation saw considerably more high velocity material.  H 1-59 is a very compact object, so if there was some slight flexure between the H$\alpha$ and [\ion{O}{3}]$\lambda 5007$ observations (the latter was obtained first), the H$\alpha$ observation could have been off-center, which could explain why the H$\alpha$ line width is considerably narrower than the [\ion{O}{3}]$\lambda 5007$ line width.  

The great majority of the Milky Way objects in Fig. \ref{fig_o3vsha} define a tight relationship.  The solid line indicates the locus of identical line widths in H$\alpha$ and [\ion{O}{3}]$\lambda 5007$.  For H$\alpha$ line widths above 20\,km/s, the extragalactic planetary nebulae follow the trend defined by their counterparts in the Bulge.  For narrower line widths, the extragalactic planetary nebulae tend to fall on the low side of the Bulge distribution.  Whether this is a sampling effect from a small sample or a systematic difference is unclear at present.  Overall, however, it appears that the kinematics derived from the [\ion{O}{3}]$\lambda 5007$ line are very representative of the kinematics of the entire ionized mass in these objects, whether galactic or extragalactic.  

\begin{figure}[!t]
  \includegraphics[width=\columnwidth,angle=0]{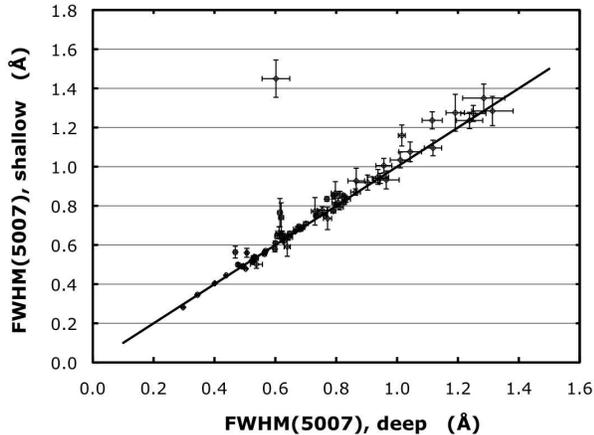}
  \caption{The observed line widths measured in the shallow and deep [\ion{O}{3}]$\lambda 5007$ spectra are in excellent agreement for our Bulge planetary nebulae.  The solid line indicates the locus of identical line widths.  This implies that the observed line width is insensitive to the S/N over the range spanned by our deep and shallow spectra.}
  \label{fig_shalvsdeep}
\end{figure}

\begin{figure}[!t]
  \includegraphics[width=\columnwidth,angle=0]{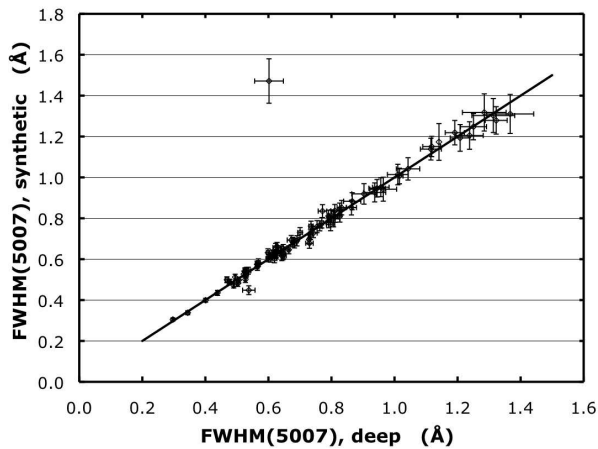}
  \caption{The observed line widths measured in the deep [\ion{O}{3}]$\lambda 5007$ spectra are in excellent agreement with those measured from the synthetic extragalactic spectra derived from them.  Again, the solid line indicates the locus of identical line widths.  The median value of the flux ratio between the deep and synthetic spectra is 110.  This extends the range in S/N over which the observed line width is insensitive to the total flux to the regime occupied by extragalactic planetary nebulae.  Therefore, the line widths measured for extragalactic planetary nebulae should be reliable.}
  \label{fig_synvsdeep}
\end{figure}

Figures \ref{fig_shalvsdeep} and \ref{fig_synvsdeep} explore whether the S/N of the observation affects the observed line width.  In both figures, the solid line is the locus of identical line widths.  In both figures, Cn 1-5 is the lone outlier, for the reasons already discussed.  Clearly, within the range of S/N spanned from the deep [\ion{O}{3}]$\lambda 5007$ spectrum to the shallow and synthetic [\ion{O}{3}]$\lambda 5007$ spectra, the observed line width is not affected.  

\begin{figure}[!t]
  \includegraphics[width=\columnwidth,angle=0]{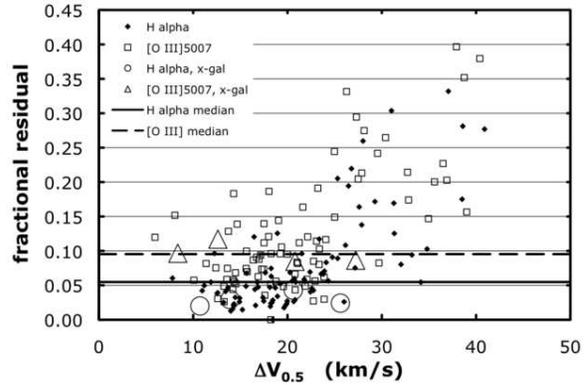}
  \caption{The fraction of the flux that is not accounted for by the Gaussian component is small.  The fraction of residual flux has median values of 5.5\% and 9.5\% for the H$\alpha$ and [\ion{O}{3}]$\lambda 5007$ line profiles, respectively.  The Gaussian component represents more than 75\% of the total flux in 89\% and 95\% of all cases for the [\ion{O}{3}]$\lambda 5007$  and H$\alpha$ lines, respectively.  Therefore, the Gaussian component adequately describes the kinematics of the majority of the ionized mass in these nebulae.}
  \label{fig_frac_res}
\end{figure}

Finally, Fig. \ref{fig_frac_res} presents the residual flux as a fraction of the flux in the Gaussian component as a function of the line width.  There are two fundamental lessons.  First, the median fraction of the residual flux is relatively small, 5.5\% and 9.5\% of the flux in the Gaussian component for the H$\alpha$ and [\ion{O}{3}]$\lambda 5007$ lines, respectively.  Alternatively, the flux represented by the Gaussian component is greater than 75\% of the total flux in 94\% and 89\% of all cases for the H$\alpha$ and [\ion{O}{3}]$\lambda 5007$ lines, respectively.  Second, for line widths exceeding about 25\,km/s, the fraction of the residual flux correlates loosely with the line width, for both the H$\alpha$ and [\ion{O}{3}]$\lambda 5007$ lines.  

\section{Discussion}\label{sec_discussion}

Our sample of Bulge planetary nebulae were selected in a way that we hoped would yield a sample of planetary nebulae whose properties were similar to those of bright extragalactic planetary nebulae in environments without star formation.  Of particular importance are the criteria of large absolute H$\beta$ luminosities and large [\ion{O}{3}]$\lambda 5007/\mathrm H\beta$ ratios, as are observed in bright extragalactic planetary nebulae in such environments \citep{richeretal2008}.  For this sample, \citet{richeretal2008} found that the more evolved objects (with hotter central stars) have systematically larger sizes and line widths as well as lower [\ion{S}{2}] densities and H$\beta$ luminosities.   

We now consider the issues that motivated this study.  How well do the kinematics measured in the [\ion{O}{3}]$\lambda 5007$ line represent the kinematics of the entire ionized shell?  How does the limited S/N of observations of extragalactic planetary nebulae affect measurements of the intrinsic line width?  Finally, does the limited S/N of observations of extragalactic planetary nebulae significantly limit our understanding of their kinematics?    

First, Fig. \ref{fig_o3vsha} indicates that the [\ion{O}{3}]$\lambda 5007$ and H$\alpha$ line widths are very similar.  The comparison of kinematics in H$\alpha$ and [\ion{O}{3}]$\lambda 5007$ is telling, since H$\alpha$ samples all of the ionized mass whereas [\ion{O}{3}]$\lambda 5007$ samples only part of it.  Likely, this near equality of the [\ion{O}{3}]$\lambda 5007$ and H$\alpha$ line widths is a result of our selection criteria. Planetary nebulae that are luminous in [\ion{O}{3}]$\lambda 5007$ are likely to have O$^{2+}$ zones that occupy a large fraction of the total volume occupied by the ionized mass.  It is not so surprising then that the kinematics of the O$^{2+}$ zone are very similar to the kinematics of the entire ionized mass.  Therefore, the line width derived from the [\ion{O}{3}]$\lambda 5007$ line for bright extragalactic planetary nebulae should be an accurate reflection of the line width of the entire ionized shell.  

In Fig. \ref{fig_o3vsha}, the differences between the galactic and extragalactic planetary nebulae are minimal.  Also, though the sample is small, there is no obvious difference between the planetary nebulae in NGC 6822, a star-forming dwarf irregular galaxy, and those in Fornax, Sagittarius, and M32, galaxies that have no current star formation.  Since NGC 6822 is currently forming stars, its bright planetary nebulae could conceivably be derived from more massive progenitor stars than those of the planetary nebulae in the other three galaxies (or the bulge of the Milky Way).  That the planetary nebulae in NGC 6822 are not anomalous compared to those in the other galaxies would seem to imply that the [\ion{O}{3}]$\lambda 5007$ line widths are equally representative of kinematics of the entire ionized mass for planetary nebulae in all galaxies.  

Second, it is clear from Figs. \ref{fig_shalvsdeep} and \ref{fig_synvsdeep} that the line width is not very sensitive to the S/N, at least over the range of S/N probed here.  The flux ratio between our deep and synthetic extragalactic spectra has a median value of 110.  Therefore, typical observations of extragalactic planetary nebulae should have sufficient S/N that the derived line width should be reliable.

Third, a Gaussian profile, characterized by the measured line width, is an adequate description of the kinematics of the great majority of the ionized mass.  The deviations from a Gaussian profile in our deep spectra of planetary nebulae in the Milky Way bulge are small, having median values of 5.5\% and 9.5\% of the total flux in the lines of H$\alpha$ and [\ion{O}{3}]$\lambda 5007$, respectively (Fig. \ref{fig_frac_res}).  If we suppose a uniform temperature throughout each object, these fractions also represent the mass fraction whose kinematics deviate significantly from a Gaussian line profile.  It is not surprising that the H$\alpha$ profiles are more nearly Gaussian since its greater thermal width obscures kinematic detail.  The extragalactic planetary nebulae in Fornax and Sagittarius do not differ from their Milky Way counterparts as respects the fractional residual flux.   

The correlation between the fractional residual flux and the line width is not surprising.  However, both our observational technique or greater kinematic complexity could contribute to the larger residual flux in the objects with larger line widths.  Theoretically, we expect the nebular shells to be accelerated while the central star emits a substantial wind \citep{schonberneretal2007, villaveretal2002}, an effect that has been found for this sample \citep{richeretal2008} and that is also obvious for the three planetary nebulae in the Sagittarius dwarf spheroidal \citep{zijlstraetal2006}.  For this sample, there is a loose correlation between line width and diameter, so, even for a fixed geometry, such as a spherical thin shell, the larger objects would be better resolved by our spectrograph slit (fixed width) and would have line profiles that are less Gaussian \citep[e.g.,][]{gesickizijlstra2000}.  If, furthermore, the more evolved objects are more spatially inhomogeneous, the deviations from a Gaussian profile will be even more pronounced.  Clearly, the fractional residuals we measure for Bulge planetary nebulae should be upper limits to those that would be observed (when feasible) for extragalactic planetary nebulae, i.e., truly spatially unresolved observations should find deviations from a Gaussian profile that are even smaller than those that we observe here.  

The foregoing should not be interpreted as indicating that bright planetary nebulae have simple kinematics.  H~1-67 is a good example of the contrary (Fig. \ref{fig_example_analysis}).  The two-dimensional spectrum clearly presents complex kinematics, but this complexity is not obvious in the spatially-unresolved, one-dimensional spectrum.  A Gaussian is a good description of the great majority of the emission (the fractional residual flux is only 17\% in H$\alpha$; Table \ref{table_kin_ha}).  Nonetheless, the line width describing this Gaussian is not a \emph{complete} description of the kinematics of \emph{all} of the ionized mass, since it does not represent the kinematics of a minority of this mass whose projected motions are more complex.  

Very generally, the above results indicate that the observations that will be available for extragalactic planetary nebulae provide reliable information regarding the typical outflow velocity of their ionized mass.  Indeed, the direct comparison of the kinematics of Bulge and extragalactic planetary nebulae in Fig. \ref{fig_o3vsha} is very good.  We expect that the line profiles for extragalactic planetary nebulae should be close to Gaussian in shape and that the line width measured from spectra of moderate S/N in the [\ion{O}{3}]$\lambda 5007$ line should accurately reflect the typical projected bulk motion of the great majority of the ionized gas in these objects.  It will be difficult to study fine details of the kinematics, such as jets, of extragalactic planetary nebulae.

Finally, we recall that the issue of the detailed interpretation of the spatially unresolved line profiles remains.  This problem is complex and we have not attempted to resolve it here.  It will be most profitably attacked once models are developed that include hydrodynamics and photoionization self-consistently and succeed in reproducing the properties of spatially-resolved data, ideally in an ab initio fashion.  

\section{Conclusions}\label{sec_conclusions}

We have measured line widths for a large sample of planetary nebulae in the Milky Way bulge in the lines of H$\alpha$ and [\ion{O}{3}]$\lambda 5007$ using the Manchester Echelle Spectrograph at the 2.1m telescope of the OAN-SPM.  The selection criteria for this sample were chosen so as to yield a sample of objects whose properties simulate those of bright extragalactic planetary nebulae in environments without star formation \citep{richeretal2008}.  We have also obtained similar data for a small sample of the brightest planetary nebulae in the nearby dwarf galaxies Fornax, Sagittarius, M32, and NGC 6822.  Our goal is to use our high S/N observations of galactic planetary nebulae to simulate observations of extragalactic planetary nebulae and determine what information can be determined reliably when observing the latter.  

Comparing the line widths measured in the H$\alpha$ and [\ion{O}{3}]$\lambda 5007$ lines, we find very similar values.  Therefore, the line width measured for the [\ion{O}{3}]$\lambda 5007$ line is a good reflection of the typical projected outflow velocity of the entire ionized mass.  This result holds for both galactic and extragalactic planetary nebulae.  Next, we find that the line widths for our deep spectra are in excellent agreement with those for our shallow or simulated extragalactic spectra for our Bulge planetary nebulae.  Therefore, the modest S/N expected for observations of extragalactic planetary nebulae should not be an impediment to studying their kinematics.  Finally, we find that departures from a Gaussian shape for the line profiles (deep spectra) of Bulge planetary nebulae are small, typically amounting to 5.5\% and 9.5\% of the total flux in the H$\alpha$ and [\ion{O}{3}]$\lambda 5007$ lines, respectively.  Alternatively, the Gaussian profile is representative of at least 75\% of the emission in 94\% and 89\% of all cases for the H$\alpha$ and [\ion{O}{3}]$\lambda 5007$ lines, respectively.  So, approximating the line shape as Gaussian is an adequate approximation for most of the flux.

Therefore, the Gaussian line profile and its characteristic width provide an adequate description of the typical projected outflow velocity of most of the mass of the entire ionized shell.  This velocity is neither equivalent to the true nebular expansion velocity (the velocity of the outer shock) nor a complete description of the kinematics of all of the ionized mass.  The interpretation of the line profiles in terms of the kinematics of extragalactic planetary nebulae will be very challenging.

Based upon the above results, it is clear that neither the use of the [\ion{O}{3}]$\lambda 5007$ line nor its likely modest S/N is an impediment to the measurement of the kinematics of the ionized mass in extragalactic planetary nebulae.  To the extent that direct comparisons are possible, our results for the bright planetary nebulae in Fornax, Sagittarius, M32, and NGC 6822 do not differ from those for the planetary nebulae in the Milky Way Bulge.  We therefore conclude that the line width measured for the [\ion{O}{3}]$\lambda 5007$ line in bright extragalactic planetary nebulae is an accurate reflection of the typical projected outflow velocity for their entire ionized mass.  These results should therefore be a useful reference for the work that will be necessary to enable the detailed interpretation of the kinematics of extragalactic planetary nebulae.

We gratefully acknowledge financial support during this project from CONACyT grants 37214, 43121, and 49447 and from DGAPA-UNAM grants 108406-2, 108506-2, 112103, and 116908-3.

\clearpage\onecolumn
\begin{landscape}
  \tablecols{9}
\tabcaption{Deep and Shallow [\ion{O}{3}]$\lambda$5007 Spectra \label{table_kin_o3}}

\def\ColumnHeaders{
 & & & \multicolumn{4}{c}{[\ion{O}{3}]$\lambda$5007 deep spectrum} & \multicolumn{2}{c}{[\ion{O}{3}]$\lambda$5007 shallow spectrum} \\
\cmidrule(lr){4-7}\cmidrule(lr){8-9}
   object   & PN G       & Run     &   Flux ($10^3$\,ADU) &   FWHM (\AA)         & $ \Delta V_{0.5}$ (km/s) & Residual  & Flux ($10^3$\,ADU) &   FWHM (\AA)      
}

\begin{longtable}{lllcccccc}
  \toprule
  \ColumnHeaders\\ \midrule
  \endfirsthead
  
  \tabcaptioncontinued
  \toprule
  \ColumnHeaders\\ \midrule
  \endhead
  
  \bottomrule
  \endfoot

Bl 3-13  & 000.9-02.0 & 2006 Jun & $ 268.8	 \pm   3.1 $ & $ 0.6063\pm 0.0079 $ & $ 16.93\pm 0.24 $ & $ 0.088 $ & $ 1 9.75\pm 0.34 $ & $ 0.646 \pm 0.012  $ \\
Cn 1-5   & 002.2-09.4 & 2004 Jun & $ 4809.6 \pm 155.1 $ & $ 1.251 \pm 0.041  $ & $ 36.9 \pm 1.2  $ & $ 0.203 $ & $ 322.43\pm 10.15$ & $ 1.272 \pm 0.041  $ \\
Cn 2-1   & 356.2-04.4 & 2004 Jun & $ 2084.6 \pm  20.3 $ & $ 0.6826\pm 0.0074 $ & $ 19.36\pm 0.22 $ & $ 0.106 $ & $ 390.78\pm 3.67 $ & $ 0.6806\pm 0.0071 $ \\
H 1-1    & 343.4+11.9 & 2004 Jun & $ 778.9	 \pm  26.1 $ & $ 1.238 \pm 0.043  $ & $ 36.5 \pm 1.3  $ & $ 0.227 $ & $ 159.75\pm 5.15 $ & $  1.236\pm 0.042  $ \\
H 1-11   & 002.6+08.2 & 2006 Jun & $ 906.3	 \pm  14.1 $ & $ 0.625 \pm 0.011  $ & $ 17.53\pm 0.32 $ & $ 0.140 $ & $  46.09\pm 0.84 $ & $ 0.642 \pm 0.013  $ \\
H 1-14   & 001.7+05.7 & 2005 Jul & $ 484.3	 \pm  24.6 $ & $ 1.3134\pm  0.068 $ & $ 38.8 \pm 2.0  $ & $ 0.352 $ & $  19.19\pm 1.09 $ & $ 1.284 \pm 0.075  $ \\
H 1-16   & 000.1+04.3 & 2005 May & $ 540.5	 \pm   8.1 $ & $ 0.743 \pm  0.012 $ & $ 21.25\pm 0.36 $ & $ 0.112 $ & $  17.97\pm 0.41 $ & $ 0.761 \pm 0.019  $ \\
H 1-17   & 358.3+03.0 & 2005 Jul & $ 221.2	 \pm   2.4 $ & $ 0.6150\pm 0.0072 $ & $ 17.20\pm 0.22 $ & $ 0.094 $ & $   1.97\pm 0.17 $ & $ 0.766 \pm 0.072  $ \\
H 1-18   & 357.6+02.6 & 2004 Jun & $ 267.5	 \pm   1.8 $ & $ 0.5062\pm 0.0039 $ & $ 13.66\pm 0.12 $ & $ 0.075 $ & $   4.60\pm 0.17 $ & $   0.56\pm 0.023  $ \\
H 1-20   & 358.9+03.2 & 2003 Jun & $ 361.7	 \pm   2.0 $ & $ 0.6905\pm 0.0042 $ & $ 19.53\pm 0.13 $ & $ 0.048 $ & $  38.95\pm 0.54 $ & $ 0.691 \pm 0.011  $ \\
H 1-23   & 357.6+01.7 & 2005 May & $ 264.2	 \pm   5.3 $ & $ 0.754 \pm 0.016  $ & $ 21.62\pm 0.49 $ & $ 0.164 $ & $  23.46\pm 0.49 $ & $ 0.778 \pm 0.018  $ \\
H 1-27   & 005.0+04.4 & 2003 Jun & $ 564.1	 \pm   5.5 $ & $ 0.7004\pm 0.0075 $ & $ 19.94\pm 0.22 $ & $ 0.095 $ & $  60.02\pm 0.64 $ & $ 0.7093\pm 0.0083 $ \\
H 1-30   & 352.0-04.6 & 2006 Jun & $ 334.5	 \pm   2.6 $ & $ 0.6000\pm 0.0052 $ & $ 16.73\pm 0.16 $ & $ 0.091 $ & $  18.36\pm 0.27 $ & $ 0.610 \pm 0.010  $ \\
H 1-31   & 355.1-02.9 & 2005 May & $ 1395.9 \pm   7.3 $ & $ 0.6646\pm 0.0038 $ & $ 18.79\pm 0.11 $ & $ 0.056 $ & $  77.48\pm 0.51 $ & $ 0.6691\pm 0.0048 $ \\
H 1-32   & 355.6-02.7 & 2005 May & $ 1202.3 \pm   8.1 $ & $ 0.4385\pm 0.0033 $ & $ 11.37\pm 0.10 $ & $ 0.082 $ & $ 114.59\pm 0.79 $ & $ 0.4451\pm 0.0035 $ \\
H 1-33   & 355.7-03.0 & 2004 Jun & $ 558.4	 \pm   2.1 $ & $ 0.5218\pm 0.0022 $ & $ 14.17\pm 0.07 $ & $ 0.044 $ & $  57.63\pm 0.34 $ & $ 0.5187\pm 0.0035 $ \\
H 1-40   & 359.7-02.6 & 2005 May & $ 376.0	 \pm   7.5 $ & $ 0.643 \pm  0.014 $ & $ 18.09\pm 0.42 $ & $ 0.187 $ & $  27.95\pm 0.60 $ & $ 0.640 \pm  0.015 $ \\
H 1-41   & 356.7-04.8 & 2006 Jun & $ 1906.4 \pm  30.2 $ & $ 0.832 \pm 0.014  $ & $ 24.05\pm 0.43 $ & $ 0.117 $ & $ 140.35\pm 2.47 $ & $ 0.834 \pm 0.016  $ \\
H 1-42   & 357.2-04.5 & 2006 Jul & $ 128.8	 \pm   1.7 $ & $ 0.5362\pm 0.0080 $ & $ 14.65\pm 0.24 $ & $ 0.139 $ & $ 285.75\pm 3.66 $ & $  0.537\pm 0.0075 $ \\
H 1-45   & 002.0-02.0 & 2005 Jul & $ 249.8	 \pm   2.6 $ & $ 1.016 \pm 0.011  $ & $ 29.70\pm 0.34 $ & $ 0.082 $ & $   5.93\pm 0.26 $ & $ 1.159 \pm 0.054  $ \\
H 1-50   & 358.7-05.2 & 2004 Jun & $ 422.0	 \pm   1.9 $ & $ 0.7291\pm 0.0038 $ & $ 20.82\pm 0.11 $ & $ 0.043 $ & $                $ & $                  $ \\
H 1-54   & 002.1-04.2 & 2007 Aug & $ 981.5	 \pm   4.6 $ & $ 0.5304\pm 0.0028 $ & $ 14.50\pm 0.08 $ & $ 0.052 $ & $  89.04\pm 0.49 $ & $ 0.5394\pm 0.0033 $ \\
H 1-56   & 001.7-04.6 & 2007 Aug & $ 941.9	 \pm   7.1 $ & $ 0.5650\pm 0.0048 $ & $ 15.67\pm 0.14 $ & $ 0.070 $ & $  60.84\pm 0.46 $ & $ 0.5637\pm 0.0047 $ \\
H 1-59   & 003.8-04.3 & 2005 May & $ 387.4	 \pm  11.1 $ & $ 1.115 \pm 0.033  $ & $ 32.76\pm 0.99 $ & $ 0.214 $ & $  10.63\pm 0.36 $ & $ 1.237 \pm 0.044  $ \\
H 1-60   & 004.2-04.3 & 2005 May & $ 775.7	 \pm  13.5 $ & $ 0.672 \pm 0.013  $ & $ 19.04\pm 0.38 $ & $ 0.144 $ & $  38.60\pm 0.68 $ & $ 0.681 \pm  0.013 $ \\
H 1-67   & 009.8-04.6 & 2005 Jul & $ 1324.2           $ & $                  $ & $  0.00\pm 0.00 $ & $ 0.000 $ & $  35.02\pm 1.20 $ & $ 0.506 \pm 0.019  $ \\
H 2-10   & 358.2+03.5 & 2004 Jun & $ 210.7	 \pm   0.9 $ & $ 0.7907\pm 0.0038 $ & $ 22.75\pm 0.11 $ & $ 0.027 $ & $  17.33\pm 0.29 $ & $ 0.851 \pm 0.016  $ \\
H 2-11   & 000.7+04.7 & 2005 Jul & $ 22.7	 \pm   0.2 $ & $ 0.4684\pm 0.0055 $ & $ 12.39\pm 0.16 $ & $ 0.000 $ & $   3.36\pm 0.16 $ & $ 0.564 \pm 0.030  $ \\
H 2-18   & 006.3+04.4 & 2004 Jun & $ 496.7	 \pm  12.1 $ & $  1.118\pm 0.028  $ & $ 32.84\pm 0.85 $ & $ 0.174 $ & $  17.56\pm 0.61 $ & $ 1.096 \pm 0.040  $ \\
Hb 8     & 003.8-17.1 & 2004 Jun & $ 1207.7 \pm   6.4 $ & $ 0.6465\pm 0.0037 $ & $ 18.21\pm 0.11 $ & $ 0.052 $ & $ 271.81\pm 1.22 $ & $ 0.6433\pm 0.0033 $ \\
He 2-250 & 000.7+03.2 & 2003 Jun & $ 63.8	 \pm   1.0 $ & $ 0.797 \pm 0.013  $ & $ 22.98\pm 0.40 $ & $ 0.056 $ & $   8.00\pm 0.51 $ & $  0.865\pm 0.059  $ \\
Hf 2-1   & 355.4-04.0 & 2005 May & $ 1115.8 \pm  58.9 $ & $ 1.285 \pm 0.069  $ & $ 37.9 \pm 2.0  $ & $ 0.397 $ & $  62.17\pm 3.25 $ & $ 1.350 \pm 0.072  $ \\
K 5-1    & 000.4+04.4 & 2006 Jul & $ 53.2	 \pm   0.6 $ & $ 0.6191\pm 0.0078 $ & $ 17.34\pm 0.23 $ & $ 0.074 $ & $   2.16\pm 0.19 $ & $ 0.740 \pm 0.075  $ \\
K 5-11   & 002.3+02.2 & 2006 Jul & $ 24.4	 \pm   0.3 $ & $ 0.6465\pm 0.0087 $ & $ 18.22\pm 0.26 $ & $ 0.000 $ & $                $ & $                  $ \\
K 5-12   & 353.5-03.3 & 2006 Jul & $ 246.0	 \pm   6.1 $ & $ 1.208 \pm 0.031  $ & $ 35.59\pm 0.93 $ & $ 0.200 $ & $                $ & $                  $ \\
K 5-14   & 003.9+02.6 & 2007 Aug & $ 270.4	 \pm   2.1 $ & $ 0.5986\pm 0.0052 $ & $ 16.50\pm 0.16 $ & $ 0.087 $ & $   9.16\pm 0.19 $ & $ 0.579 \pm 0.014  $ \\
K 5-17   & 004.3+02.1 & 2007 Aug & $ 273.1	 \pm   9.2 $ & $ 1.043 \pm 0.037  $ & $ 30.5 \pm 1.1  $ & $ 0.265 $ & $   9.51\pm 0.42 $ & $ 1.076 \pm 0.050  $ \\
K 5-19   & 005.1+02.0 & 2007 Aug & $ 56.0	 \pm   1.4 $ & $ 1.191 \pm 0.030  $ & $ 35.08\pm 0.91 $ & $ 0.147 $ & $   3.49\pm 0.25 $ & $ 1.275 \pm 0.094  $ \\
K 5-20   & 356.8-03.0 & 2007 Aug & $ 75.5	 \pm   1.1 $ & $ 0.639 \pm 0.010  $ & $ 18.02\pm 0.30 $ & $ 0.121 $ & $   2.20\pm 0.16 $ & $ 0.591 \pm 0.049  $ \\
K 5-3    & 002.6+05.5 & 2006 Jul & $ 121.8	 \pm   4.0 $ & $ 0.538 \pm 0.020  $ & $ 14.70\pm 0.59 $ & $ 0.384 $ & $   5.32\pm 0.19 $ & $ 0.501 \pm 0.020  $ \\
K 5-4    & 351.9-01.9 & 2006 Jul & $ 363.3	 \pm   1.1 $ & $ 0.4946\pm 0.0016 $ & $ 13.28\pm 0.05 $ & $ 0.026 $ & $  20.57\pm 0.18 $ & $ 0.4922\pm 0.0049 $ \\
K 5-5    & 001.5+03.6 & 2006 Jul & $ 22.3	 \pm   0.3 $ & $ 0.612 \pm 0.011  $ & $ 17.12\pm 0.31 $ & $ 0.063 $ & $   3.99\pm 0.17 $ & $ 0.662 \pm 0.031  $ \\
K 5-6    & 003.6+04.9 & 2006 Jul & $ 62.1	 \pm   1.8 $ & $ 0.866 \pm 0.027  $ & $ 25.09\pm 0.82 $ & $ 0.245 $ & $   4.73\pm 0.31 $ & $ 0.927 \pm 0.065  $ \\
K 5-7    & 003.1+04.1 & 2006 Jul & $ 61.5	 \pm   1.6 $ & $ 1.323 \pm 0.035  $ & $ 39.1 \pm 1.0  $ & $ 0.157 $ & $                $ & $                  $ \\
K 5-9    & 355.54-1.4 & 2006 Jul & $ 22.2	 \pm   0.9 $ & $ 0.964 \pm 0.043  $ & $ 28.1 \pm 1.3  $ & $ 0.275 $ & $  14.19\pm 0.65 $ & $ 0.932 \pm  0.046 $ \\
M 1-19   & 351.1+04.8 & 2005 May & $ 1142.8 \pm  10.2 $ & $ 0.5640\pm 0.0056 $ & $ 15.56\pm 0.17 $ & $ 0.074 $ & $  48.00\pm 0.46 $ & $ 0.5540\pm 0.0059 $ \\
M 1-20   & 006.1+08.3 & 2004 Jun & $ 888.4	 \pm   8.0 $ & $ 0.2969\pm 0.0031 $ & $  5.97\pm 0.09 $ & $ 0.120 $ & $  59.92\pm 0.62 $ & $ 0.2815\pm 0.0034 $ \\
M 1-29   & 359.1-01.7 & 2004 Jun & $ 1106.0 \pm   8.0 $ & $ 0.7693\pm 0.0060 $ & $ 22.08\pm 0.18 $ & $ 0.059 $ & $  47.62\pm 0.63 $ & $  0.835\pm 0.012  $ \\
M 1-31   & 006.4+02.0 & 2005 Jul & $ 641.6	 \pm   2.9 $ & $ 0.4899\pm 0.0025 $ & $ 13.11\pm 0.07 $ & $ 0.049 $ & $  14.30\pm 0.23 $ & $ 0.4902\pm 0.0088 $ \\
M 1-35   & 003.9-02.3 & 2007 Aug & $ 1456.2 \pm  13.3 $ & $ 0.7336\pm 0.0073 $ & $ 21.03\pm 0.22 $ & $ 0.082 $ & $  50.22\pm 0.48 $ & $ 0.7499\pm 0.0078 $ \\
M 1-42   & 002.7-04.8 & 2003 Jun & $ 1423.3 \pm  10.4 $ & $ 0.6772\pm 0.0055 $ & $ 19.21\pm 0.16 $ & $ 0.096 $ & $ 133.90\pm 1.05 $ & $ 0.6946\pm 0.0061 $ \\
M 1-48   & 013.4-03.9 & 2005 Jul & $ 612.4	 \pm   4.7 $ & $ 0.5301\pm 0.0046 $ & $ 14.45\pm 0.14 $ & $ 0.068 $ & $  34.17\pm 0.40 $ & $ 0.5352\pm 0.0069 $ \\
M 2-13   & 011.1+11.5 & 2006 Jun & $ 925.7	 \pm   4.1 $ & $ 0.4008\pm 0.0020 $ & $ 10.04\pm 0.06 $ & $ 0.057 $ & $  59.89\pm 0.41 $ & $ 0.4043\pm 0.0032 $ \\
M 2-15   & 011.0+06.2 & 2006 Jun & $ 738.5	 \pm  25.6 $ & $ 0.937 \pm  0.035 $ & $ 27.3 \pm 1.0  $ & $ 0.295 $ & $  37.69\pm 1.40 $ & $ 0.948 \pm 0.038  $ \\
M 2-16   & 357.4-03.2 & 2004 Jun & $ 1014.5 \pm   9.1 $ & $ 0.7905\pm 0.0077 $ & $ 22.74\pm 0.23 $ & $ 0.085 $ & $ 107.34\pm 0.93 $ & $ 0.7727\pm 0.0073 $ \\
M 2-20   & 000.4-01.9 & 2006 Jul & $ 862.4	 \pm   2.6 $ & $ 0.8262\pm 0.0027 $ & $ 23.86\pm 0.08 $ & $ 0.030 $ & $  48.77\pm 0.30 $ & $ 0.8514\pm 0.0056 $ \\
M 2-21   & 000.7-02.7 & 2005 May & $ 2083.6 \pm  54.2 $ & $ 0.945 \pm 0.026  $ & $ 27.52\pm 0.78 $ & $ 0.205 $ & $  93.91\pm 2.43 $ & $ 0.940 \pm 0.026  $ \\
M 2-22   & 357.4-04.6 & 2007 Aug & $ 840.5	 \pm  33.5 $ & $ 0.903 \pm 0.039  $ & $ 26.3 \pm 1.2  $ & $ 0.332 $ & $  14.50\pm 0.56 $ & $ 0.919 \pm 0.038  $ \\
M 2-23   & 002.2-02.7 & 2004 Jun & $ 2029.5 \pm  10.5 $ & $ 0.5258\pm 0.0030 $ & $ 14.30\pm 0.09 $ & $ 0.058 $ & $ 161.07\pm 0.96 $ & $ 0.5132\pm 0.0034 $ \\
M 2-26   & 003.6-02.3 & 2006 Jul & $ 204.8	 \pm   6.4 $ & $ 1.011 \pm  0.033 $ & $ 29.55\pm 0.99 $ & $ 0.242 $ & $  12.06\pm 0.45 $ & $ 1.034 \pm 0.040  $ \\
M 2-27   & 359.9-04.5 & 2004 Jun & $ 766.9	 \pm   5.0 $ & $ 0.6230\pm 0.0044 $ & $ 17.46\pm 0.13 $ & $ 0.073 $ & $  84.78\pm 0.65 $ & $ 0.6244\pm 0.0053 $ \\
M 2-29   & 004.0-03.0 & 2004 Jun & $ 596.8	 \pm   4.0 $ & $ 0.4771\pm 0.0036 $ & $ 12.68\pm 0.11 $ & $ 0.075 $ & $  25.44\pm 0.42 $ & $ 0.4996\pm 0.0092 $ \\
M 2-30   & 003.7-04.6 & 2004 Jun & $ 1648.5 \pm  28.2 $ & $ 0.864 \pm 0.016  $ & $ 25.02\pm 0.47 $ & $ 0.149 $ & $ 333.00\pm 5.66 $ & $ 0.871 \pm 0.016  $ \\
M 2-31   & 006.0-03.6 & 2004 Jun & $ 947.8	 \pm   6.4 $ & $ 0.8266\pm 0.0060 $ & $ 23.87\pm 0.18 $ & $ 0.062 $ & $  79.56\pm 1.13 $ & $ 0.820 \pm 0.013  $ \\
M 2-33   & 002.0-06.2 & 2007 Aug & $ 1051.5 \pm  12.1 $ & $ 0.3441\pm 0.0046 $ & $  7.98\pm 0.14 $ & $ 0.152 $ & $ 107.01\pm 1.28 $ & $ 0.3456\pm 0.0048 $ \\
M 2-39   & 008.1-04.7 & 2006 Jul & $ 889.4	 \pm   9.5 $ & $ 0.5258\pm 0.0063 $ & $ 14.31\pm 0.19 $ & $ 0.129 $ & $  58.55\pm 0.73 $ & $ 0.5212\pm 0.0072 $ \\
M 2-4    & 349.8+04.4 & 2007 Aug & $ 934.8	 \pm   5.7 $ & $ 0.5025\pm 0.0035 $ & $ 13.58\pm 0.10 $ & $ 0.059 $ & $ 138.62\pm 0.66 $ & $ 0.4785\pm 0.0026 $ \\
M 2-8    & 352.1+05.1 & 2006 Jun & $ 1367.9 \pm   4.2 $ & $ 0.6275\pm 0.0021 $ & $ 17.61\pm 0.06 $ & $ 0.035 $ & $  83.45\pm 0.63 $ & $ 0.6172\pm 0.0051 $ \\
M 3-10   & 358.2+03.6 & 2004 Jun & $ 2168.0 \pm  18.7 $ & $ 0.7351\pm 0.0069 $ & $ 21.01\pm 0.21 $ & $ 0.084 $ & $                $ & $                  $ \\
M 3-14   & 355.4-02.4 & 2004 Jun & $ 823.5	 \pm   8.1 $ & $ 0.8005\pm 0.0087 $ & $ 23.06\pm 0.26 $ & $ 0.114 $ & $  90.15\pm 0.95 $ & $ 0.8066\pm 0.0095 $ \\
M 3-15   & 006.8+04.1 & 2004 Jun & $ 471.4	 \pm   4.1 $ & $  0.647\pm 0.0062 $ & $ 18.23\pm 0.19 $ & $ 0.096 $ & $  24.53\pm 0.35 $ & $ 0.659 \pm 0.010  $ \\
M 3-16   & 359.1-02.3 & 2005 May & $ 222.6	 \pm   2.8 $ & $ 0.6229\pm 0.0085 $ & $ 17.46\pm 0.25 $ & $ 0.112 $ & $  15.46\pm 0.41 $ & $ 0.652 \pm 0.019  $ \\
M 3-20   & 002.1-02.2 & 2007 Aug & $ 843.4	 \pm   4.6 $ & $ 0.7629\pm 0.0046 $ & $ 21.89\pm 0.14 $ & $ 0.049 $ & $  47.46\pm 0.26 $ & $ 0.7531\pm 0.0045 $ \\
M 3-21   & 355.1-06.9 & 2004 Jun & $ 1346.5 \pm  12.0 $ & $ 0.5682\pm 0.0056 $ & $ 15.69\pm 0.17 $ & $ 0.100 $ & $ 535.43\pm 4.08 $ & $   0.57\pm 0.0049 $ \\
M 3-26   & 004.8-05.0 & 2005 Sep & $ 697.5	           $ & $  1.141           $ & $ 33.54         $ & $ 0.467 $ & $  11.84    1.51 $ & $ 0.1822           $ \\
M 3-32   & 009.4-09.8 & 2005 Jul & $ 1309.1 \pm  19.3 $ & $ 0.811 \pm 0.013  $ & $ 23.37\pm 0.38 $ & $ 0.103 $ & $  87.49\pm 1.32 $ & $ 0.793 \pm 0.013  $ \\
M 3-33   & 009.6-10.6 & 2004 Jun & $ 3092.1 \pm  72.3 $ & $ 0.806 \pm 0.020  $ & $ 23.24\pm 0.61 $ & $ 0.191 $ & $ 107.12\pm 2.61 $ & $ 0.817 \pm 0.021  $ \\
M 3-38   & 356.9+04.4 & 2004 Jun & $ 321.3	 \pm   4.2 $ & $ 0.5289\pm 0.0077 $ & $ 14.41\pm 0.23 $ & $ 0.183 $ & $  15.59\pm 0.26 $ & $ 0.5348\pm 0.0098 $ \\
M 3-42   & 357.5+03.2 & 2003 Jun & $ 88.9	 \pm   6.0 $ & $ 0.602 \pm 0.046  $ & $ 16.7 \pm 1.4  $ & $ 0.000 $ & $  16.25\pm 1.03 $ & $ 1.450 \pm 0.095  $ \\
M 3-45   & 359.7-01.8 & 2005 Jul & $ 243.5	 \pm   2.6 $ & $ 0.8117\pm 0.0095 $ & $ 23.41\pm 0.28 $ & $ 0.080 $ & $  23.46\pm 0.43 $ & $ 0.857 \pm 0.017  $ \\
M 3-54   & 018.6-02.2 & 2006 Jul & $ 387.9	 \pm   9.9 $ & $ 0.957 \pm 0.026  $ & $ 27.89\pm 0.79 $ & $ 0.213 $ & $  17.76\pm 0.61 $ & $ 1.005 \pm 0.037  $ \\
M 4-3    & 357.2+07.4 & 2005 May & $ 946.8	 \pm   4.0 $ & $ 0.5271\pm 0.0025 $ & $ 14.35\pm 0.07 $ & $ 0.055 $ & $  49.79\pm 0.34 $ & $ 0.5323\pm 0.0040 $ \\
M 4-6    & 358.6+01.8 & 2004 Jun & $ 79.6	 \pm   1.3 $ & $ 0.772 \pm 0.014  $ & $ 22.15\pm 0.40 $ & $ 0.120 $ & $   4.53\pm 0.32 $ & $ 0.736 \pm 0.057  $ \\
M 4-7    & 358.5-02.5 & 2006 Jun & $ 65.9	 \pm   1.0 $ & $ 0.730 \pm 0.012  $ & $ 20.86\pm 0.36 $ & $ 0.066 $ & $   3.56\pm 0.30 $ & $ 0.772 \pm 0.070  $ \\
PC 12    & 000.1+17.2 & 2005 May & $ 151.2	 \pm   0.5 $ & $ 0.4749\pm 0.0019 $ & $ 12.61\pm 0.06 $ & $ 0.031 $ & $                $ & $                  $ \\
Te 1580  & 002.6+02.1 & 2007 Aug & $ 67.5	 \pm   3.6 $ & $ 1.367 \pm 0.075  $ & $ 40.4 \pm 2.2  $ & $ 0.380 $ & $                $ & $                  $ \\
\end{longtable}

  \label{table_kin_o3}
  \clearpage
  \tablecols{9}
\tabcaption{Deep H$\alpha$ and Simulated Extragalactic Spectra \label{table_kin_ha}}

\def\ColumnHeaders{
   & & \multicolumn{4}{c}{Deep H$\alpha$ spectrum} 
   & & \multicolumn{2}{c}{[\ion{O}{3}]$\lambda$5007 sim x-gal spectrum} \\
  \cmidrule(lr){3-6}\cmidrule(lr){8-9}
  object   & PN G   & 
  Flux ($10^3$\,ADU)  &   FWHM (\AA)  & $ \Delta V_{0.5}$ (km/s) & Residual  & 
  $I(6560)/I(\mathrm H\alpha)$ & 
  Flux ($10^3$\,ADU) &   FWHM (\AA)   
}

\begin{longtable}{llccccccc}
  \toprule
  \ColumnHeaders\\ \midrule
  \endfirsthead
  
  \tabcaptioncontinued
  \toprule
  \ColumnHeaders\\ \midrule
  \endhead
  
  \bottomrule
  \endfoot

Bl 3-13  & 000.9-02.0 & $ 911.8	\pm  2.4 $ & $ 0.9427\pm 0.0028 $ & $ 16.05\pm 0.06 $ & $ 0.025 $ &                      & $ 5.21\pm 0.16 $ & $ 0.607 \pm 0.021  $ \\
Cn 1-5   & 002.2-09.4 & $ 1899.7	\pm  4.3 $ & $ 1.0384\pm 0.0026 $ & $ 18.87\pm 0.06 $ & $ 0.029 $ &                      & $ 5.45\pm 0.27 $ & $  1.248\pm 0.064  $ \\
Cn 2-1   & 356.2-04.4 & $ 175.5	\pm  0.7 $ & $ 0.9255\pm 0.0039 $ & $ 15.51\pm 0.09 $ & $ 0.027 $ &                      & $ 5.08\pm 0.15 $ & $ 0.678 \pm  0.023 $ \\
H 1-1    & 343.4+11.9 & $ 1291.7	\pm 14.2 $ & $  1.617\pm 0.0186 $ & $ 34.04\pm 0.43 $ & $ 0.103 $ & $ 0.0105\pm 0.0088 $ & $ 5.17\pm 0.28 $ & $ 1.205 \pm 0.067  $ \\
H 1-11   & 002.6+08.2 & $ 1245.2	\pm  7.0 $ & $ 0.9219\pm 0.0057 $ & $ 15.40\pm 0.13 $ & $ 0.054 $ & $ 0.0039\pm 0.0046 $ & $ 5.23\pm 0.15 $ & $ 0.626 \pm 0.019  $ \\
H 1-14   & 001.7+05.7 & $ 664.7	\pm 12.5 $ & $ 1.8033\pm 0.0355 $ & $ 38.62\pm 0.81 $ & $ 0.175 $ & $ 0.0094\pm 0.0153 $ & $ 5.36\pm 0.33 $ & $ 1.303 \pm 0.083  $ \\
H 1-16   & 000.1+04.3 & $ 1447.8	\pm  3.9 $ & $ 1.0257\pm  0.003 $ & $ 18.51\pm 0.07 $ & $ 0.023 $ & $ 0.0100\pm 0.0022 $ & $ 5.19\pm 0.15 $ & $  0.728\pm 0.023  $ \\
H 1-17   & 358.3+03.0 & $ 1022.5	\pm  3.4 $ & $ 0.9635\pm 0.0035 $ & $ 16.67\pm 0.08 $ & $ 0.032 $ &                      & $ 5.29\pm 0.15 $ & $ 0.622 \pm 0.019  $ \\
H 1-18   & 357.6+02.6 & $ 1218.8	\pm  3.5 $ & $ 0.8105\pm 0.0026 $ & $ 11.67\pm 0.06 $ & $ 0.041 $ & $ 0.0003\pm 0.0023 $ & $ 5.09\pm 0.13 $ & $ 0.498 \pm 0.015  $ \\
H 1-20   & 358.9+03.2 & $ 960.1	\pm  2.4 $ & $ 0.9361\pm 0.0025 $ & $ 15.85\pm 0.06 $ & $ 0.019 $ &                      & $ 4.71\pm 0.16 $ & $ 0.690 \pm  0.025 $ \\
H 1-23   & 357.6+01.7 & $ 502.3	\pm  3.6 $ & $ 1.0266\pm  0.008 $ & $ 18.54\pm 0.18 $ & $ 0.069 $ &                      & $ 5.20\pm 0.18 $ & $ 0.760 \pm 0.030  $ \\
H 1-27   & 005.0+04.4 & $ 1264.4	\pm  3.2 $ & $ 1.0238\pm 0.0029 $ & $ 18.48\pm 0.07 $ & $ 0.026 $ &                      & $ 5.23\pm 0.16 $ & $ 0.730 \pm 0.024  $ \\
H 1-30   & 352.0-04.6 & $ 665.5	\pm  2.8 $ & $ 0.9616\pm 0.0044 $ & $ 16.62\pm 0.10 $ & $ 0.047 $ & $ 0.0101\pm 0.0034 $ & $ 5.29\pm 0.15 $ & $ 0.605 \pm 0.018  $ \\
H 1-31   & 355.1-02.9 & $ 904.8	\pm  3.1 $ & $ 0.9932\pm 0.0037 $ & $ 17.56\pm 0.08 $ & $ 0.034 $ & $ 0.0020\pm 0.0028 $ & $ 5.12\pm 0.14 $ & $ 0.646 \pm 0.019  $ \\
H 1-32   & 355.6-02.7 & $ 2318.3	\pm  7.0 $ & $ 0.7685\pm 0.0026 $ & $ 10.08\pm 0.06 $ & $ 0.039 $ &                      & $ 5.09\pm 0.13 $ & $ 0.435 \pm 0.012  $ \\
H 1-33   & 355.7-03.0 & $ 1339.9	\pm  2.7 $ & $ 0.7904\pm 0.0018 $ & $ 10.93\pm 0.04 $ & $ 0.013 $ &                      & $ 5.22\pm 0.13 $ & $  0.527\pm 0.015  $ \\
H 1-40   & 359.7-02.6 & $ 163.5	\pm  2.1 $ & $ 1.0137\pm 0.0142 $ & $ 18.16\pm 0.32 $ & $ 0.126 $ &                      & $ 5.14\pm 0.18 $ & $ 0.616 \pm 0.024  $ \\
H 1-41   & 356.7-04.8 & $ 1655.8	\pm 12.0 $ & $ 1.1607\pm 0.0091 $ & $ 22.30\pm 0.21 $ & $ 0.066 $ & $ 0.0148\pm 0.0058 $ & $ 5.34\pm 0.20 $ & $ 0.852 \pm 0.035  $ \\
H 1-42   & 357.2-04.5 & $ 1558.4	\pm  8.9 $ & $ 0.9351\pm 0.0059 $ & $ 15.81\pm 0.13 $ & $ 0.068 $ &                      & $ 5.20\pm 0.15 $ & $ 0.544 \pm 0.017  $ \\
H 1-45   & 002.0-02.0 & $ 1429.3	\pm  9.7 $ & $ 1.5897\pm 0.0113 $ & $ 33.37\pm 0.26 $ & $ 0.055 $ & $ 0.0223\pm 0.0055 $ & $ 5.27\pm 0.19 $ & $ 1.007 \pm 0.038  $ \\
H 1-50   & 358.7-05.2 & $ 2058.0	\pm  3.3 $ & $ 1.0263\pm 0.0018 $ & $ 18.53\pm 0.04 $ & $ 0.021 $ & $ 0.0074\pm 0.0013 $ & $ 5.09\pm 0.15 $ & $  0.696\pm 0.023  $ \\
H 1-54   & 002.1-04.2 & $ 3220.8	\pm 14.6 $ & $ 0.9029\pm 0.0045 $ & $ 14.81\pm 0.10 $ & $ 0.047 $ &                      & $ 5.36\pm 0.13 $ & $ 0.545 \pm  0.015 $ \\
H 1-56   & 001.7-04.6 & $ 1157.9	\pm  1.6 $ & $ 0.8261\pm 0.0013 $ & $ 12.25\pm 0.03 $ & $ 0.015 $ &                      & $ 5.35\pm 0.15 $ & $ 0.563 \pm 0.018  $ \\
H 1-59   & 003.8-04.3 & $ 302.9	\pm  3.4 $ & $ 1.2366\pm 0.0147 $ & $ 24.33\pm 0.34 $ & $ 0.091 $ & $ 0.0408\pm 0.0092 $ & $ 5.33\pm 0.24 $ & $ 1.138 \pm 0.053  $ \\
H 1-60   & 004.2-04.3 & $ 680.2	\pm  4.9 $ & $  0.956\pm 0.0076 $ & $ 16.45\pm 0.17 $ & $ 0.063 $ &                      & $ 5.29\pm 0.16 $ & $ 0.693 \pm 0.023  $ \\
H 1-67   & 009.8-04.6 & $ 1278.4	\pm 25.0 $ & $ 1.4737\pm 0.0297 $ & $ 30.46\pm 0.68 $ & $ 0.169 $ & $ 0.0395\pm 0.0160 $ & $ 4.27\pm 0.21 $ & $ 0.657 \pm 0.036  $ \\
H 2-10   & 358.2+03.5 & $ 222.9	\pm  1.7 $ & $ 1.1028\pm 0.0093 $ & $ 20.70\pm 0.21 $ & $ 0.055 $ &                      & $ 5.32\pm 0.17 $ & $ 0.811 \pm 0.029  $ \\
H 2-11   & 000.7+04.7 & $ 64.5	\pm  0.4 $ & $ 0.7697\pm 0.0058 $ & $ 10.13\pm 0.13 $ & $ 0.038 $ &                      & $ 5.31\pm 0.14 $ & $ 0.499 \pm 0.014  $ \\
H 2-18   & 006.3+04.4 & $ 717.5	\pm  9.4 $ & $ 1.4733\pm 0.0204 $ & $ 30.44\pm 0.47 $ & $ 0.125 $ & $ 0.0043\pm 0.0105 $ & $ 5.34\pm 0.22 $ & $  1.151\pm 0.050  $ \\
Hb 8     & 003.8-17.1 & $ 1553.2	\pm  3.2 $ & $ 0.9662\pm 0.0022 $ & $ 16.75\pm 0.05 $ & $ 0.024 $ &                      & $ 4.92\pm 0.14 $ & $ 0.617 \pm 0.020  $ \\
He 2-250 & 000.7+03.2 & $ 287.4	\pm  5.1 $ & $ 1.1849\pm 0.0211 $ & $ 22.96\pm 0.48 $ & $ 0.117 $ & $ 0.0313\pm 0.0153 $ & $ 4.94\pm 0.19 $ & $ 0.771 \pm 0.032  $ \\
Hf 2-1   & 355.4-04.0 & $ 759.4	\pm 24.6 $ & $ 1.7739\pm  0.059 $ & $ 37.91\pm 1.35 $ & $ 0.281 $ & $ 0.0876\pm 0.0266 $ & $ 5.23\pm 0.35 $ & $ 1.318 \pm 0.091  $ \\
K 5-1    & 000.4+04.4 & $ 264.7	\pm  0.7 $ & $ 0.8709\pm 0.0027 $ & $ 13.76\pm 0.06 $ & $ 0.021 $ &                      & $ 5.17\pm 0.16 $ & $ 0.602 \pm 0.020  $ \\
K 5-11   & 002.3+02.2 & $ 96.3	\pm  0.5 $ & $ 0.9577\pm 0.0054 $ & $ 16.51\pm 0.12 $ & $ 0.000 $ &                      & $ 5.17\pm 0.15 $ & $ 0.624 \pm 0.020  $ \\
K 5-12   & 353.5-03.3 & $ 255.2	\pm  2.6 $ & $  1.548\pm 0.0166 $ & $ 32.33\pm 0.38 $ & $ 0.094 $ & $ 0.0396\pm 0.0083 $ & $ 4.92\pm 0.26 $ & $ 1.193 \pm 0.065  $ \\
K 5-14   & 003.9+02.6 & $ 155.1	\pm  1.2 $ & $ 1.0314\pm 0.0087 $ & $ 18.69\pm 0.20 $ & $ 0.075 $ & $ 0.0114\pm 0.0063 $ & $ 4.99\pm 0.14 $ & $ 0.632 \pm 0.019  $ \\
K 5-17   & 004.3+02.1 & $ 509.9	\pm  9.1 $ & $ 1.4345\pm 0.0271 $ & $ 29.48\pm 0.62 $ & $ 0.172 $ & $ 0.0120\pm 0.0145 $ & $ 4.60\pm 0.23 $ & $ 1.042 \pm 0.055  $ \\
K 5-19   & 005.1+02.0 & $ 111.5	\pm  1.2 $ & $ 1.4087\pm 0.0157 $ & $ 28.81\pm 0.36 $ & $ 0.075 $ & $ 0.0517\pm 0.0086 $ & $ 5.38\pm 0.25 $ & $ 1.219 \pm 0.060  $ \\
K 5-20   & 356.8-03.0 & $ 161.9	\pm  0.9 $ & $ 0.9118\pm 0.0054 $ & $ 15.09\pm 0.12 $ & $ 0.044 $ &                      & $ 5.34\pm 0.15 $ & $ 0.646 \pm 0.020  $ \\
K 5-3    & 002.6+05.5 & $ 234.7	\pm  6.6 $ & $ 1.3682\pm  0.041 $ & $ 27.77\pm 0.94 $ & $ 0.260 $ & $ 0.0356\pm 0.0227 $ & $ 4.56\pm 0.19 $ & $ 0.448 \pm 0.021  $ \\
K 5-4    & 351.9-01.9 & $ 1422.2	\pm  2.1 $ & $ 0.8187\pm 0.0013 $ & $ 11.97\pm 0.03 $ & $ 0.016 $ &                      & $ 5.15\pm 0.13 $ & $ 0.512 \pm 0.014  $ \\
K 5-5    & 001.5+03.6 & $ 119.9	\pm  0.5 $ & $ 0.9265\pm 0.0047 $ & $ 15.54\pm 0.11 $ & $ 0.025 $ &                      & $ 5.28\pm 0.16 $ & $ 0.625 \pm 0.021  $ \\
K 5-6    & 003.6+04.9 & $ 86.5	\pm  1.1 $ & $ 1.3015\pm 0.0182 $ & $ 26.04\pm 0.42 $ & $ 0.089 $ & $ 0.0245\pm 0.0105 $ & $ 5.25\pm 0.23 $ & $ 0.883 \pm 0.042  $ \\
K 5-7    & 003.1+04.1 & $ 55.6	\pm  0.7 $ & $ 1.6188\pm 0.0227 $ & $ 32.49\pm 0.52 $ & $ 0.080 $ & $ 0.0294\pm 0.0108 $ & $ 5.10\pm 0.26 $ & $ 1.279 \pm 0.067  $ \\
K 5-9    & 355.54-1.4 & $ 156.7	\pm  4.0 $ & $ 1.2893\pm 0.0354 $ & $ 25.72\pm 0.81 $ & $ 0.220 $ & $ 0.0185\pm 0.0207 $ & $ 4.94\pm 0.29 $ & $ 0.942 \pm 0.059  $ \\
M 1-19   & 351.1+04.8 & $ 1619.8	\pm  3.8 $ & $ 0.8781\pm 0.0022 $ & $ 14.00\pm 0.05 $ & $ 0.021 $ &                      & $ 5.22\pm 0.13 $ & $ 0.575 \pm 0.016  $ \\
M 1-20   & 006.1+08.3 & $ 3164.6	\pm 13.7 $ & $ 0.7527\pm 0.0037 $ & $  9.43\pm 0.08 $ & $ 0.055 $ &                      & $ 5.19\pm 0.11 $ & $ 0.3048\pm 0.0073 $ \\
M 1-29   & 359.1-01.7 & $ 2186.2	\pm  6.5 $ & $ 1.0319\pm 0.0033 $ & $ 18.69\pm 0.08 $ & $ 0.027 $ & $ 0.0154\pm 0.0024 $ & $ 5.36\pm 0.16 $ & $ 0.778 \pm 0.025  $ \\
M 1-31   & 006.4+02.0 & $ 2958.7	\pm 14.8 $ & $ 0.7954\pm 0.0044 $ & $ 11.11\pm 0.10 $ & $ 0.049 $ &                      & $ 5.22\pm 0.12 $ & $ 0.471 \pm 0.012  $ \\
M 1-35   & 003.9-02.3 & $ 3276.4	\pm 18.3 $ & $ 1.0215\pm 0.0063 $ & $ 18.40\pm 0.14 $ & $ 0.056 $ &                      & $ 5.42\pm 0.17 $ & $ 0.721 \pm 0.025  $ \\
M 1-42   & 002.7-04.8 & $ 3097.7	\pm 10.2 $ & $ 1.0485\pm 0.0038 $ & $ 19.18\pm 0.09 $ & $ 0.055 $ & $ 0.0056\pm 0.0026 $ & $ 5.31\pm 0.16 $ & $ 0.684 \pm  0.023 $ \\
M 1-48   & 013.4-03.9 & $ 974.1	\pm  2.5 $ & $ 0.8262\pm 0.0024 $ & $ 12.23\pm 0.05 $ & $ 0.026 $ & $ 0.0057\pm 0.0021 $ & $ 5.03\pm 0.14 $ & $ 0.541 \pm 0.017  $ \\
M 2-13   & 011.1+11.5 & $ 1556.5	\pm  5.0 $ & $ 0.7047\pm 0.0025 $ & $  7.25\pm 0.06 $ & $ 0.042 $ &                      & $ 5.24\pm 0.11 $ & $ 0.3992\pm 0.0094 $ \\
M 2-15   & 011.0+06.2 & $ 1080.0	\pm 22.1 $ & $ 1.2705\pm 0.0278 $ & $ 25.23\pm 0.64 $ & $ 0.194 $ & $ 0.0160\pm 0.0164 $ & $ 5.28\pm 0.25 $ & $ 0.927 \pm 0.046  $ \\
M 2-16   & 357.4-03.2 & $ 1508.6	\pm  7.3 $ & $  1.109\pm 0.0057 $ & $ 20.87\pm 0.13 $ & $ 0.041 $ & $ 0.0093\pm 0.0040 $ & $ 5.22\pm 0.16 $ & $ 0.780 \pm 0.026  $ \\
M 2-20   & 000.4-01.9 & $ 1417.2	\pm  4.3 $ & $ 1.2428\pm 0.0041 $ & $ 24.50\pm 0.09 $ & $ 0.026 $ &                      & $ 5.33\pm 0.17 $ & $  0.845\pm 0.023  $ \\
M 2-21   & 000.7-02.7 & $ 1497.2	\pm 18.2 $ & $ 1.2804\pm 0.0164 $ & $ 25.49\pm 0.37 $ & $ 0.108 $ & $ 0.0148\pm 0.0098 $ & $ 5.02\pm 0.21 $ & $ 0.947 \pm 0.043  $ \\
M 2-22   & 357.4-04.6 & $ 481.5	\pm 10.7 $ & $ 1.2341\pm 0.0295 $ & $ 24.27\pm 0.67 $ & $ 0.205 $ & $ 0.0223\pm 0.0179 $ & $ 5.37\pm 0.27 $ & $ 0.919 \pm 0.050  $ \\
M 2-23   & 002.2-02.7 & $ 2511.1	\pm  5.7 $ & $  0.855\pm 0.0021 $ & $ 13.22\pm 0.05 $ & $ 0.023 $ &                      & $ 4.91\pm 0.14 $ & $ 0.539 \pm 0.017  $ \\
M 2-26   & 003.6-02.3 & $ 543.1	\pm  9.4 $ & $ 1.3191\pm 0.0244 $ & $ 26.50\pm 0.56 $ & $ 0.164 $ & $ 0.0196\pm 0.0140 $ & $ 5.28\pm 0.25 $ & $  1.014\pm 0.050  $ \\
M 2-27   & 359.9-04.5 & $ 2440.0	\pm  7.8 $ & $ 0.9625\pm 0.0034 $ & $ 16.64\pm 0.08 $ & $ 0.046 $ &                      & $ 4.96\pm 0.14 $ & $  0.624\pm  0.020 $ \\
M 2-29   & 004.0-03.0 & $ 2212.0	\pm  6.6 $ & $ 0.7864\pm 0.0026 $ & $ 10.77\pm 0.06 $ & $ 0.060 $ &                      & $ 5.27\pm 0.13 $ & $ 0.489 \pm 0.013  $ \\
M 2-30   & 003.7-04.6 & $ 2473.0	\pm 21.5 $ & $ 1.1838\pm 0.0108 $ & $ 22.91\pm 0.25 $ & $ 0.071 $ & $ 0.0261\pm 0.0069 $ & $ 4.82\pm 0.19 $ & $ 0.853 \pm 0.036  $ \\
M 2-31   & 006.0-03.6 & $ 2982.6	\pm 15.8 $ & $ 1.1907\pm 0.0068 $ & $ 23.10\pm 0.16 $ & $ 0.056 $ &                      & $ 4.92\pm 0.16 $ & $ 0.811 \pm 0.029  $ \\
M 2-33   & 002.0-06.2 & $ 1398.1	\pm  5.6 $ & $ 0.6988\pm 0.0031 $ & $  6.98\pm 0.07 $ & $ 0.050 $ &                      & $ 5.18\pm 0.12 $ & $ 0.3379\pm 0.0092 $ \\
M 2-39   & 008.1-04.7 & $ 2037.2	\pm  9.5 $ & $ 1.0341\pm 0.0053 $ & $ 18.76\pm 0.12 $ & $ 0.096 $ &                      & $ 5.31\pm 0.14 $ & $ 0.542 \pm 0.016  $ \\
M 2-4    & 349.8+04.4 & $ 1660.4	\pm  7.6 $ & $ 0.8088\pm 0.0041 $ & $ 11.63\pm 0.09 $ & $ 0.047 $ &                      & $ 5.03\pm 0.13 $ & $ 0.480 \pm 0.014  $ \\
M 2-8    & 352.1+05.1 & $ 2316.7	\pm  4.2 $ & $ 0.9742\pm 0.0019 $ & $ 17.00\pm 0.04 $ & $ 0.028 $ & $ 0.0104\pm 0.0015 $ & $ 5.34\pm 0.15 $ & $ 0.662 \pm  0.020 $ \\
M 3-10   & 358.2+03.6 & $ 4224.1	\pm 31.0 $ & $ 1.0963\pm 0.0085 $ & $ 20.52\pm 0.19 $ & $ 0.059 $ & $ 0.0094\pm 0.0061 $ & $ 5.40\pm 0.16 $ & $ 0.762 \pm 0.024  $ \\
M 3-14   & 355.4-02.4 & $ 1998.8	\pm 12.1 $ & $ 1.1585\pm 0.0075 $ & $ 22.23\pm 0.17 $ & $ 0.066 $ & $ 0.0089\pm 0.0050 $ & $ 5.24\pm 0.18 $ & $ 0.811 \pm 0.030  $ \\
M 3-15   & 006.8+04.1 & $ 1305.8	\pm  3.9 $ & $ 0.9783\pm 0.0033 $ & $ 17.12\pm 0.08 $ & $ 0.052 $ &                      & $ 5.30\pm 0.16 $ & $ 0.651 \pm 0.021  $ \\
M 3-16   & 359.1-02.3 & $ 548.6	\pm  3.6 $ & $ 1.0279\pm 0.0074 $ & $ 18.58\pm 0.17 $ & $ 0.060 $ &                      & $ 5.40\pm 0.15 $ & $  0.657\pm 0.021  $ \\
M 3-20   & 002.1-02.2 & $ 848.9	\pm  6.5 $ & $ 1.1206\pm 0.0093 $ & $ 21.20\pm 0.21 $ & $ 0.069 $ &                      & $ 5.27\pm 0.15 $ & $  0.763\pm 0.025  $ \\
M 3-21   & 355.1-06.9 & $ 450.3	\pm  2.8 $ & $ 0.8936\pm 0.0062 $ & $ 14.49\pm 0.14 $ & $ 0.068 $ &                      & $ 5.18\pm 0.14 $ & $ 0.584 \pm 0.018  $ \\
M 3-26   & 004.8-05.0 & $ 928.6	\pm 31.3 $ & $ 1.4618\pm 0.0519 $ & $ 30.16\pm 1.19 $ & $ 0.304 $ & $ 0.0202\pm 0.0271 $ & $ 5.35\pm 0.40 $ & $ 1.173 \pm 0.090  $ \\
M 3-32   & 009.4-09.8 & $ 3166.9	\pm 29.4 $ & $ 1.1757\pm 0.0118 $ & $ 22.70\pm 0.27 $ & $ 0.083 $ & $ 0.0103\pm 0.0075 $ & $ 5.32\pm 0.18 $ & $ 0.837 \pm 0.031  $ \\
M 3-33   & 009.6-10.6 & $ 2168.7	\pm 22.1 $ & $ 1.0977\pm 0.0122 $ & $ 20.55\pm 0.28 $ & $ 0.096 $ & $ 0.0149\pm 0.0082 $ & $ 4.92\pm 0.23 $ & $ 0.798 \pm 0.040  $ \\
M 3-38   & 356.9+04.4 & $ 442.8	\pm  3.7 $ & $ 0.9558\pm 0.0086 $ & $ 16.44\pm 0.20 $ & $ 0.120 $ & $ 0.0108\pm 0.0068 $ & $ 5.00\pm 0.15 $ & $ 0.526 \pm 0.017  $ \\
M 3-42   & 357.5+03.2 & $ 240.8  \pm 14.6 $ & $ 1.7103\pm 0.0955 $ & $ 36.35\pm 2.18 $ & $ 0.332 $ & $ 0.0384\pm 0.0562 $ & $ 5.11\pm 0.36 $ & $ 1.47  \pm 0.11   $ \\
M 3-45   & 359.7-01.8 & $ 422.8  \pm  2.3 $ & $ 1.1658\pm 0.0068 $ & $ 22.43\pm 0.16 $ & $ 0.043 $ & $ 0.0197\pm 0.0044 $ & $ 5.19\pm 0.15 $ & $ 0.806 \pm 0.025  $ \\
M 3-54   & 018.6-02.2 & $ 728.4	\pm 10.3 $ & $ 1.3395\pm   0.02 $ & $ 27.03\pm 0.46 $ & $ 0.138 $ & $ 0.0315\pm 0.0114 $ & $ 4.84\pm 0.22 $ & $ 0.952 \pm 0.046  $ \\
M 4-3    & 357.2+07.4 & $ 1918.1	\pm  4.6 $ & $ 0.8502\pm 0.0022 $ & $ 13.07\pm 0.05 $ & $ 0.031 $ &                      & $ 5.06\pm 0.13 $ & $ 0.501 \pm 0.015  $ \\
M 4-6    & 358.6+01.8 & $ 255.3	\pm  1.4 $ & $ 1.1311\pm 0.0067 $ & $ 21.48\pm 0.15 $ & $ 0.046 $ &                      & $ 5.21\pm 0.18 $ & $ 0.835 \pm 0.032  $ \\
M 4-7    & 358.5-02.5 & $ 175.2	\pm  0.9 $ & $ 1.0558\pm  0.006 $ & $ 19.38\pm 0.14 $ & $ 0.017 $ &                      & $ 4.86\pm 0.16 $ & $ 0.678 \pm 0.025  $ \\
PC 12    & 000.1+17.2 & $ 1820.5	\pm  3.7 $ & $ 0.8346\pm 0.0019 $ & $ 12.53\pm 0.04 $ & $ 0.033 $ &                      & $ 5.24\pm 0.12 $ & $  0.487\pm 0.013  $ \\
Te 1580  & 002.6+02.1 & $ 140.9	\pm  4.6 $ & $ 1.8833\pm 0.0647 $ & $ 40.58\pm 1.48 $ & $ 0.277 $ & $ 0.0136\pm 0.0269 $ & $ 4.97\pm 0.35 $ & $ 1.310 \pm 0.095  $ \\
\end{longtable}

  \label{table_kin_ha}
\end{landscape}

\end{document}